\documentclass{aa}

\usepackage{orcidlink}
\usepackage[varg]{txfonts}
\usepackage{svg}
\usepackage{hyperref}
\hypersetup{colorlinks, allcolors=blue}
\usepackage{graphicx,amsmath}
\usepackage{color}
\usepackage{xcolor}
\usepackage{txfonts}
\usepackage{lipsum}
\usepackage{subcaption}         
\usepackage{lscape}             
\usepackage{placeins}          
                                
\usepackage{float}

\newcommand{\Msun}{{\rm\,M_\odot}}
\newcommand{\kpc}{{\rm\,kpc}}
\newcommand{\Gyr}{{\rm\,Gyr}}
\newcommand{\kms}{{\rm\,km\,s^{-1}}}
\newcommand{\Fsum}{{F_{\rm{sum}}}}
\newcommand{\Ftmax}{{F_{2,\rm{max}}}}

\newcommand{\epsdm}{\epsilon_{\rm{DM}}}
\newcommand{\epsstar}{\epsilon_{\star}}
\newcommand{\ndm}{N_{\rm{DM}}}
\newcommand{\nstar}{N_{\star}}
\newcommand{\mdm}{m_{\rm{DM}}}
\newcommand{\mstar}{m_{\star}}
\newcommand{\mratio}{m_{\rm{DM}}/m_\star}

\begin{document} 

   \title{Effects of resolution and local stability on galactic disks:}
   \subtitle{Halo resolution and softening on bar formation}

    \author{SungWon Kwak \inst{\ref{aip}}\orcidlink{0000-0003-0957-6201},
            Ivan Minchev \inst{\ref{aip}}\orcidlink{0000-0002-5627-0355},
            Matthias Steinmetz \inst{\ref{aip},\ref{uniP}}\orcidlink{0000-0001-6516-7459},
            \and 
            Sukyoung K. Yi  \inst{\ref{yonsei}}\orcidlink{0000-0002-4556-2619} 
            }

    \institute{Leibniz-Instit\"ut f\"ur Astrophysik Potsdam (AIP), An der Sternwarte 16, 14482, Potsdam, Germany\label{aip} \\
    \email{skwak@aip.de}
    \and Universit\"at Potsdam, Institut f\"ur Physik und Astronomie, Karl-Liebknecht-Str. 24-25, 14476, Potsdam, Germany \label{uniP}
    \and Department of Astronomy and Yonsei University Observatory, Yonsei University, Seoul, 03722, Republic of Korea \label{yonsei}}

   \date{\today}

  \abstract{
Using N-body simulations, we examined the impact of dark matter (DM) halo resolution and gravitational softening on bar formation. We generated isolated disk-halo systems with fixed stellar disk parameters, varying the number of halo particles, softening lengths, and halo concentration to modulate disk stability via the central DM\ fraction. The effects of DM resolution ($\mratio=1$, 10, and 100) on bar formation are less pronounced in more unstable disks, in which the overall evolutionary path is similar except that the lowest DM resolution model suffers gradual bar weakening. Irrespective of the halo resolution, large softening, $\epsdm$, flattens the central halo density profile within the softening scale, impeding angular momentum transfer to the nascent bar and preventing bar formation in more stable models. In unstable models with $\epsdm=0.96 \, \kpc$, a small bar still emerges due to enhanced initial instability and a larger seed perturbation, yet the bar strength $F_2$ remains capped around $0.3$ owing to unresolved central dynamical friction. Despite the destabilizing effect of the reduced central DM fractions, our results indicate that deficient central angular momentum exchange can still suppress bar growth. Furthermore, halo softening influences buckling instability, as larger values ($\epsdm=0.30$ and $0.60 \, \kpc$) inhibit central vertical heating, exacerbating radial-vertical velocity dispersion anisotropies and triggering stronger buckling. We recommend adopting $  \mratio \le 10  $, $  \mstar \leq 10^4 \, \Msun  $, and $  \epsdm < 0.30\,\mathrm{kpc}$ when investigating the formation and evolution of non-axisymmetric structures in Milky Way-mass galaxies. 
}

   \titlerunning{Halo resolution and softening on bar formation}
   \authorrunning{Kwak et al.}
   \maketitle
   \nolinenumbers
%
%-------------------------------------------------------------------

\section{Introduction}\label{sec:intro}
Numerical effects, including particle numbers and gravitational softening lengths, have long posed challenges in galaxy simulations by introducing artifacts that bias outcomes beyond physical effects \citep{athanassoula86, white88, hernquist90a, pfenniger93, romeo94, merritt96, weinberg96, romeo97, romeo98, derijcke19}. To reduce the numerical noise, different techniques (e.g., wavelet) have been proposed \citep{romeo03,romeo04}. Among early studies on two-body relaxation, \cite{steinmetz97} demonstrated that discreteness effects from massive dark matter (DM) particles cause spurious heating of baryons. It overwhelms radiative cooling when DM particle masses exceed critical thresholds and leads to artificial gas expansion or suppressed cooling. Later, \cite{moore98} and \cite{fukushige01} showed in the simulations of DM halo formation that larger softening values produce shallower central densities, emphasizing the need for sufficient resolution to preserve cuspy profiles instead of flattened cores. Similarly, \cite{power03} stressed that convergence in halo profiles requires precise tuning of the gravitational softening, time steps, force accuracy, initial redshifts, and particle counts. Poor choices yield artificially low central densities or, in some cases, spurious cusps. Gravitational softening imposes a characteristic acceleration limit below which the local mean acceleration must lie, the time step must resolve the local orbital timescale, and enough particles must be enclosed that the two-body relaxation timescale exceeds the age of the Universe. \cite{rodionov05} also conducted convergence tests on optimal softening criteria and suggested that softening lengths should be 1.5--2 times smaller than interparticle distances in dense regions to minimize irregular forces and maintain equilibrium in collisionless models such as Plummer or Hernquist spheres.

Recent studies have demonstrated how numerical artifacts, especially from low mass resolution and high DM-to-baryon particle mass ratios, significantly affect galaxy morphologies and properties in idealized and cosmological simulations. For example, energy equipartition in setups with ratios $>1$ causes spurious kinetic energy transfer from DM to baryons, artificially enlarging galaxy sizes \citep{ludlow19b}. Related research on stellar disk heating demonstrates that collisional effects from coarse DM particles raise vertical and radial velocity dispersions, thickening and expanding disks across all radii, with severity inversely proportional to DM particle count and pronounced for halos with $\lesssim 10^6$ particles \citep{ludlow21}. Here, spurious dispersions can surpass $10\%$ of the halo's virial velocity over a Hubble time, thickening stellar disks in Milky Way-mass galaxies. In full cosmological contexts, these heating artifacts produce resolution-dependent divergences in kinematics and structures, yielding hotter stellar disks, colder halos, elevated central DM densities, and oversized galaxies, while underresolving dwarf halos suppresses substructure and mergers \citep{ludlow23}. Hence, even in the modern simulation era, the numerical setup must be assigned carefully.

Considering all these numerical effects, conducting large-scale cosmological simulations with star formation and feedback to form realistic galaxies remains challenging, particularly for the formation of non-axisymmetric structures and their evolution over the Hubble time \citep{grand17,pillepich18a,pillepich18b,pillepich19,libeskin20,dubois21,ansar25}. Among them, the \textsc{NewHorizon} simulation \citep{dubois21} may be exposed to greater numerical effects as their DM particle mass is about 92 times more massive than baryonic particles. In fact, the bar instability in disks is sensitive to galaxy size, mass, thickness, and kinematics \citep{fall80,efstathiou82, athanassoula03, klypin09, kwak17}, while those properties are sensitive to the DM resolution effects. Moreover, gravitational softening may affect the angular momentum exchange, while the history of angular momentum exchange via DM-star dynamical interactions directly influences the evolution of bar properties such as bar strength, length, pattern speed, buckling instability, and overall morphology \citep{athanassoula02, athanassoula03,athanassoula14, berentzen06, frosst24, frosst26, jang24, jang25,kwak17,kwak19, lokas14, lokas16, lokas19b, lokas19,  lokas25a, lokas25b, martinez04, martinez06}.

Indeed, \cite{kaufmann07} showed that angular momentum transport, disk morphology, and radial profiles depend critically on force and mass resolution, with large softening suppressing bar instability. Such excessive softening ($\epsilon=2\,\kpc$ for both baryon and DM particles) fails to resolve the central region, where angular momentum transport plays a pivotal role in forming a bar. As \cite{bekki23} demonstrated, a weak seed bar ($F_2 < 0.1$) initially forms and subsequently grows through apsidal precession synchronization (APS). This seed forms more rapidly under conditions of higher disk mass or strong tidal perturbations. However, adopting large softening lengths and lower DM resolution is expected to reduce angular momentum transfer, especially during the early growth phase when the bar is small as originating from the very center. In \cite{kwak25} (Paper I), we emphasized the role of central angular momentum exchange in mediating the transition from the spiral to the bar phase. In a fixed potential halo, in which the angular momentum exchange is absent, this transition does not occur, thereby suppressing bar formation in models with the same disk stability. Thus, poorly resolving central dynamics between DM and stars with a low number of particles and a large softening length is expected to alter the bar instability, as observed in the \textsc{NewHorizon} simulation \citep{reddish22}.

In this paper, we investigate the effects of resolution and gravitational softening length in the DM halo on bar formation and evolution in idealized disk-halo models. All models employed identical stellar disk parameters, allowing us to isolate their numerical influences. By varying the DM concentration, we compared the relative impacts of different disk stabilities on bar formation and long-term growth. Furthermore, by examining numerical effects on vertical heating, we assessed how buckling instability strength varies with resolution and softening. In Sect. 2, we list our numerical setup, including the disk-halo models, halo structure, and $N$-body code. Section 3 presents an overview of bar evolution through resolution and softening—both in combination and separately—and visualizes the occurrence of buckling instability under different softening lengths. In Sect. 4, we discuss the underlying physical mechanisms, linking our findings to similar studies and cosmological simulations. We conclude Sect. 4 with recommendations on appropriate options for resolution and softening.

\section{Numerical setup}\label{sec:method}

\begin{table}[h!]
\caption{\label{ic} Initial conditions.}
\centering
\begin{tabular}{lccccc}
\hline\hline
Model & $c$  & $\ndm$ & $\mratio$ & $\epsdm$ \\
 & & & & [kpc] \\
\hline
r1c16        & 16 & 1.14e7 & 10 & 0.03  \\
r1c16ldm     & 16 & 1.14e6 & 100 & 0.03  \\
r1c16ldmsf06 & 16 & 1.14e6 & 100 & 0.60  \\
r1c16ldmsf1  & 16 & 1.14e6 & 100 & 0.96  \\
r1c16sf1     & 16 & 1.14e7 & 10 & 0.96  \\
\hline
r1c14        & 14 & 1.14e7 & 10 & 0.03   \\
r1c14ldm     & 14 & 1.14e6 & 100 & 0.03   \\
r1c14ldmsf06 & 14 & 1.14e6 & 100 & 0.60   \\
r1c14ldmsf1  & 14 & 1.14e6 & 100 & 0.96   \\
\hline
\end{tabular}
\tablefoot{Column (1) lists the model name. Column (2) lists the halo concentration. Columns (3) and (4) list the total number of DM halo particles and the mass ratio between the DM and
star particle. The last column lists the gravitational softening for the DM particles.}
\label{table:model}
\end{table}

\begin{figure*}[htbp]
\centering
\renewcommand{\arraystretch}{0}
\begin{tabular}{@{}c@{\hspace{0pt}}c@{\hspace{0pt}}c@{\hspace{0pt}}c@{\hspace{0pt}}c@{}}
  \rotatebox{90}{\footnotesize \qquad\qquad\qquad\quad y [kpc]} &
  \includegraphics[width=0.3\textwidth]{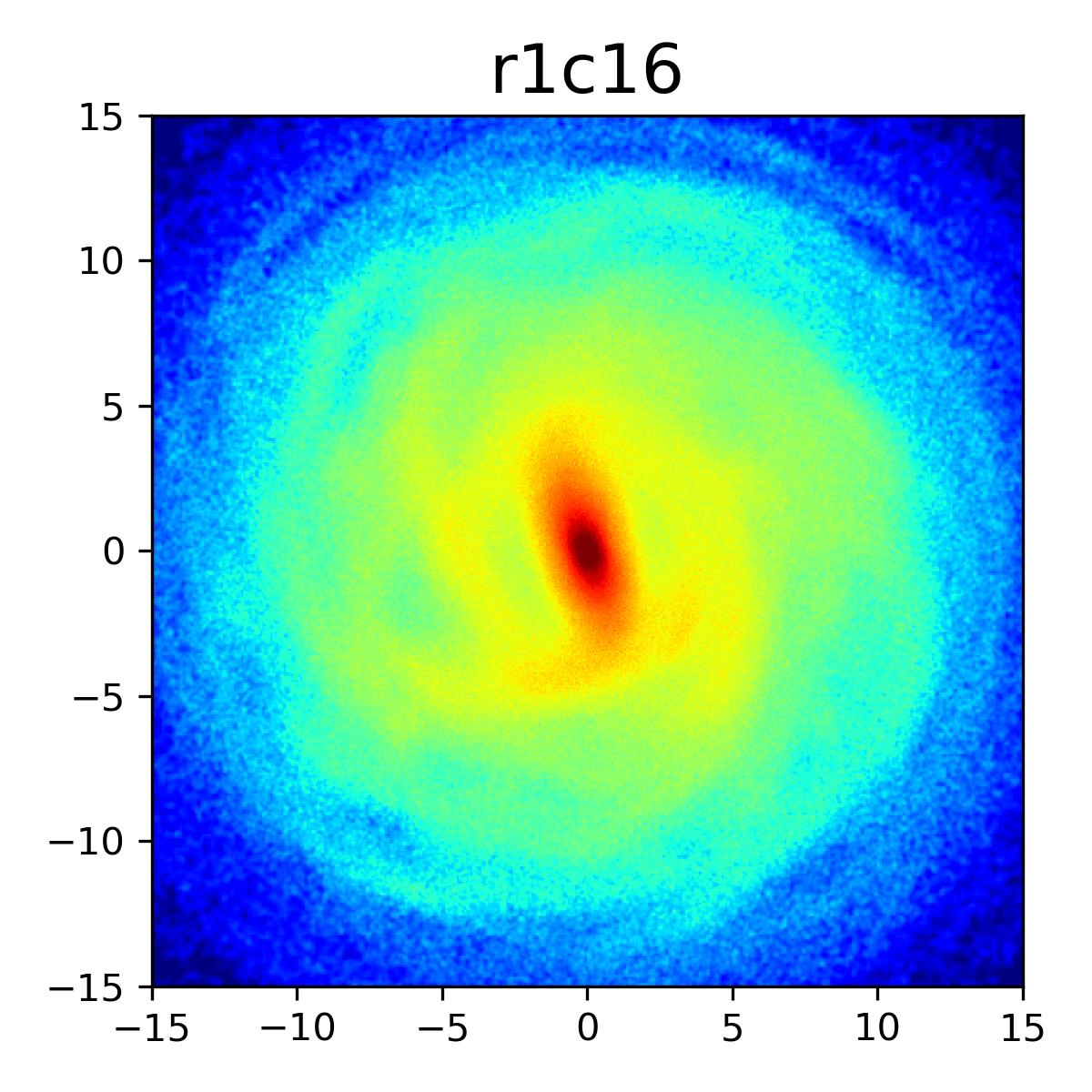} &
  \includegraphics[width=0.3\textwidth]{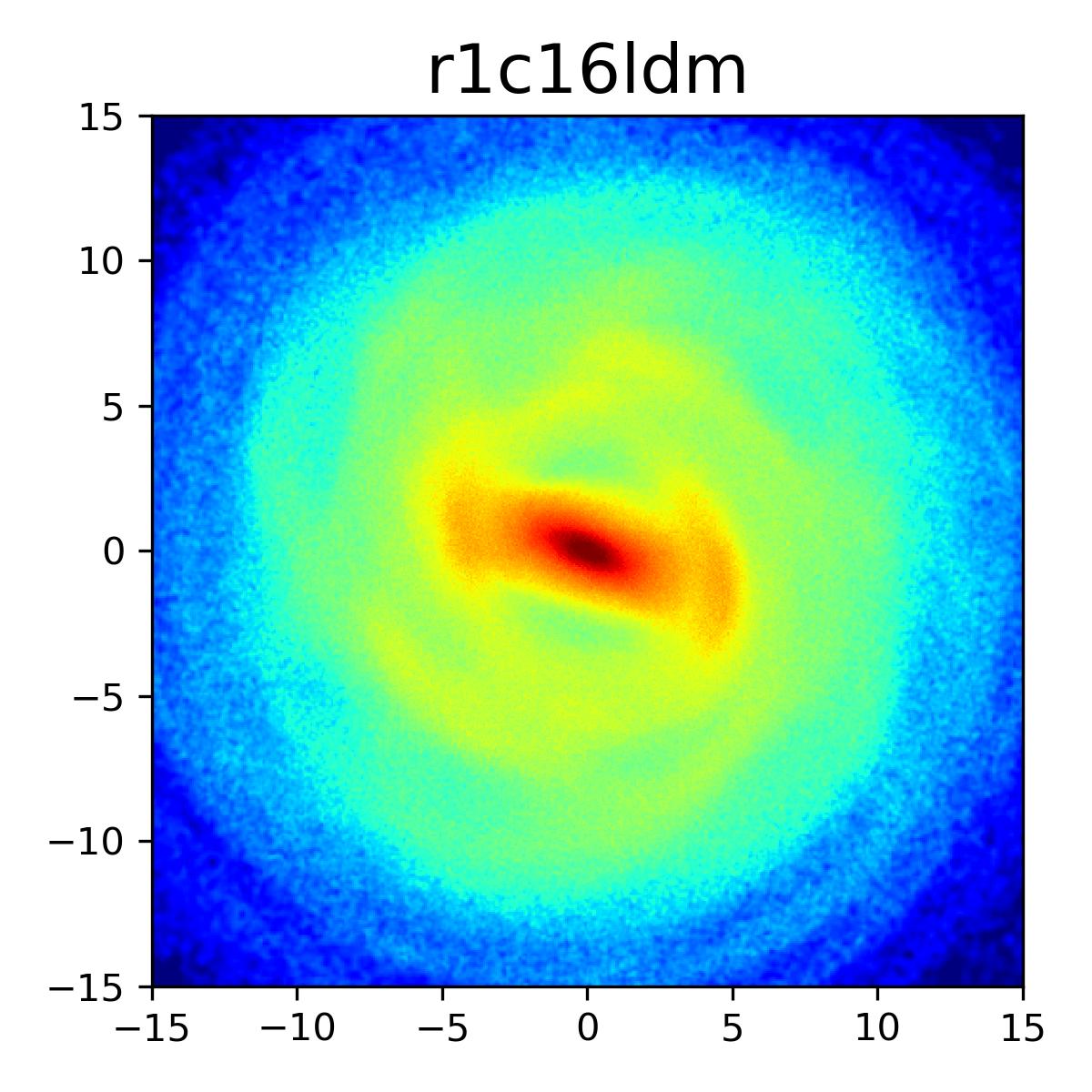} &
  \includegraphics[width=0.3\textwidth]{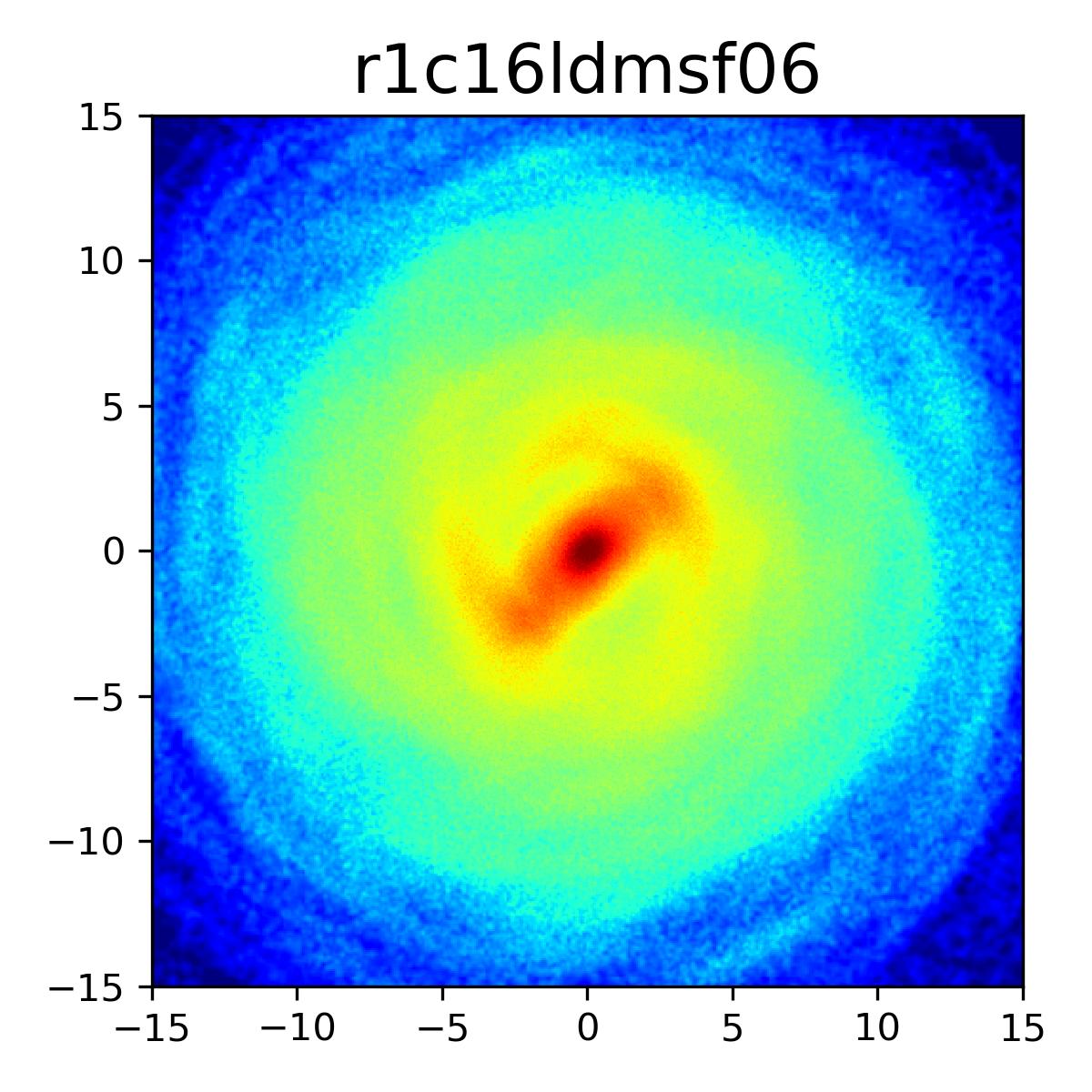} & \\[0pt]

  \rotatebox{90}{\footnotesize \qquad\qquad\qquad\quad y [kpc]} &
  \includegraphics[width=0.3\textwidth]{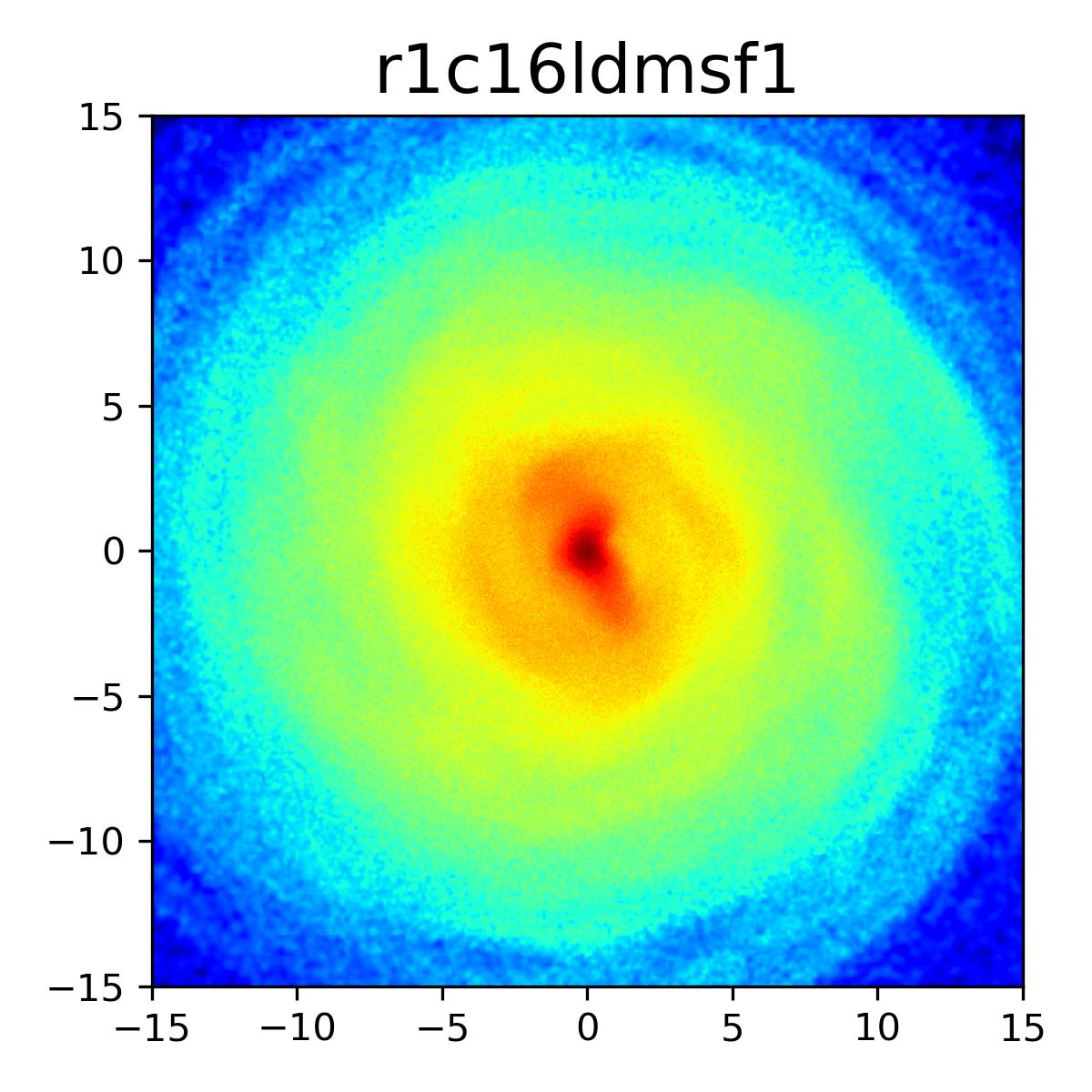} &
  \includegraphics[width=0.3\textwidth]{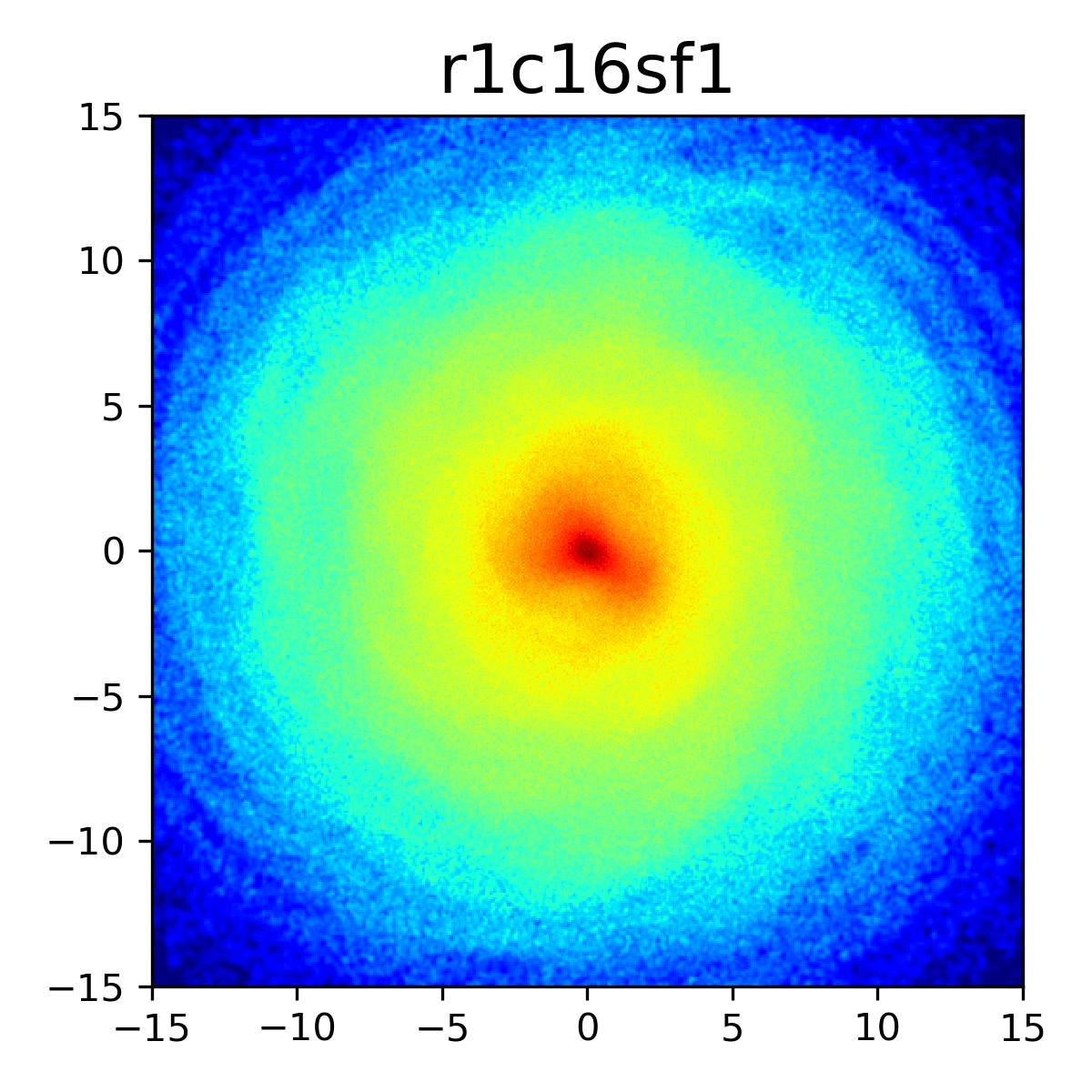} &
  \includegraphics[width=0.3\textwidth]{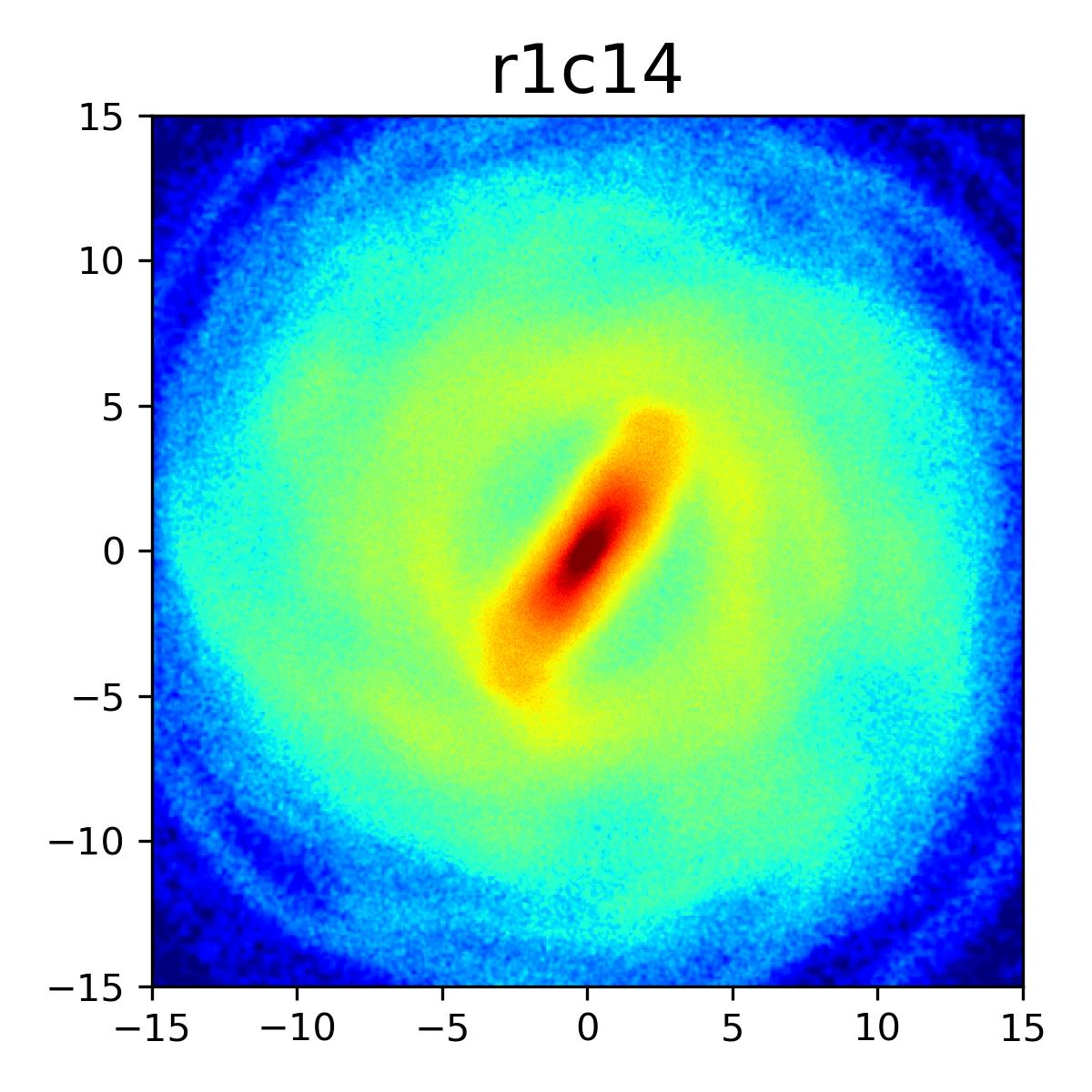} & \\[0pt]

  \rotatebox{90}{\footnotesize \qquad\qquad\qquad\quad y [kpc]} &
  \includegraphics[width=0.3\textwidth]{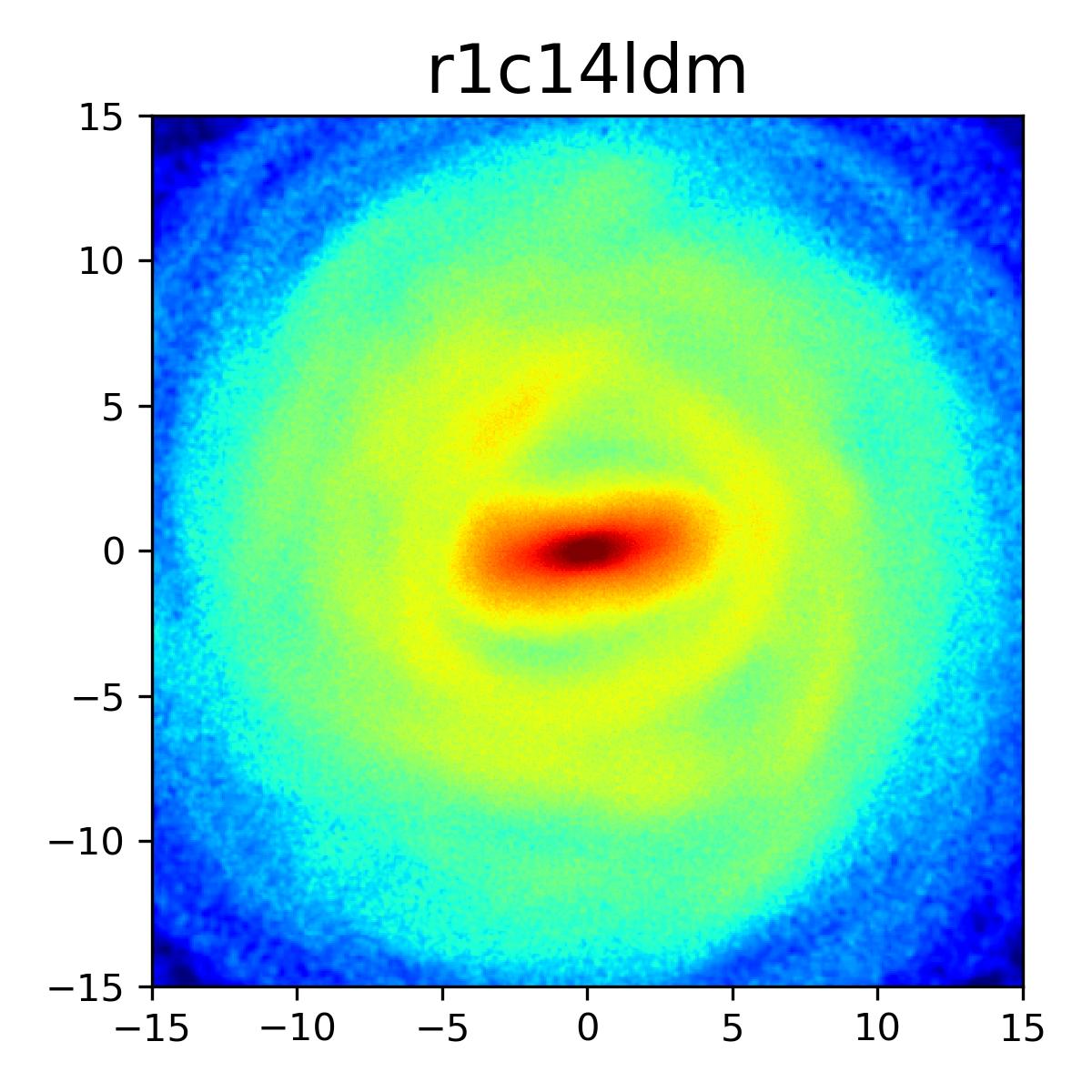} &
  \includegraphics[width=0.3\textwidth]{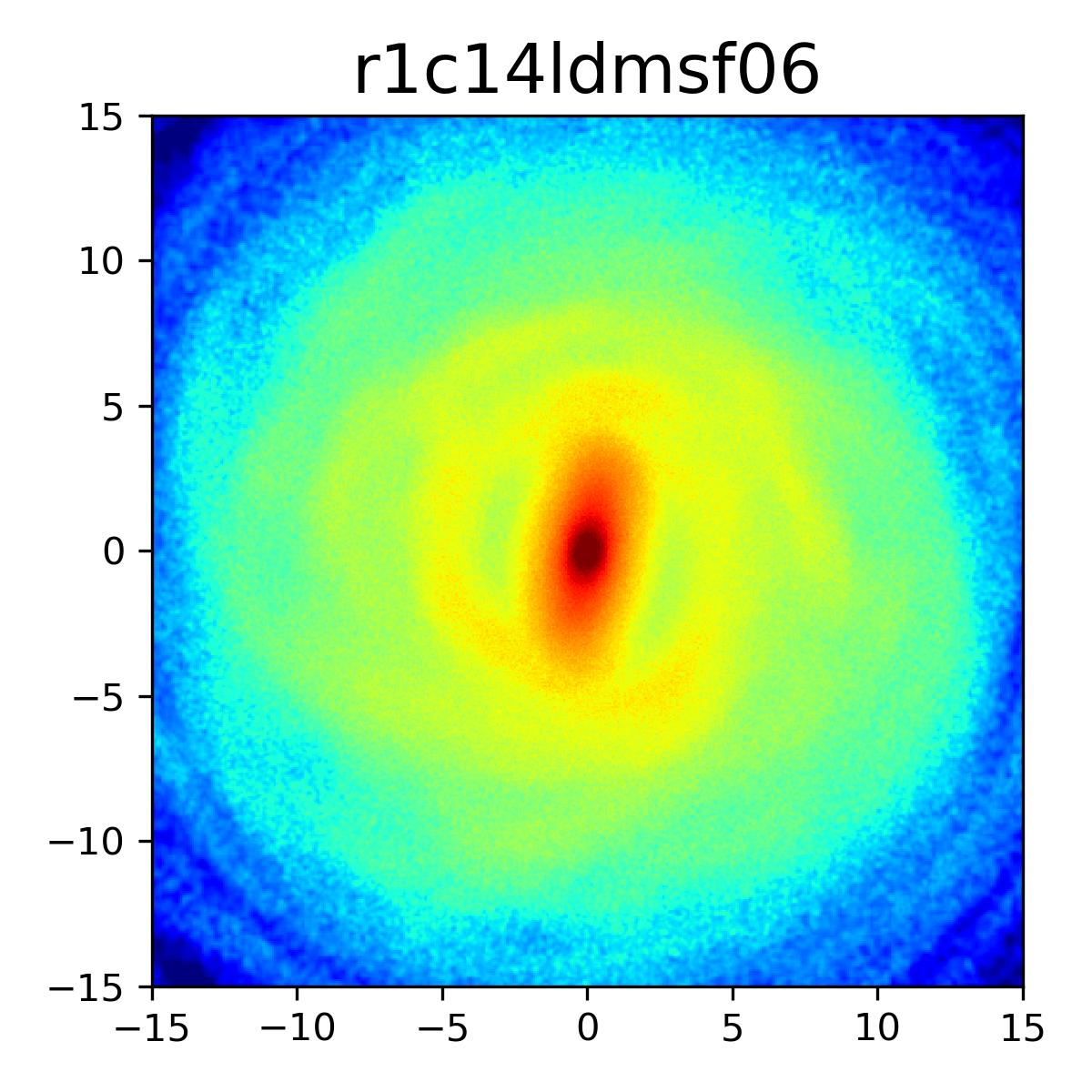} &
  \includegraphics[width=0.3\textwidth]{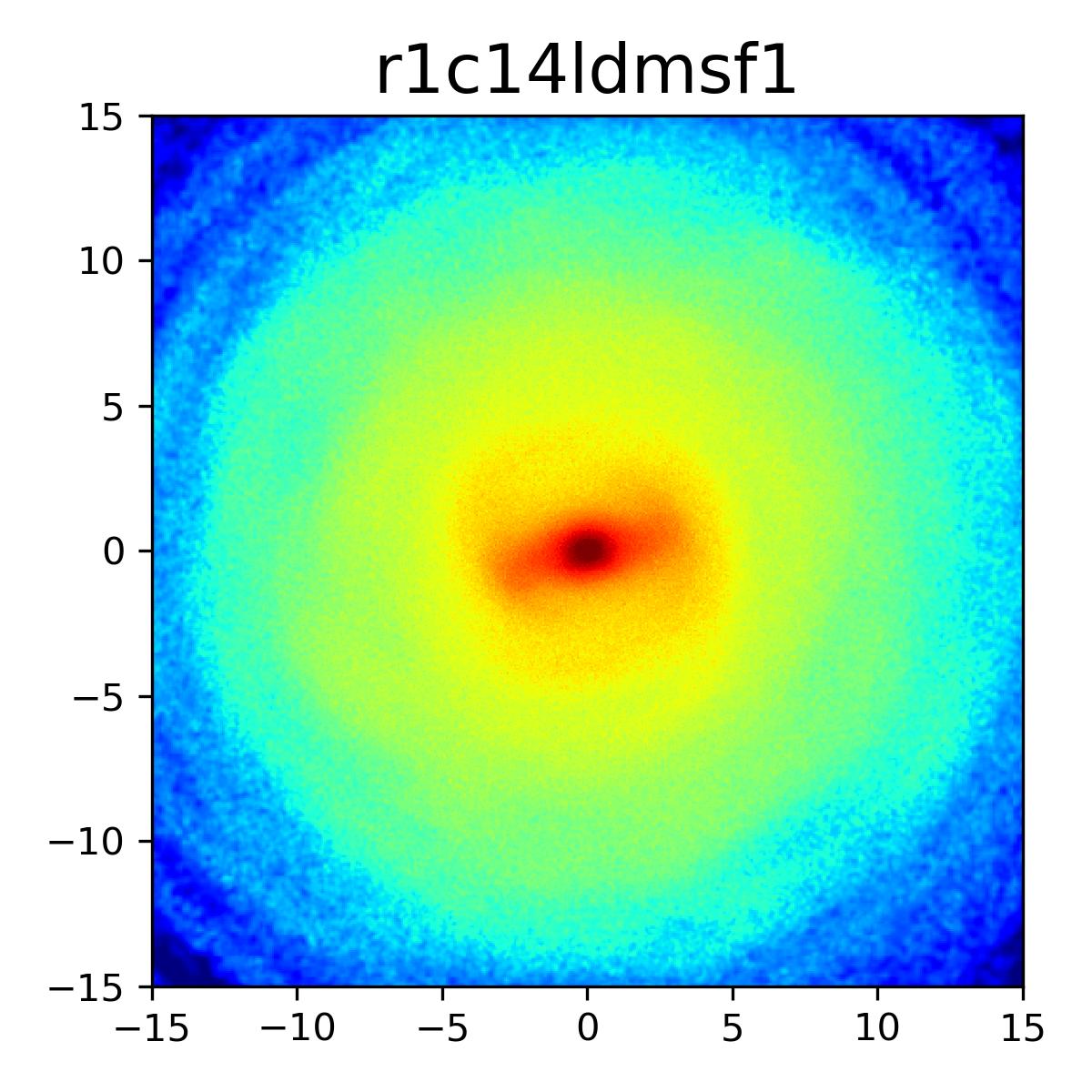} &
  \includegraphics[height=0.3\textwidth]{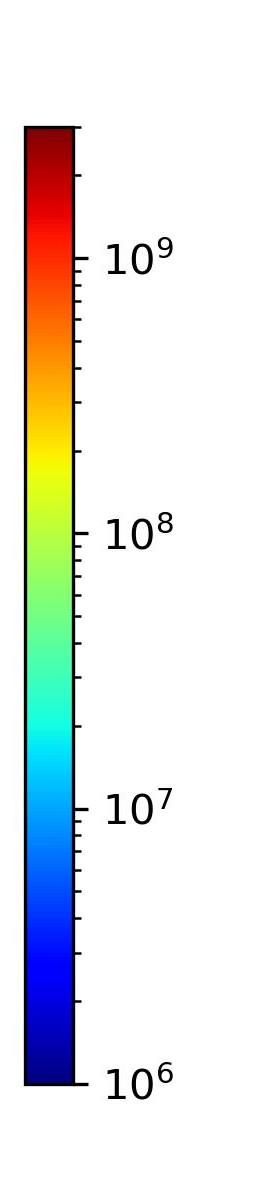} \\[0pt]

  & \footnotesize \quad x [kpc] & \footnotesize \quad x [kpc] & \footnotesize \quad x [kpc] &
\end{tabular}

\caption{Face-on projections of the stellar surface density distribution in a $30 \times 30$ kpc box at 4 Gyr for all models. The color bar indicates the surface density in units of solar masses per square kiloparsec, $\Msun\,\mathrm{kpc}^{-2}$.}
\label{fig:face}
\end{figure*}

We constructed the initial conditions in the same manner as in Paper I using the \textsc{GALIC} code \citep{yurin14}. Our galaxy models consist of a disk-halo system in isolation. The stellar disk follows the distribution
\begin{equation}\label{eq:disk}\rho_{\star} (R, z) = \frac{M_{d}}{4\pi z_{d} R_{d}^2} \exp \left( -\frac{R}{R_{d}} \right) \text{sech}^2 \left(\frac{z}{z_{d}}\right),
\end{equation}
where $R_{d}$ is the scale length, $z_{d}$ is the vertical scale height, and $M_{d}$ is the total mass of the stellar disk in cylindrical coordinates.
In all models, the parameters for the stellar disk are the same: $M_{d} = 5 \times 10^{10} \, \Msun$, $R_d = 3 \, \kpc$, and $z_d = 0.3 \, \kpc$. The resolution for the stellar disk was fixed, with $N_{\star} = 5 \times 10^6$ particles, each having a mass of $m_{\star} = 10^4 \, \Msun$. The softening length of the stellar component $\epsstar$ is 0.03 $\kpc$, which is the mean particle separation in the disk within the half-mass radius.

The DM halo follows a \cite{hernquist90} profile in spherical coordinates
\begin{equation}\label{eq:hernquist}
\rho_\text{DM} (r) = \frac{M_\text{DM}}{2\pi} \frac{a}{r(r+a)^3},
\end{equation}
where $a$ is the scale length of the halo and $M_\text{DM}$ is the total mass.
The total mass was fixed at $M_\text{DM} = 1.14 \times 10^{12} \, \Msun$ in all models.
The scale length of the DM halo varies with the concentration parameter $c$ as
\begin{equation}\label{eq:haloc}
a = \frac{r_{200}}{c} \left[2 \ln(1+c)-\frac{c}{1+c}\right]^{1/2},
\end{equation}
where $r_{200}$ is the virial radius \citep{springel05}.
We adopted two halo concentration parameters, $c=14$ and 16, that fall within the reasonable range from Aquarius and TNG simulations \citep{springel08,bose19}. The scale lengths $a$ are 22.88 and 20.68 kpc, respectively. Although we used the same parameters for the stellar disk, varying the halo concentration changes the central fraction of the DM halo, which affects the bar instability of disk galaxies \citep{kwak17,kwak19, zhou20,jang23}. Hence, our models were imposed with two different levels of stability. According to the Toomre $Q$ parameter \citep{toomre64},
\begin{equation}
Q = \frac{\kappa \sigma_R}{3.36\,G\,\Sigma_\star},
\label{eq:Toomre}
\end{equation} 
the corresponding minimum $Q$ values are 0.830 and 0.875 for the "c14" and "c16" models, respectively.

The initial conditions for our isolated disk-halo systems were generated with the GALIC code \citep{yurin14}. GALIC constructs N-body realizations in approximate collisionless equilibrium by an iterative method. Particle positions were sampled directly from the prescribed density profiles of the stellar disk and DM halo. The radial velocity dispersion profile $  \sigma_R(R)  $ was obtained by solving the axisymmetric Jeans equations for a chosen anisotropy parameter $f_R = \sigma_R^2 / \sigma_z^2$ (we adopt $  f_R = 1  $ in this work), which in turn set the desired radial profile of the Toomre stability parameter $  Q  $ of the stellar disk. Initial velocities were drawn from local Gaussian distributions consistent with these second moments and were subsequently refined iteratively until the time-averaged density response matched the target configuration to within the required tolerance.

The parameters of our models are listed in Table \ref{table:model}. In each set of our galaxy models with different stability, we varied the resolution and the gravitational softening length of the DM halo.
The "ldm" models contain $\ndm = 1.14 \times 10^6$ with $\mratio = 100$, while the rest contain $\ndm = 1.14 \times 10^7$ with $\mratio = 10$. 
For such galaxies with low DM resolution, large softening was often adopted.
We therefore evolved two additional models with $\epsdm = 0.60 \, \kpc$ and $0.96 \, \kpc$ for the "c14" and "c16" models. 
The value of $0.96 \, \kpc$ was derived from the mean particle separation within the effective radius of the DM halo in model r1c16ldm, while $0.60 \, \kpc$ was derived from the mean particle separation within $27 \, \kpc$ ($9 R_d$).
We denote models with $\epsdm=0.96 \, \kpc$ and $0.60 \, \kpc$ as "sf1" and "sf06," respectively. For example, model r1c16ldmsf1 includes a DM halo with $\ndm = 1.14 \times 10^6$, $\epsdm=0.96 \, \kpc$, and $c=16$.  
In the models without "sf1" or "sf06" denotation, the same softening length was assigned for both the disk and halo $\epsdm=0.03 \, \kpc$. In Paper I, we studied two stellar disk resolutions ("r1" and "r2") with $  N_\star = 5 \times 10^6  $ and $  5 \times 10^7  $ particles, respectively, whereas in this paper we focus on the "r1" models while varying the DM halo resolution.

To examine the softening effects separately, we applied the same large softening in the higher resolution halo in the r1c16sf1 model with $\epsdm = 0.96 \, \kpc$. 
Additionally, the effects of $\epsdm=0.30 \kpc$ and $\mratio=1$ are compared in Appendix \ref{appendix:additional} after evolving model r1c14ldmsf03 and r1c14hdm. The "hdm" stands for the high DM resolution, which contains $1.14\times10^8$ DM particles.
All models were evolved for $4 \, \Gyr$ with 400 output snapshots using the \textsc{AREPO} code \citep{weinberger20}. We adopted a hierarchical time-stepping scheme with a maximum time step of $50\,\mathrm{Myr}$.

Our models are purely collisionless systems that do not include gas or star formation. Including hydrodynamics and stellar feedback would be an interesting extension for future work. For instance, the bar-driven gas inflows can exert positive torques that counteract the angular momentum transfer from the bar to the halo and may even accelerate the bar pattern speed \citep{beane23}. Furthermore, \cite{ansar25} show that stellar feedback can significantly weaken bar strength. However, cosmological simulations, which employ a wide variety of subgrid physics prescriptions and resolutions, often suffer from the well-known ``missing bar'' problem. Therefore, in the present work we focused on isolating the effects of force and mass resolution in purely collisionless systems. We plan to revisit this topic with more realistic baryonic physics in future work.

\section{Results}\label{sec:results}
\subsection{Overview}\label{sec:overview}

\begin{figure*}[htbp]
    \centering
    \includegraphics[width=0.415\textwidth]{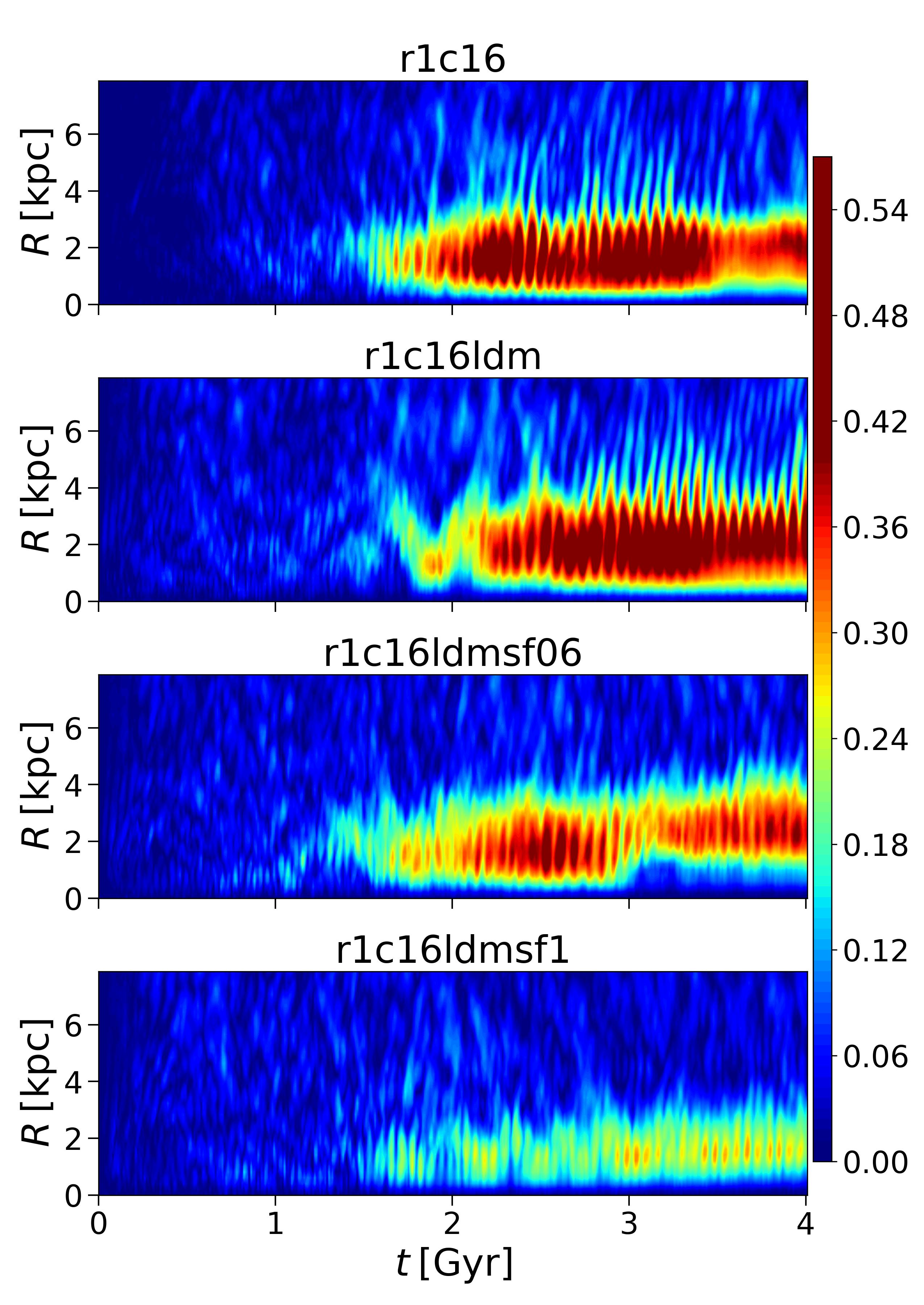}
    \includegraphics[width=0.415\textwidth]{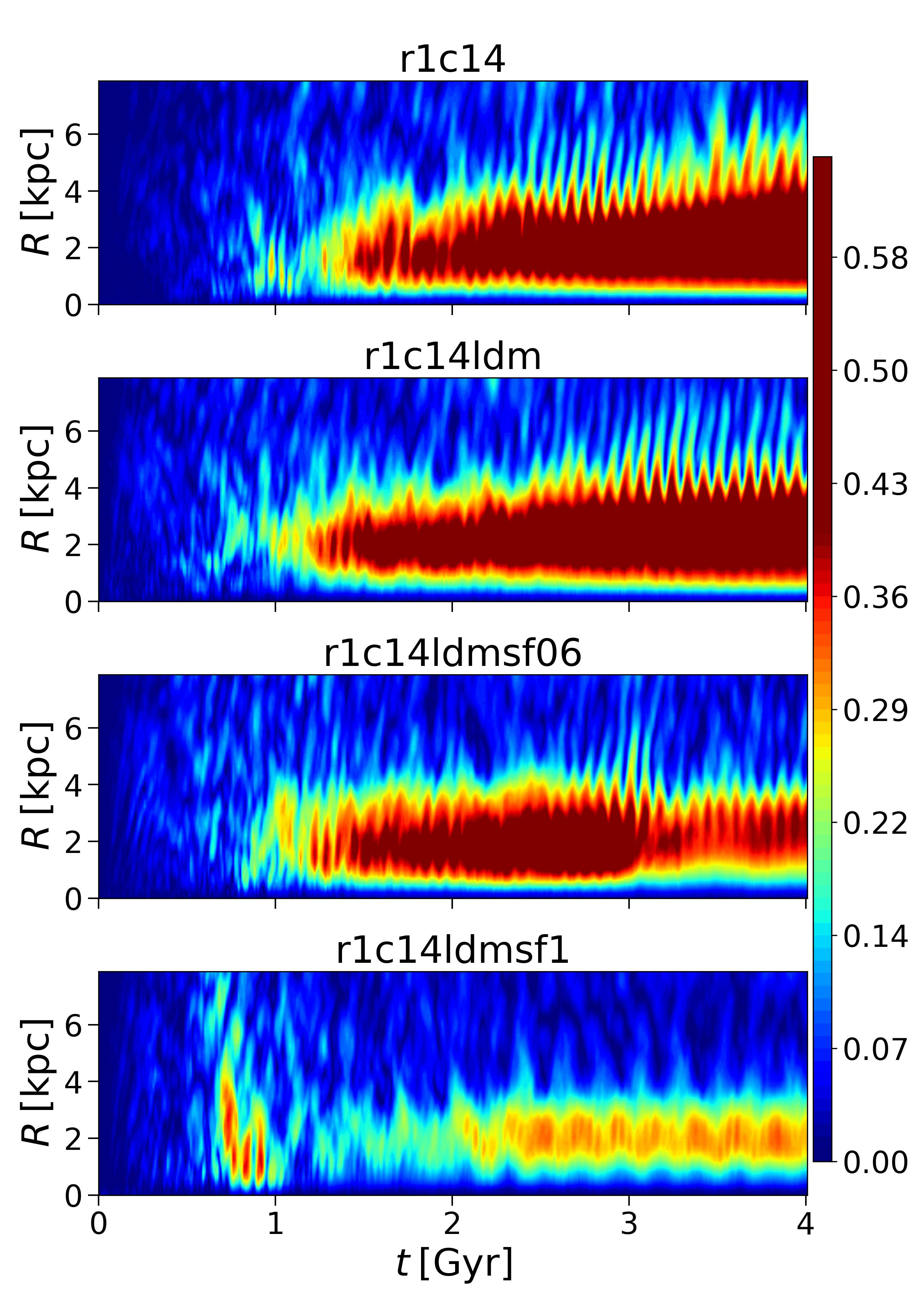}

    \caption{Fourier amplitude map $F_2(R,t)$ within 8 kpc, computed from the time evolution of the radial Fourier profiles of the $m=2$ mode. Snapshots were taken every 0.01 Gyr over a total of 4 Gyr (400 snapshots). To enable direct comparison across models, the color scale inside the color bar is fixed across all panels, such that the same color corresponds to the same amplitude value between 0 and 0.4. All values above 0.4 are shown in the same saturated red.}
    \label{fig:fmapall}
\end{figure*}

\begin{figure}[htbp]
    \centering
    \begin{tabular}{@{}cccc@{}}
        
        \raisebox{0.12\height}{\rotatebox{90}
        {\textbf{\Large
        $\Fsum \qquad\quad\quad F_6 \qquad\quad\quad\: F_5 \qquad\qquad\quad F_4 \qquad\quad\quad F_3 \qquad\qquad\quad F_2 \qquad\qquad\quad F_1$}}}
    \includegraphics[width=0.415\textwidth]{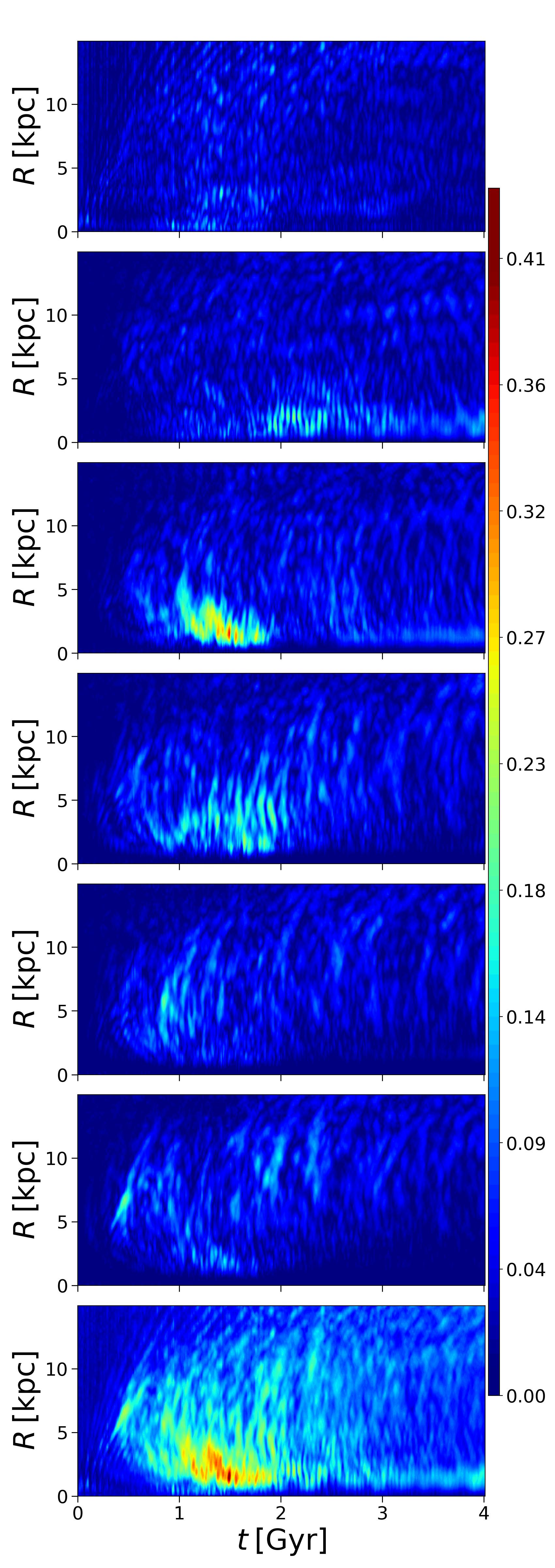}
    \end{tabular}
    \caption{Fourier amplitude map $F_m(R,t)$ calculated from the time evolution of the radial Fourier profiles for each mode and $\Fsum$ in the r1c16sf1 model. The time interval of each snapshot used is 0.01 Gyr. To enable direct comparison across models, the color scale inside the color bar is fixed across all panels, such that the same color corresponds to the same amplitude value between 0 and 0.4. All values above 0.4 are shown in the same saturated red.}
    \label{fig:fmapsf1}
\end{figure}

\begin{figure}[htbp]
    \centering
    \includegraphics[width=0.5\textwidth]{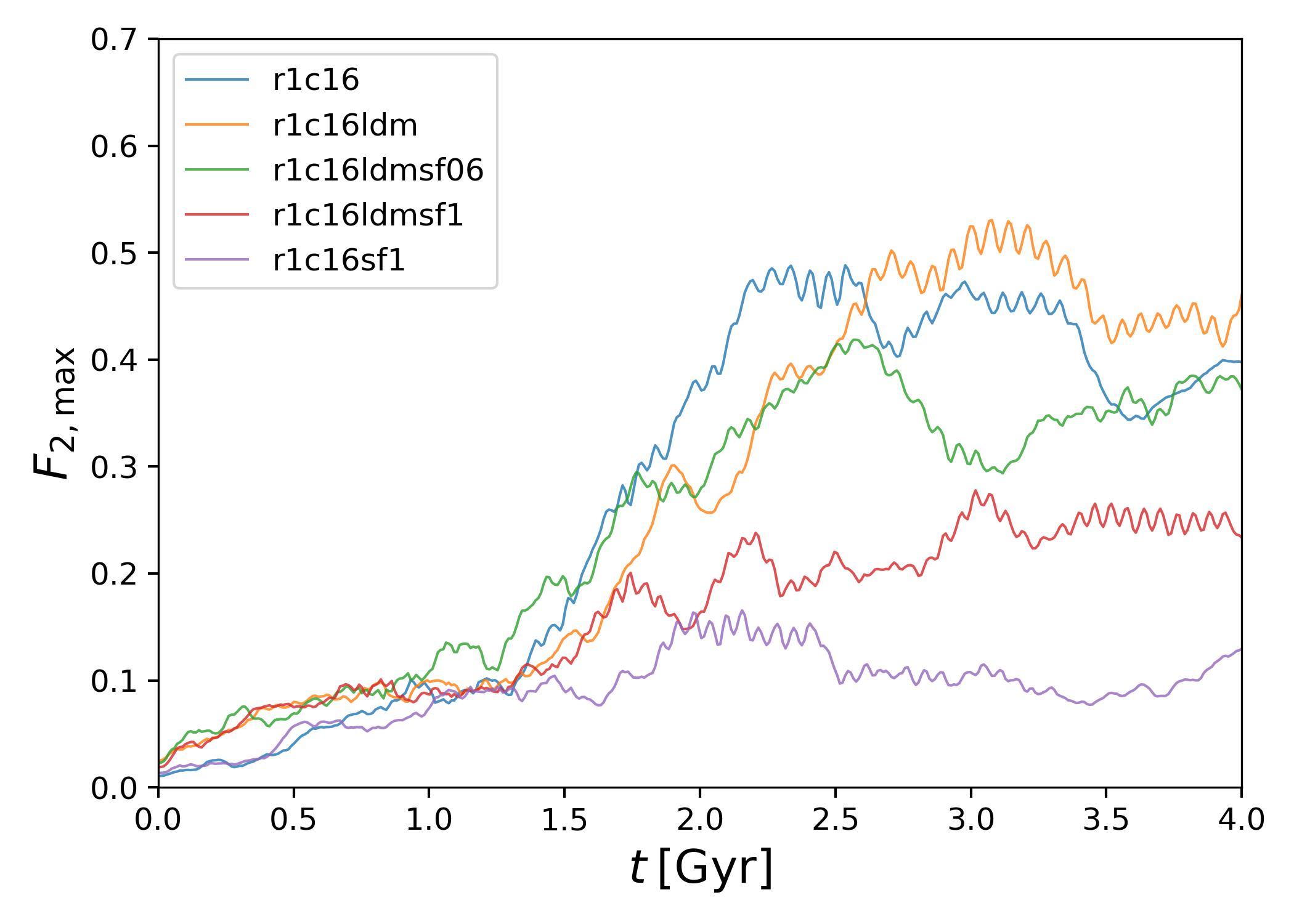}
    \includegraphics[width=0.5\textwidth]{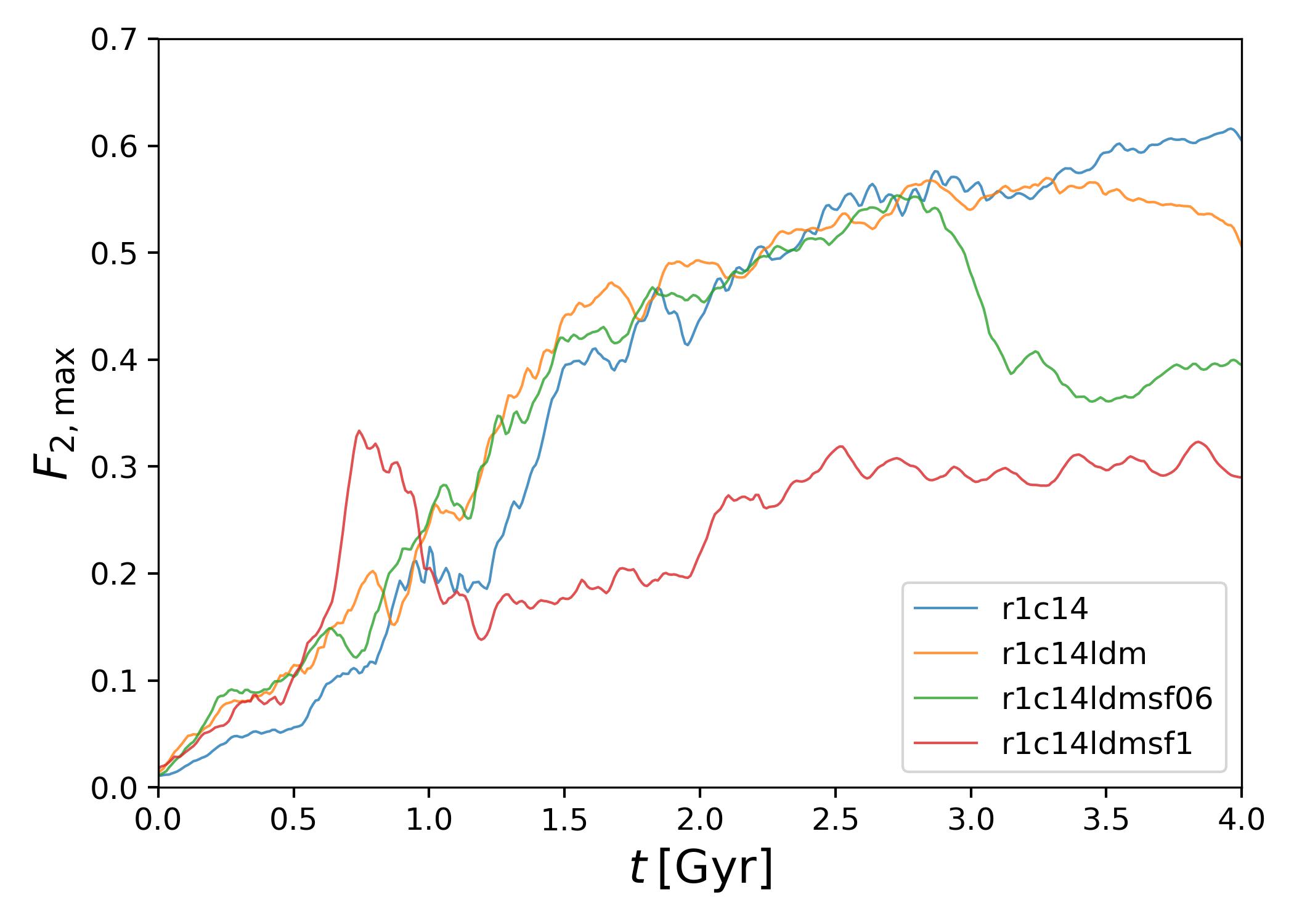}
    \caption{Time evolution of the maximum value of the Fourier mode $m=2$ in the "c16" models (top panel) and "c14" models (bottom panel), after smoothing the data by applying a moving average over nine points. The unsmoothed version is shown in Fig.~\ref{fig:f2max_raw}.}
    \label{fig:f2max}
\end{figure}

\begin{figure}[htbp]
    \centering
    \includegraphics[width=0.5\textwidth]{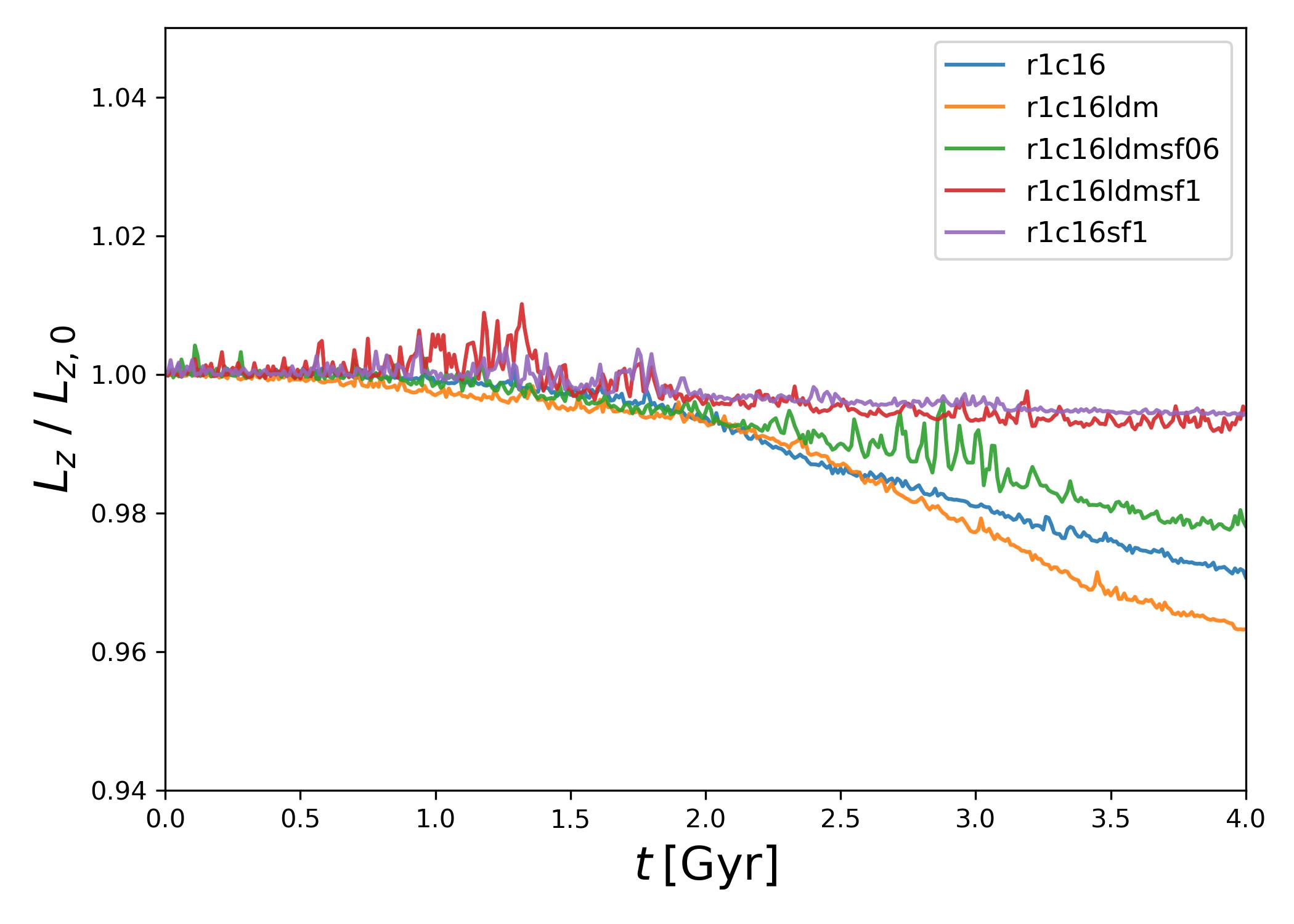}
    \includegraphics[width=0.5\textwidth]{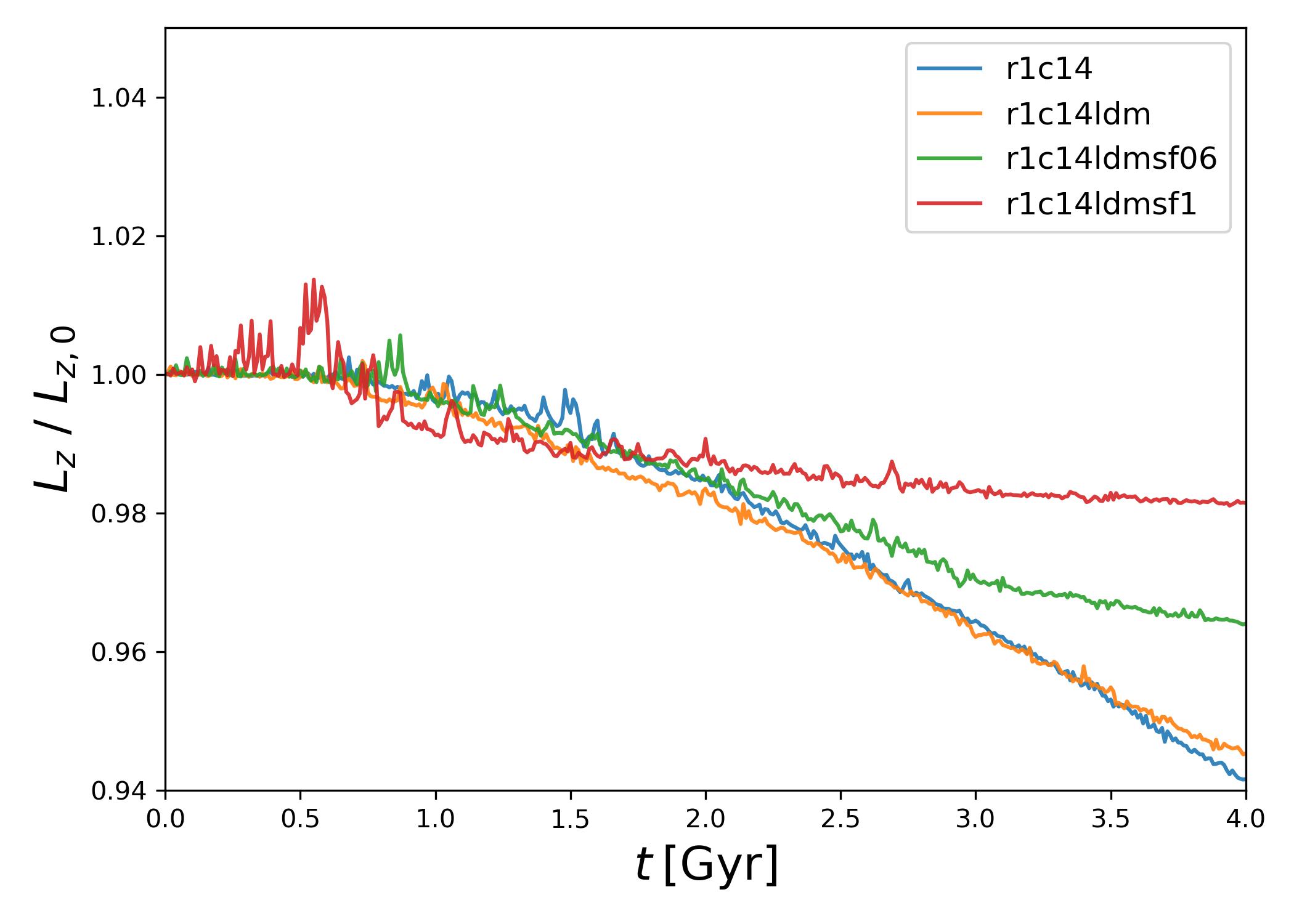}
    \caption{Time evolution of the disk angular momentum normalized to its initial value, $L_z / L_{z,0}$, over $4 \, \Gyr$. Top: "c16" models. Bottom: "c14" models. The corresponding figure including the additional r1c14ldmsf03 and r1c14hdm models is shown in Fig.~\ref{fig:f2max_lz_appendix}.}
    \label{fig:lz}
\end{figure}

\begin{figure*}[htbp]

    \centering
    \includegraphics[width=0.33\textwidth]{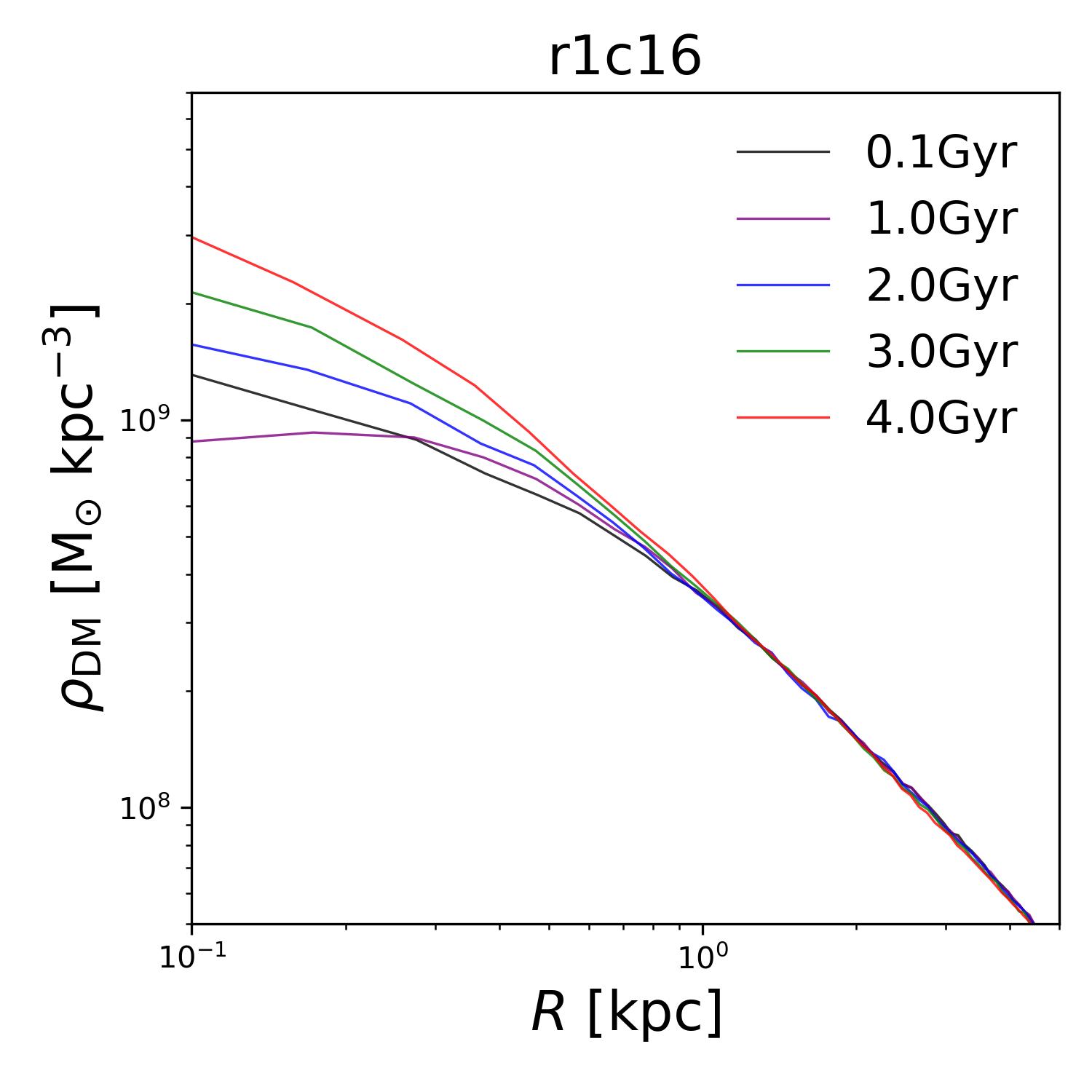}\includegraphics[width=0.33\textwidth]{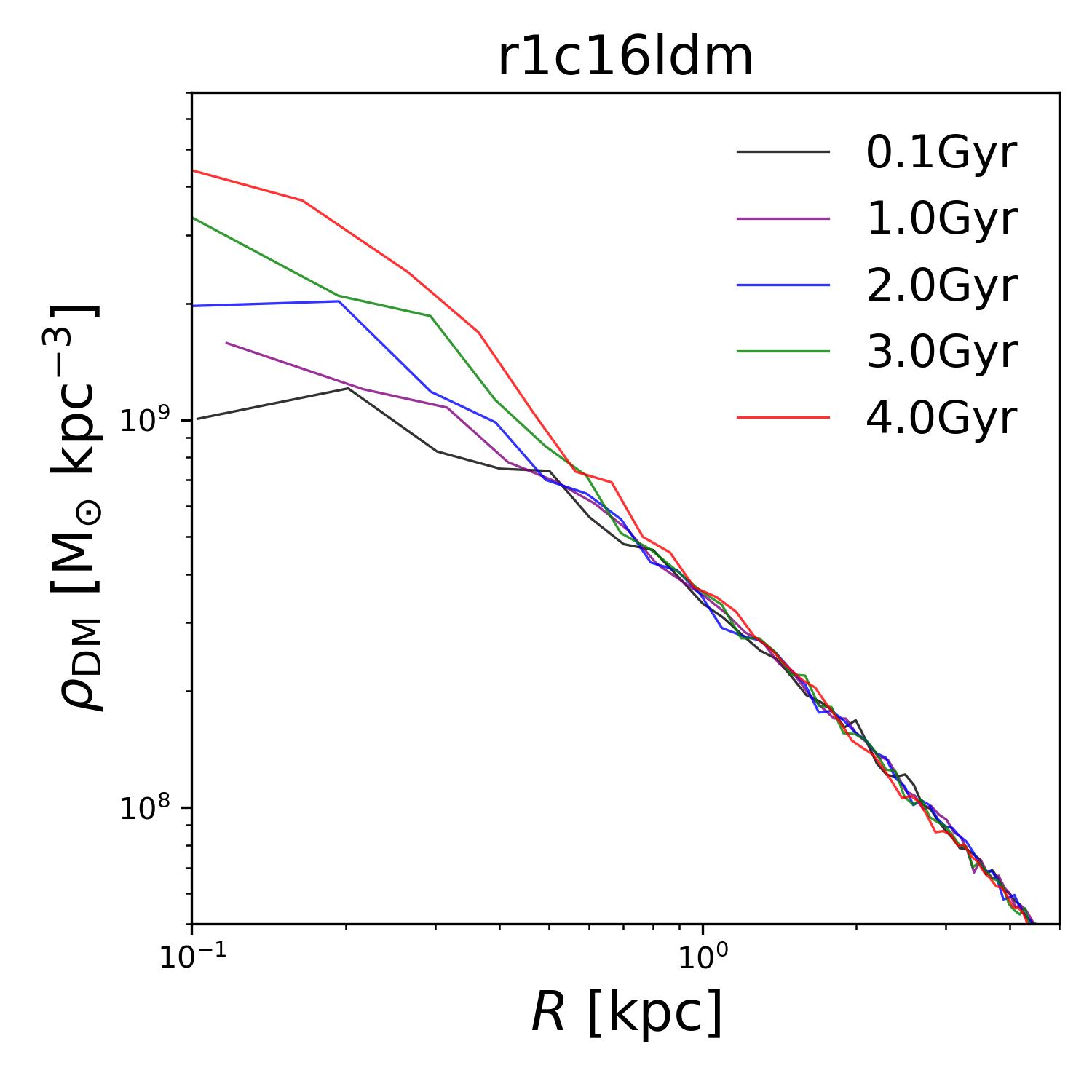}\includegraphics[width=0.33\textwidth]{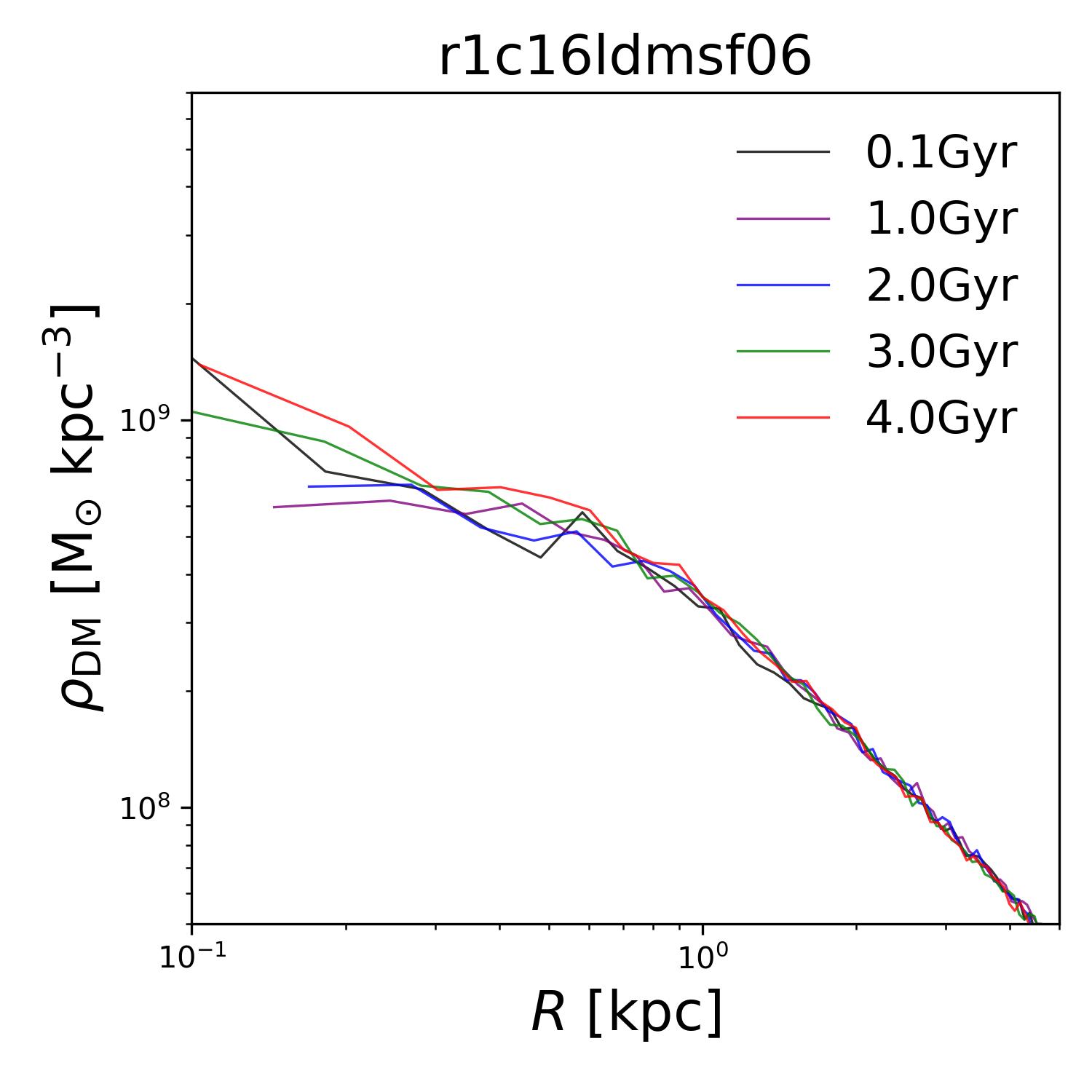}
    \includegraphics[width=0.33\textwidth]{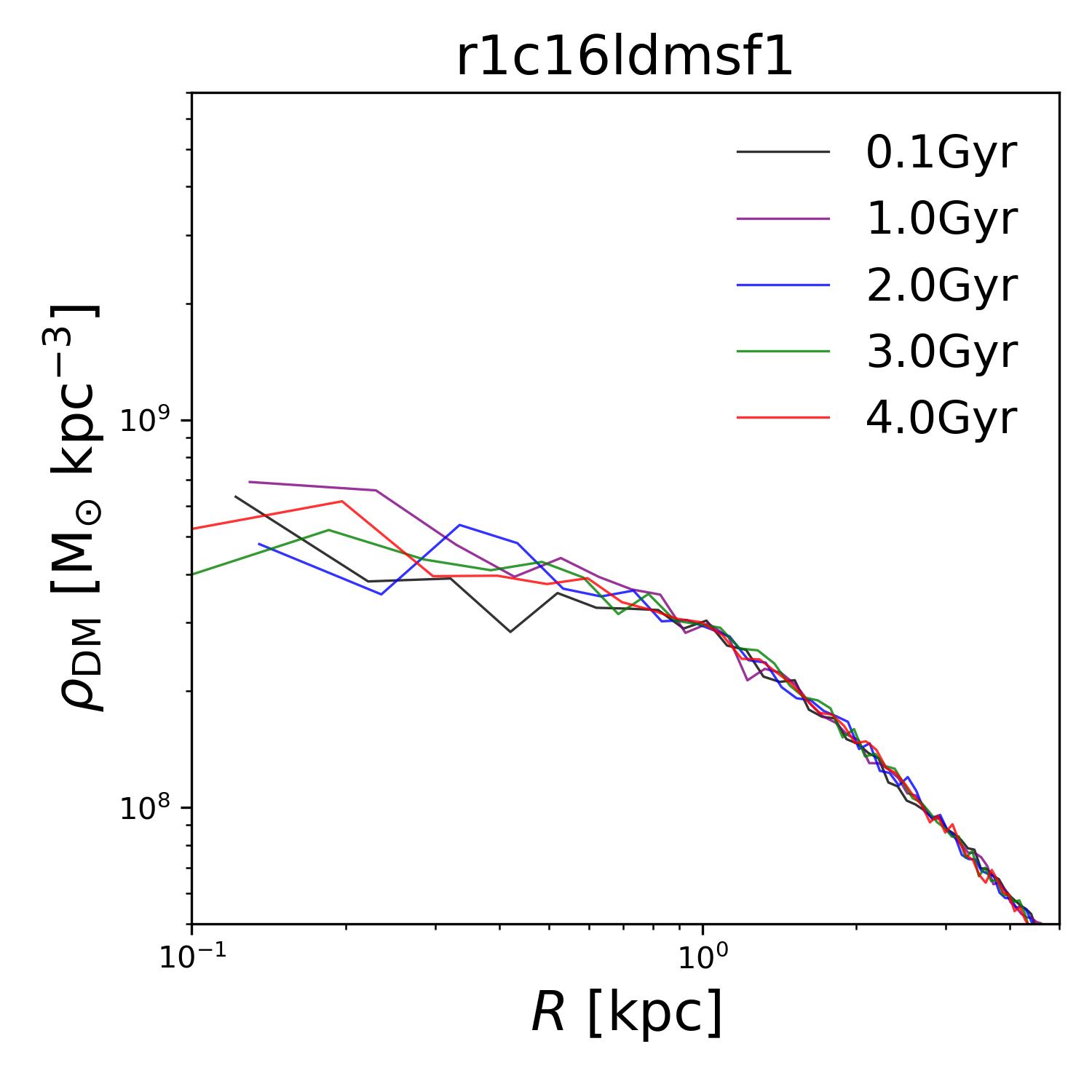}\includegraphics[width=0.33\textwidth]{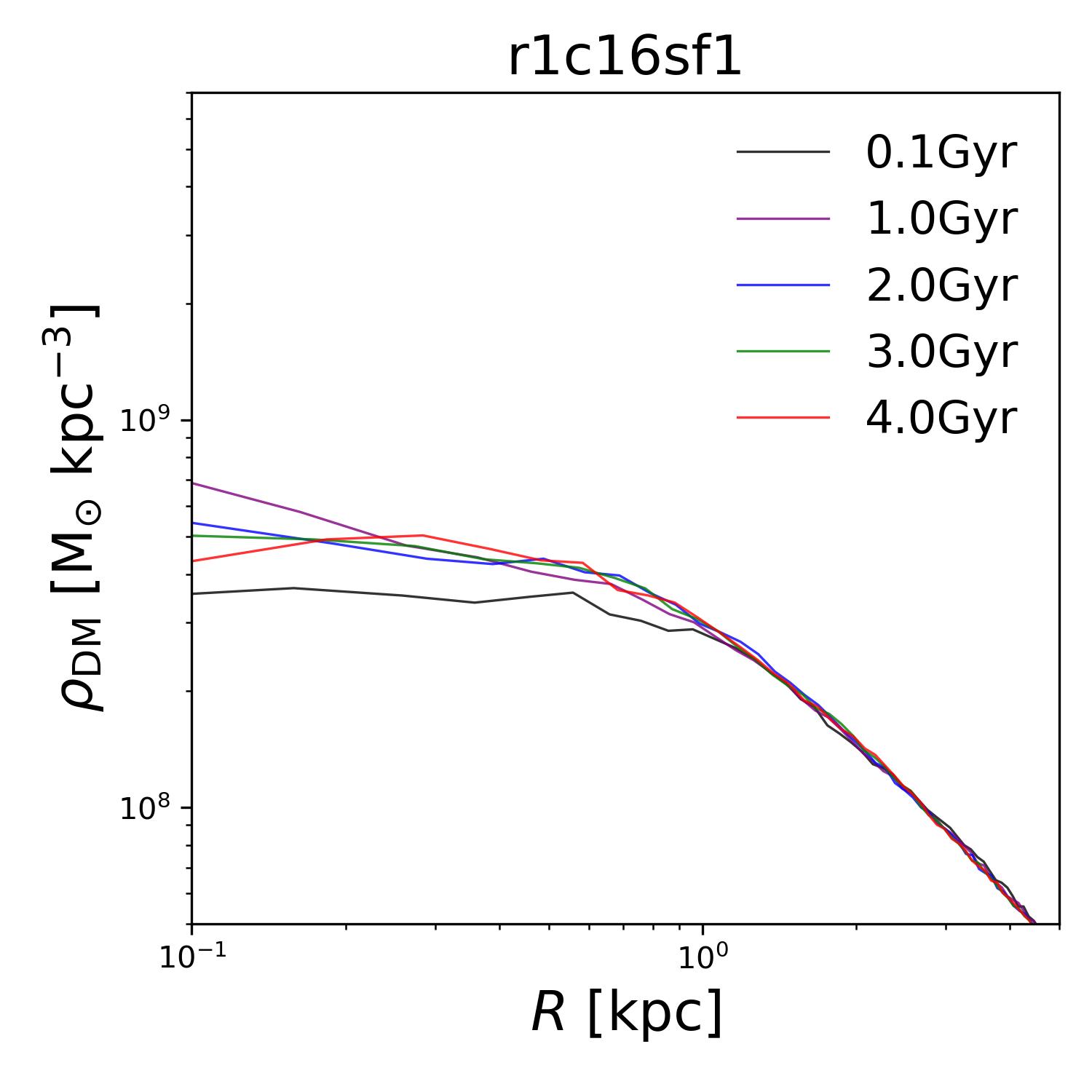}

    \caption{Radial density profiles of the DM halo in the "c16" models from $0.1 \, \Gyr$ to $4 \, \Gyr$, shown between $0.1 \, \kpc$ and $5 \, \kpc$. Both axes are logarithmic. Profiles at different times are color-coded and labeled accordingly. The corresponding profiles for the "c14" models are shown in Fig.~\ref{fig:c14density}.}
    \label{fig:den}
\end{figure*}

\begin{figure*}[htbp]
    \includegraphics[width=0.33\textwidth]{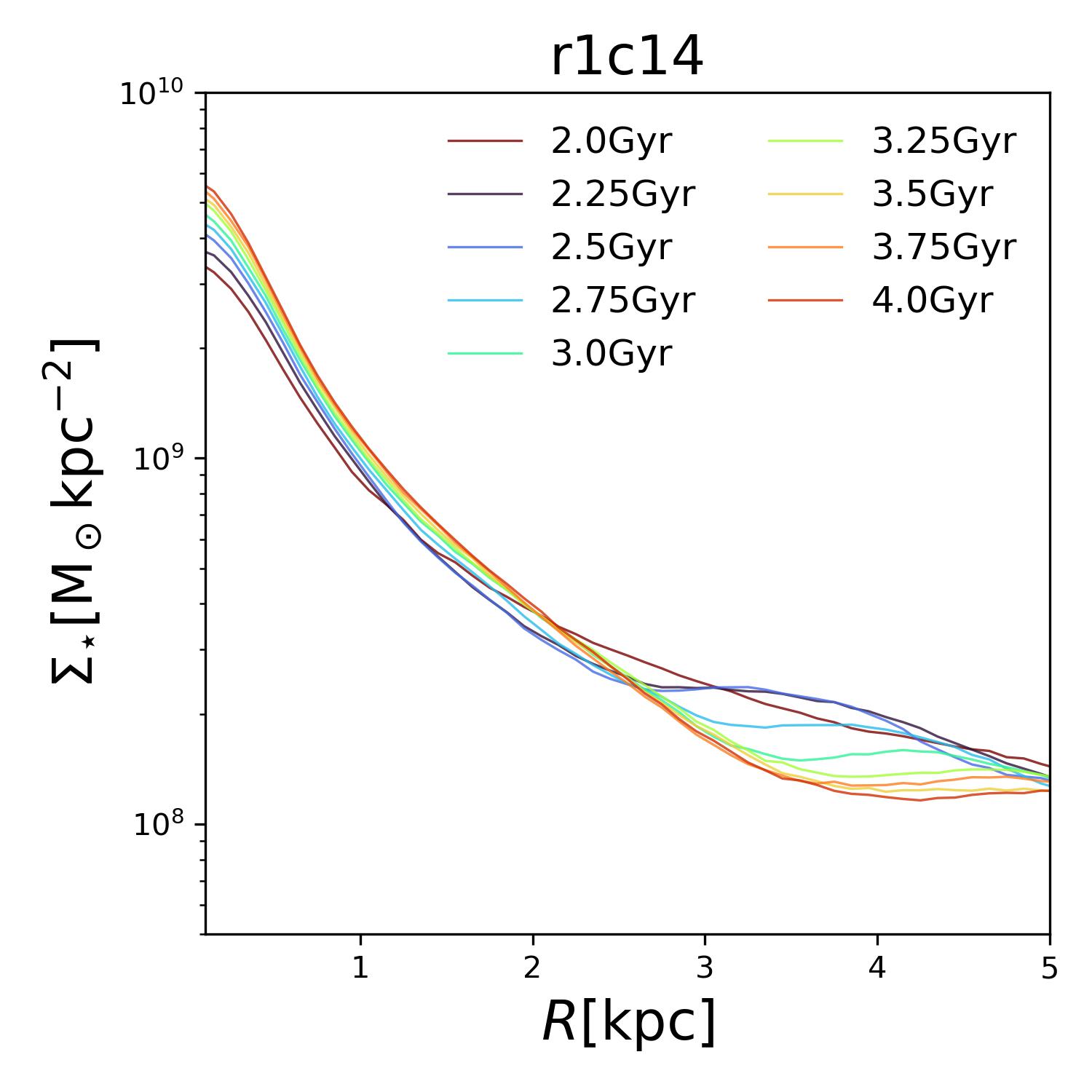}\includegraphics[width=0.33\textwidth]{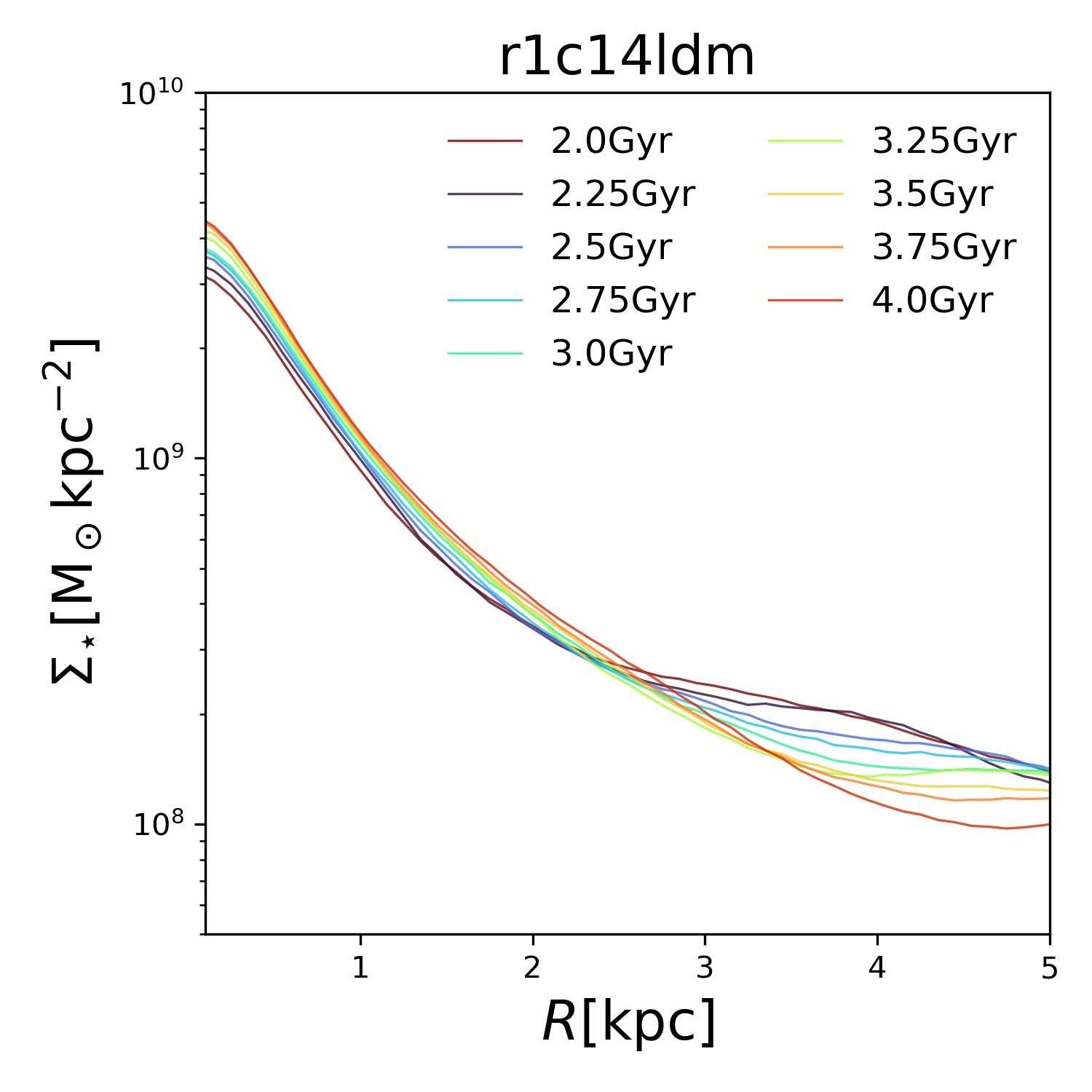}\includegraphics[width=0.33\textwidth]{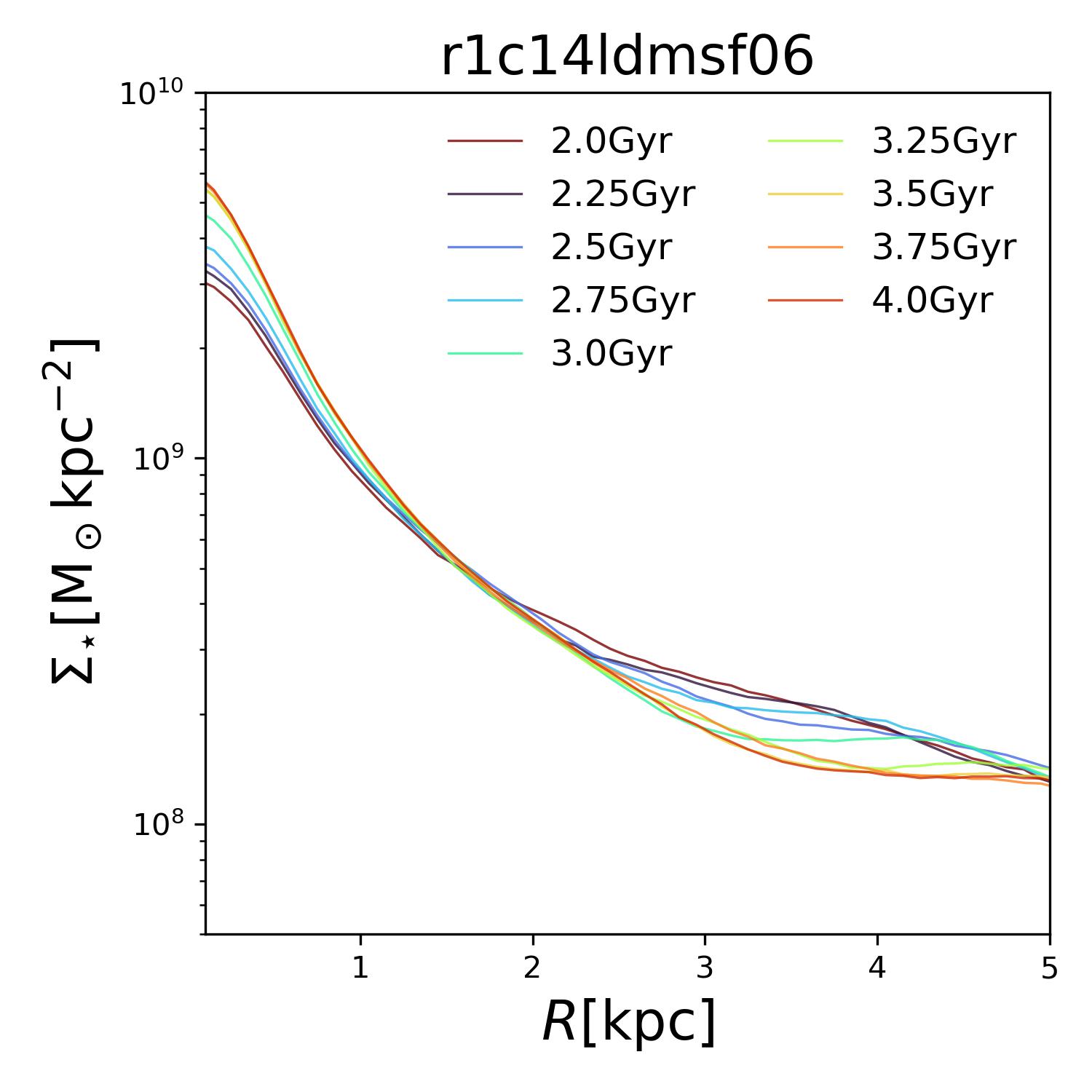}

    \includegraphics[width=0.33\textwidth]{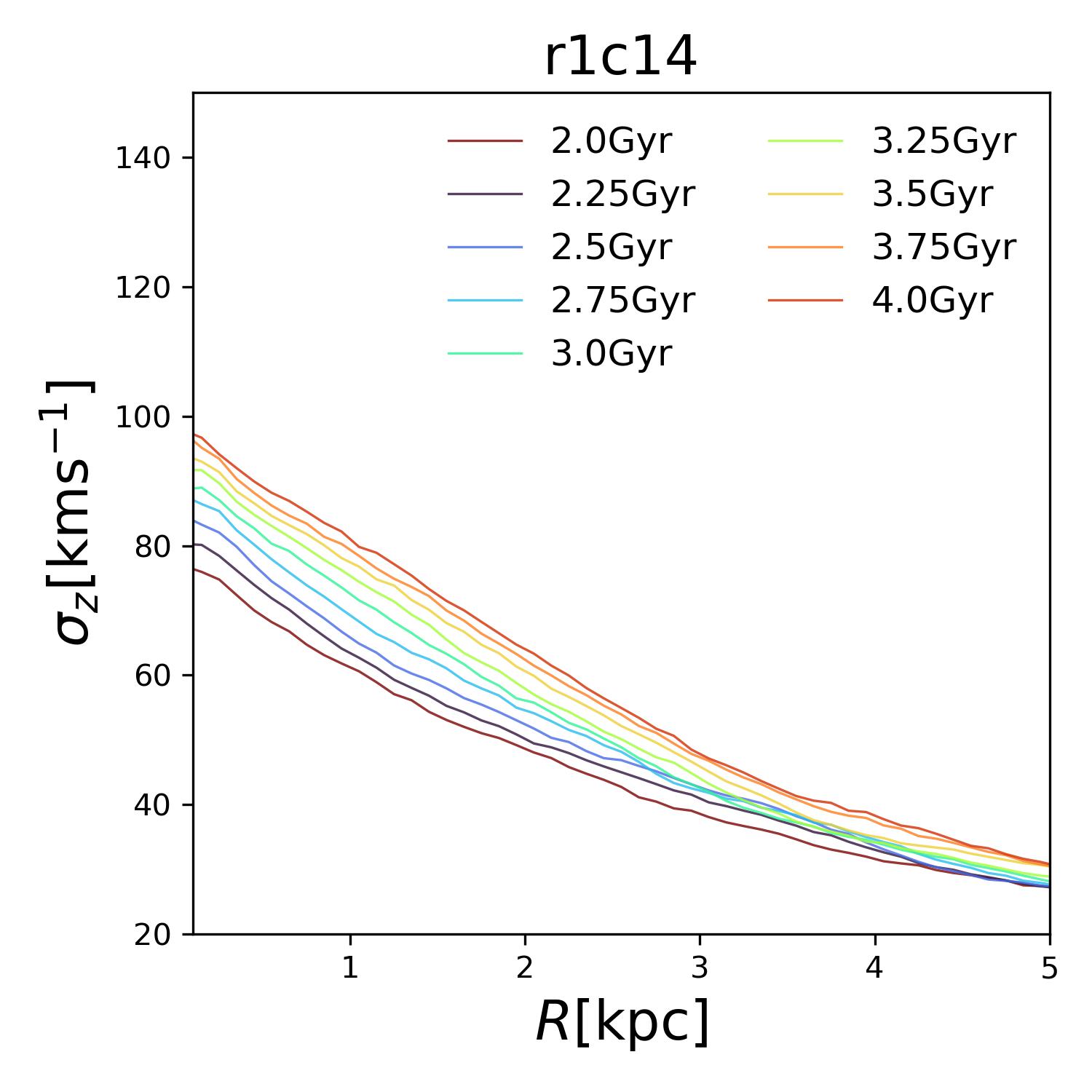}\includegraphics[width=0.33\textwidth]{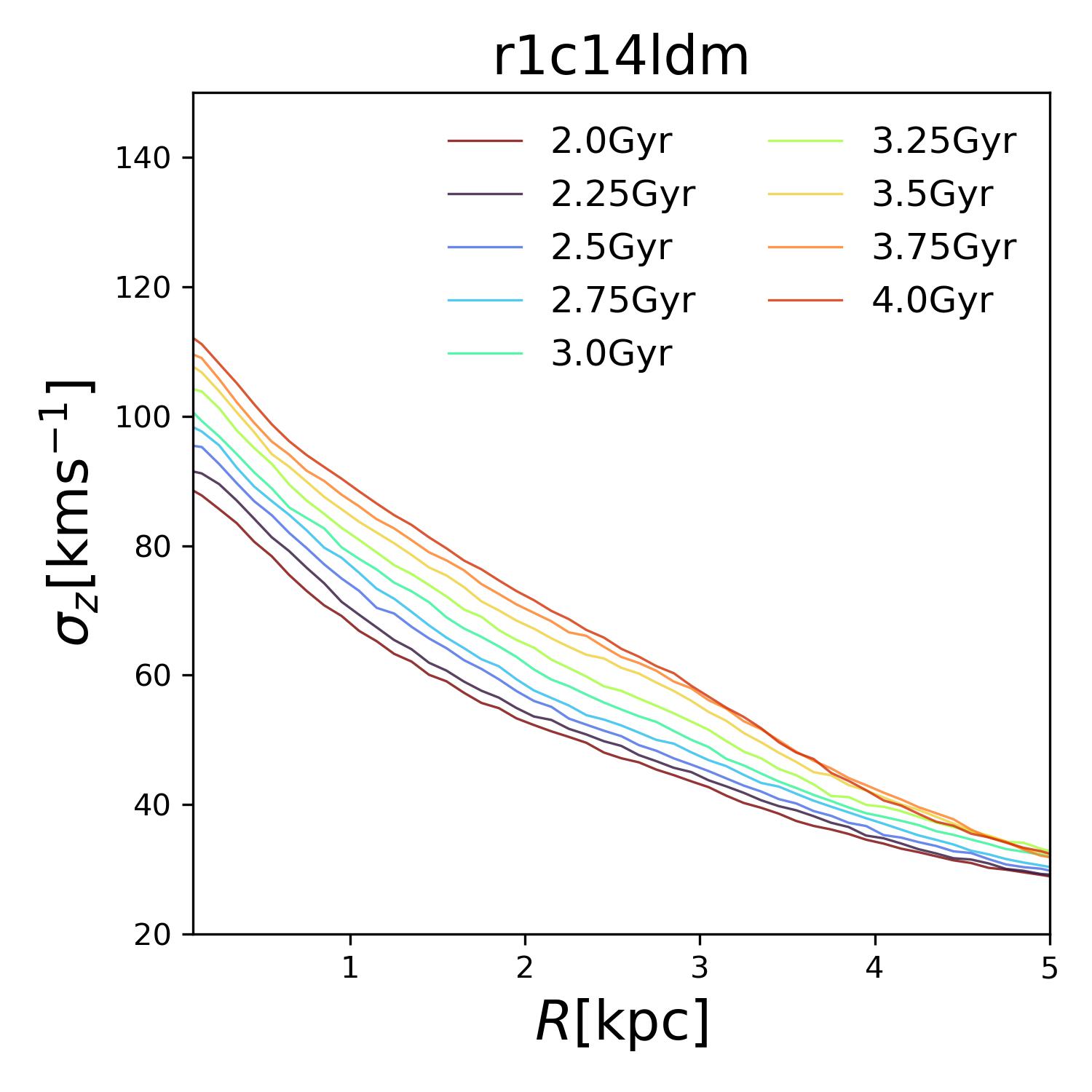}\includegraphics[width=0.33\textwidth]{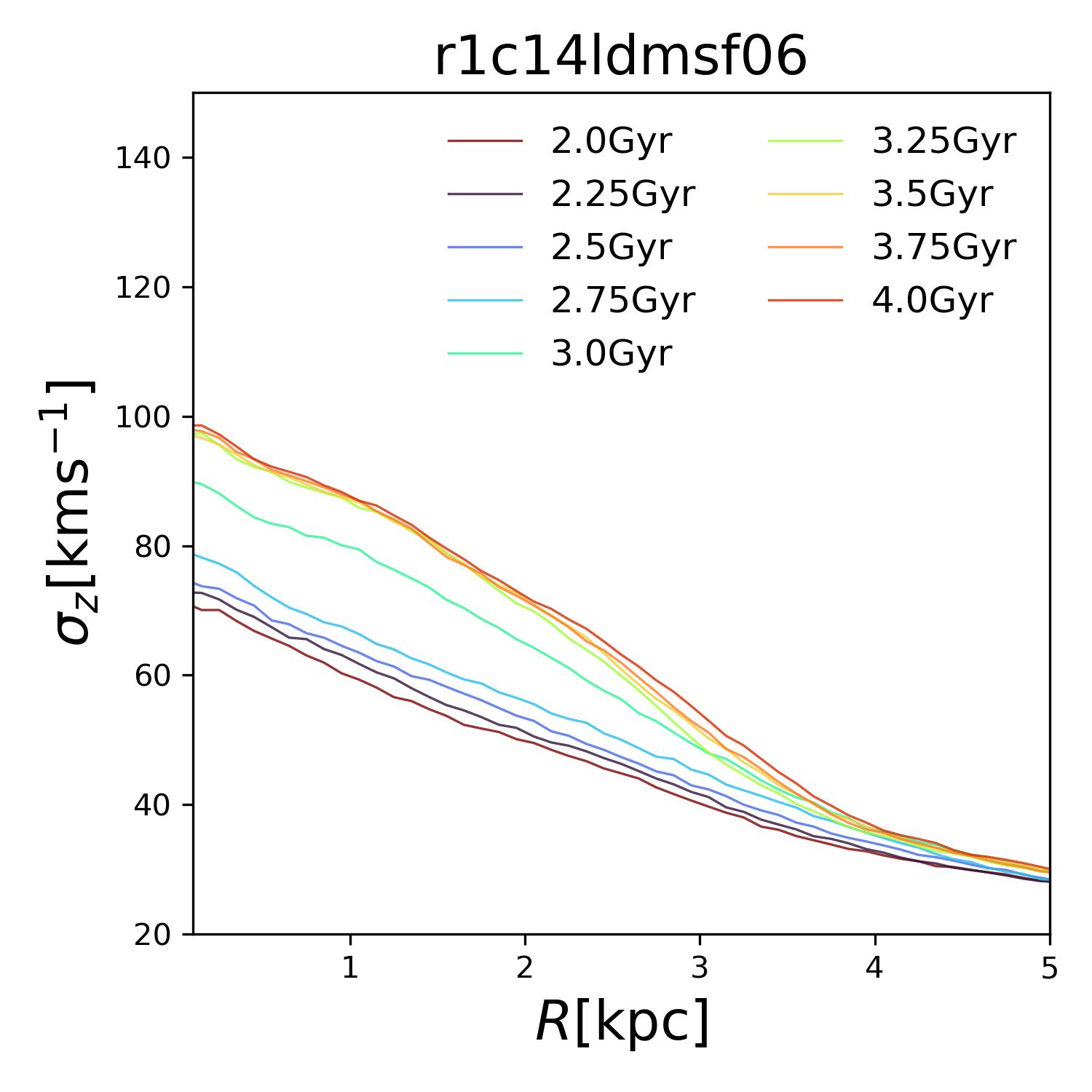}

    \includegraphics[width=0.33\textwidth]{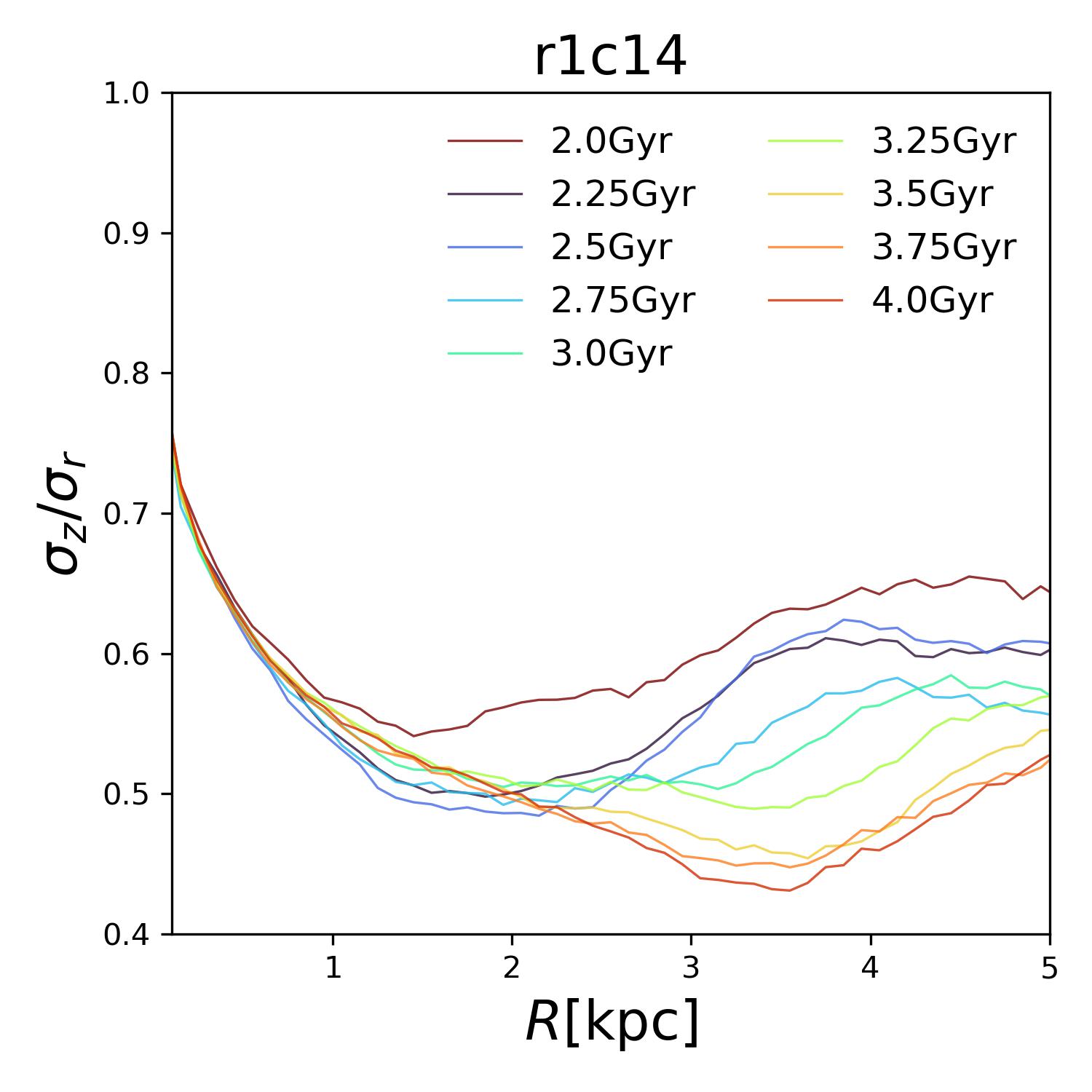}\includegraphics[width=0.33\textwidth]{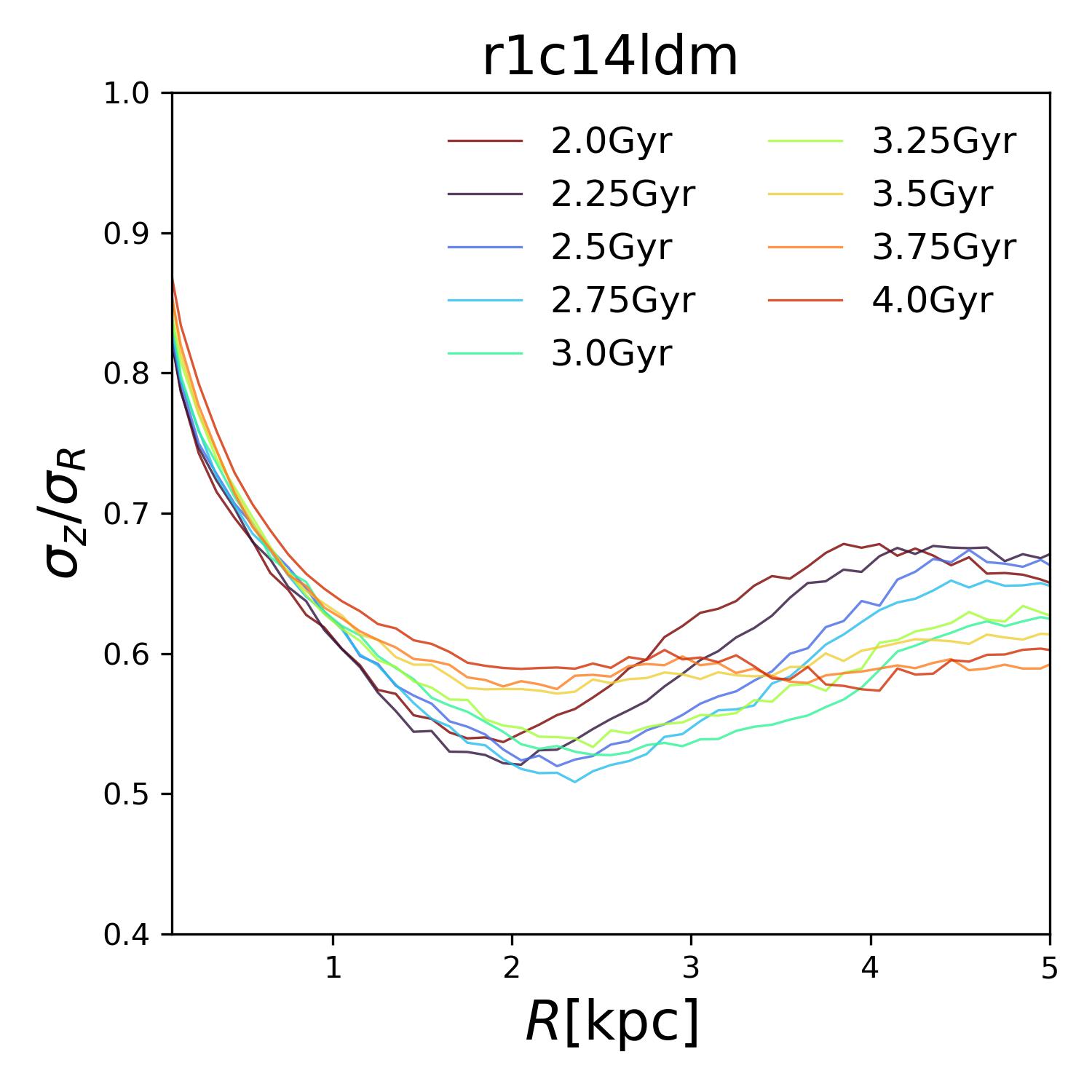}\includegraphics[width=0.33\textwidth]{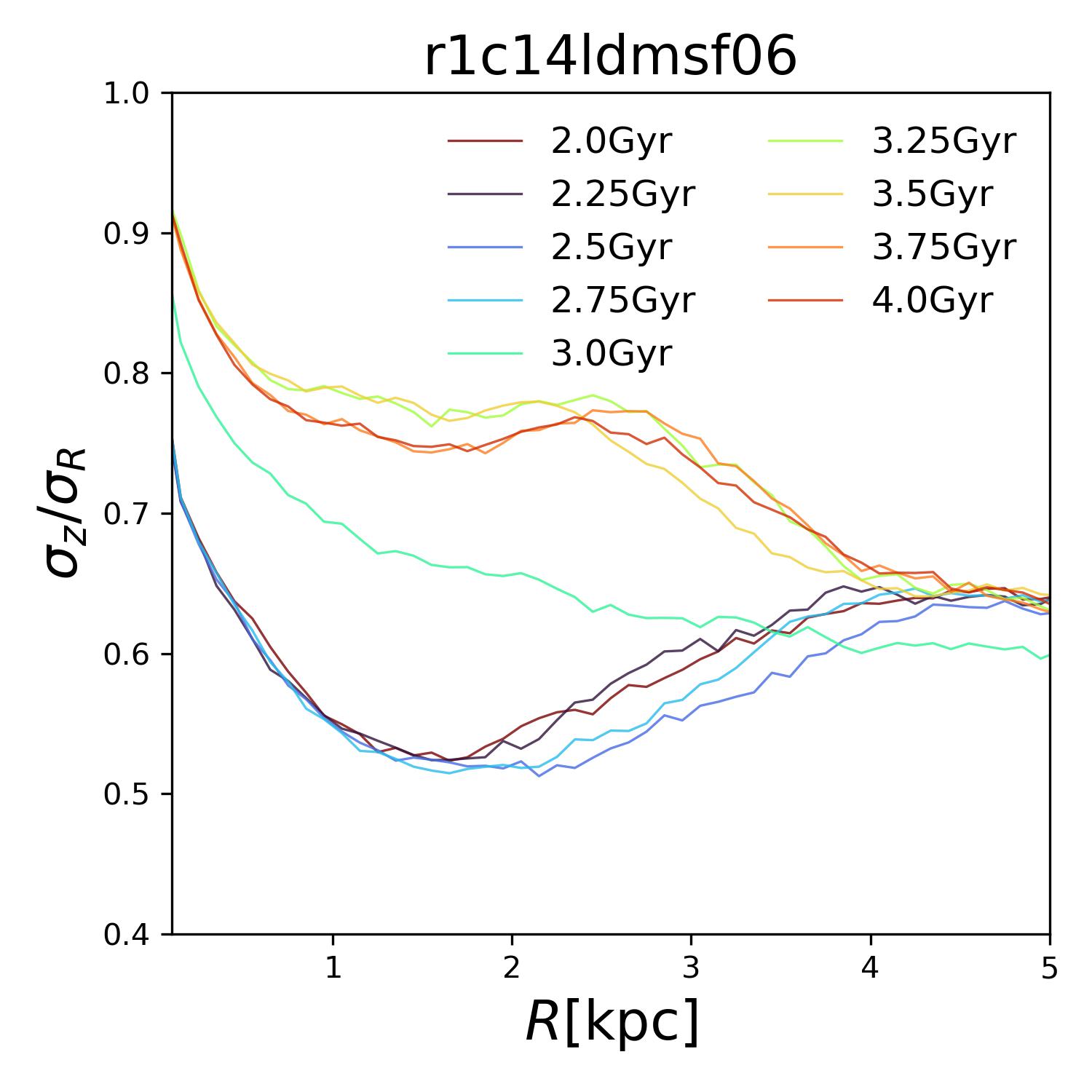}
    \caption{Radial profiles of stellar disk properties within $5 \, \kpc$, from $2 \, \Gyr$ to $4 \, \Gyr$ with a time step of $0.25 \, \Gyr$. Left, middle, and right columns: r1c14, r1c14ldm, and r1c14ldmsf06 models, respectively, all of which form a bar. Top row: Density profiles for each model. Middle row: Vertical velocity dispersion of the disk. Bottom row: Ratio between the vertical and radial velocity dispersions. The corresponding profiles for model r1c14hdm are shown in Fig.~\ref{fig:sigmarz_hdm}.}
    \label{fig:sigmarz}
\end{figure*}

\begin{figure*}[htbp]
    \centering
    \begin{tabular}{@{}ccc@{}}
        r1c14 & r1c14ldm & r1c14ldmsf06 \\
        \includegraphics[width=0.33\textwidth]{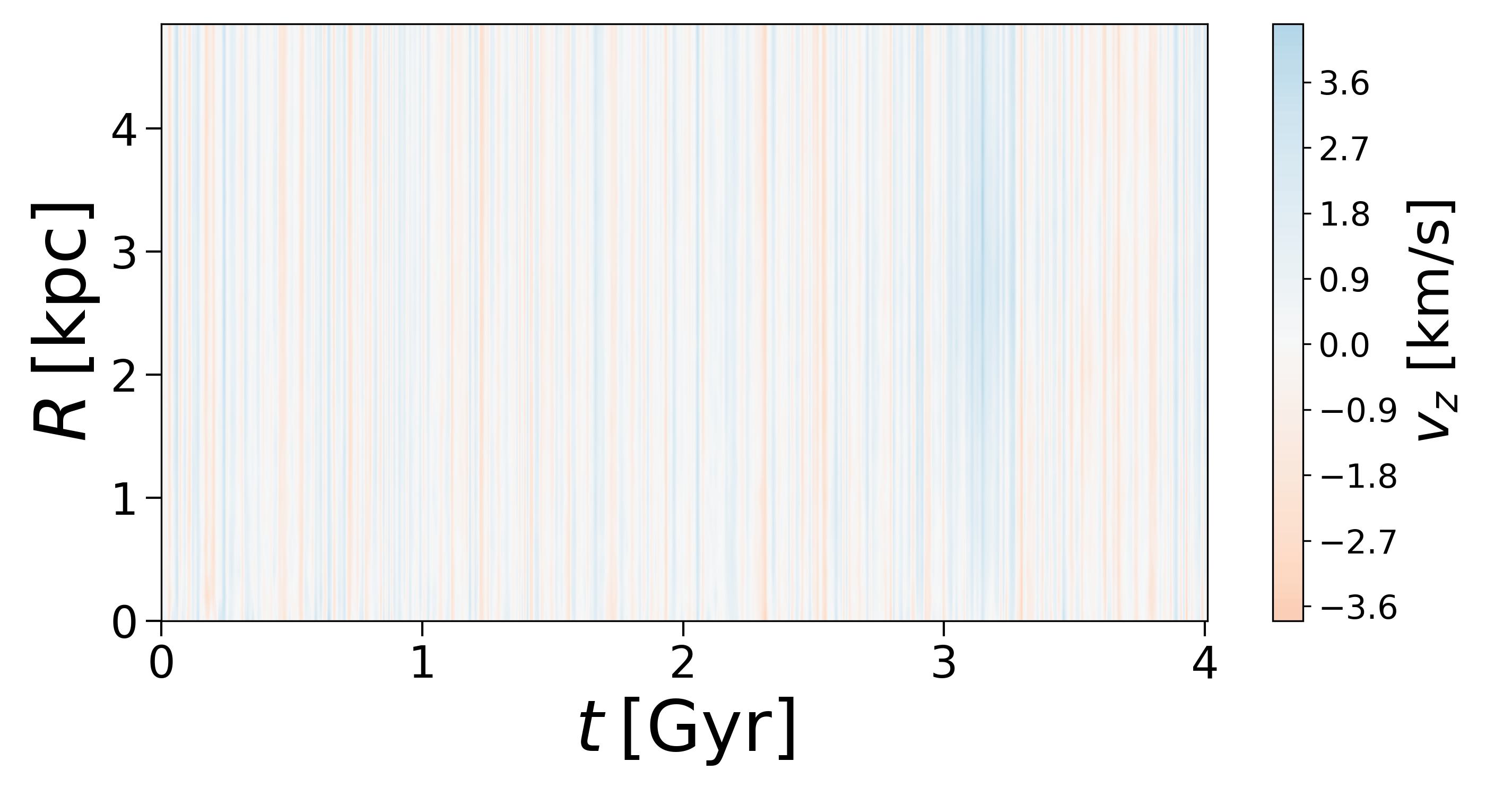} &
        \includegraphics[width=0.33\textwidth]{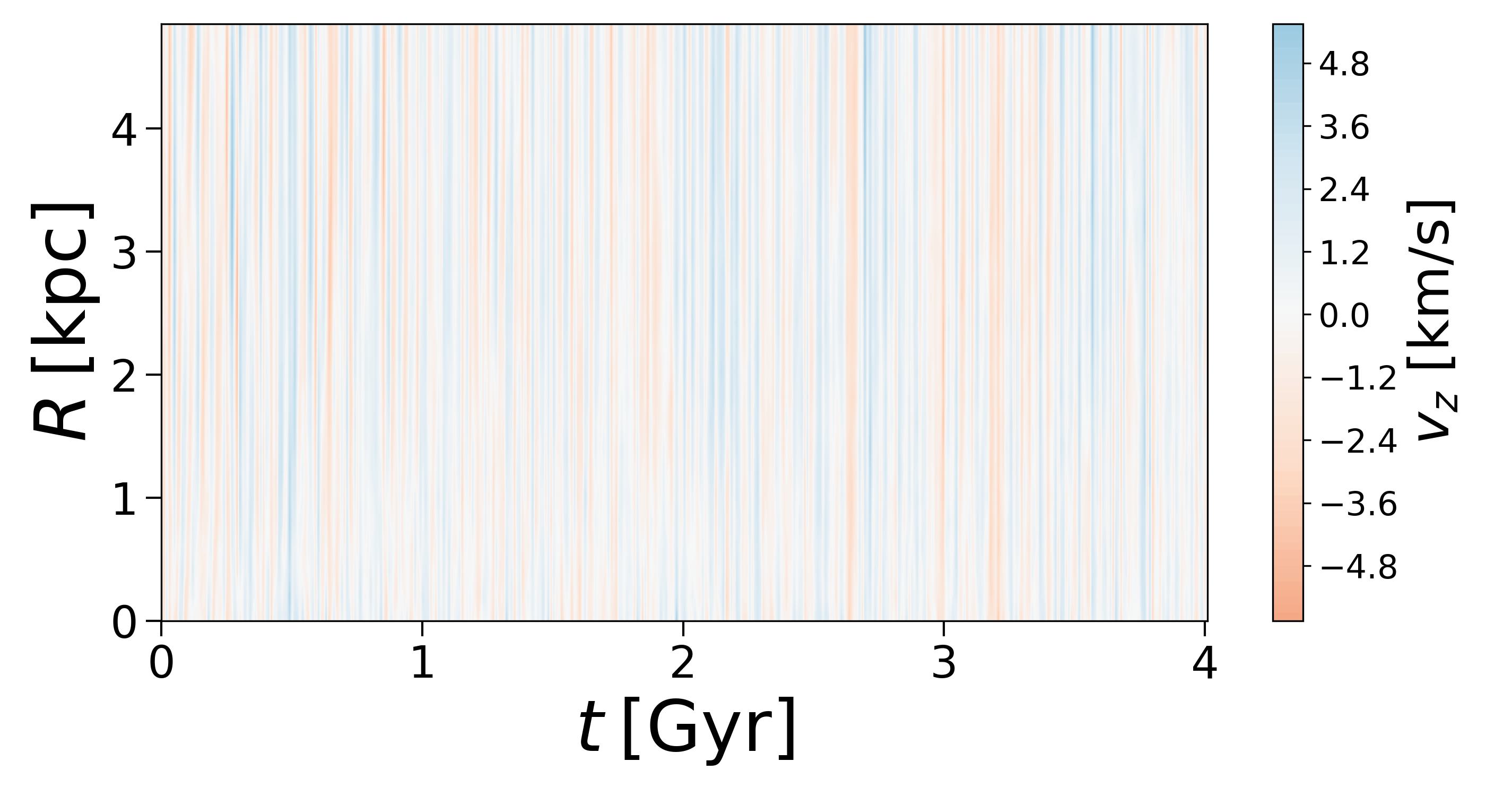} &
        \includegraphics[width=0.33\textwidth]{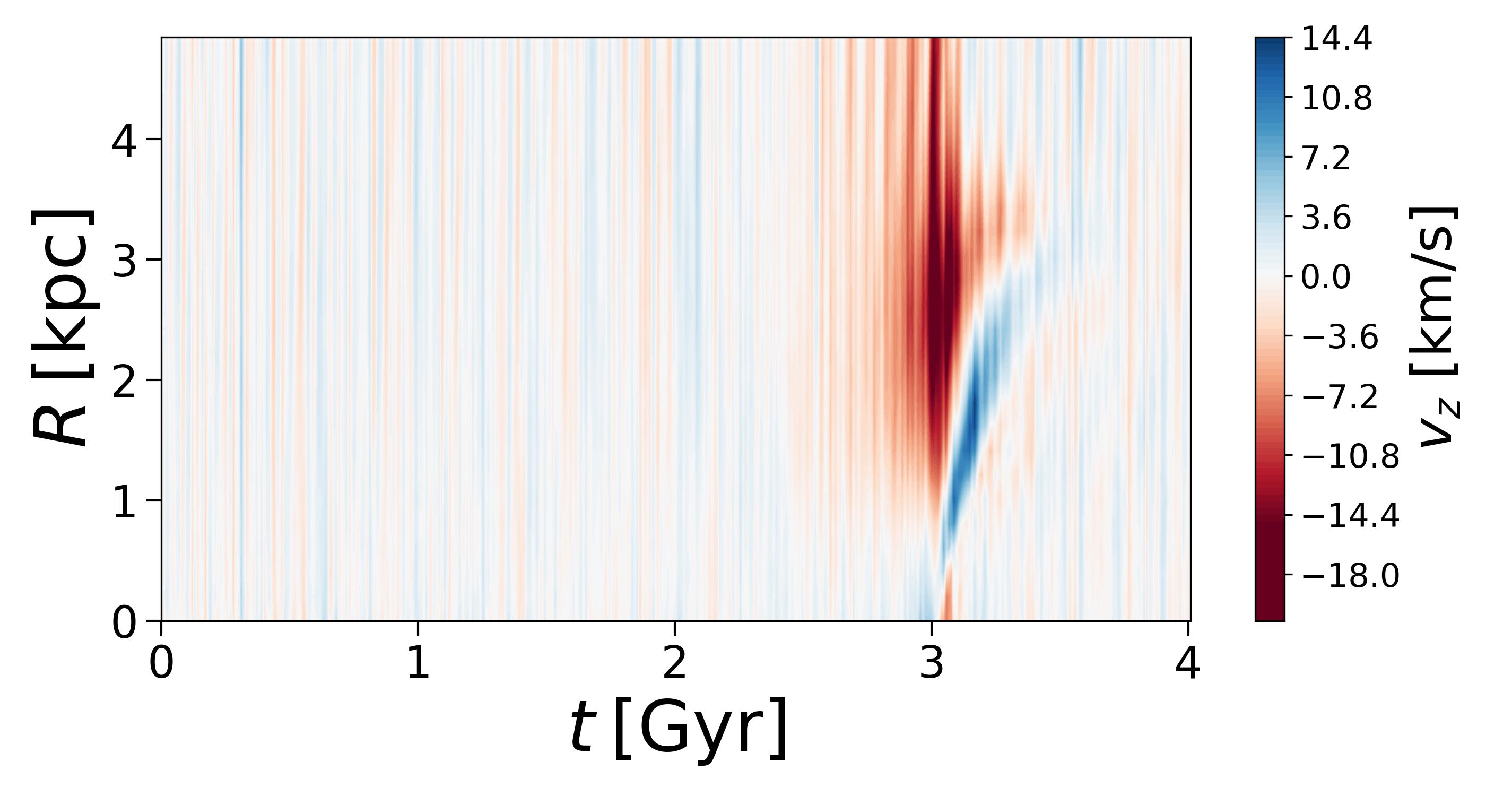} \\
        \includegraphics[width=0.33\textwidth]{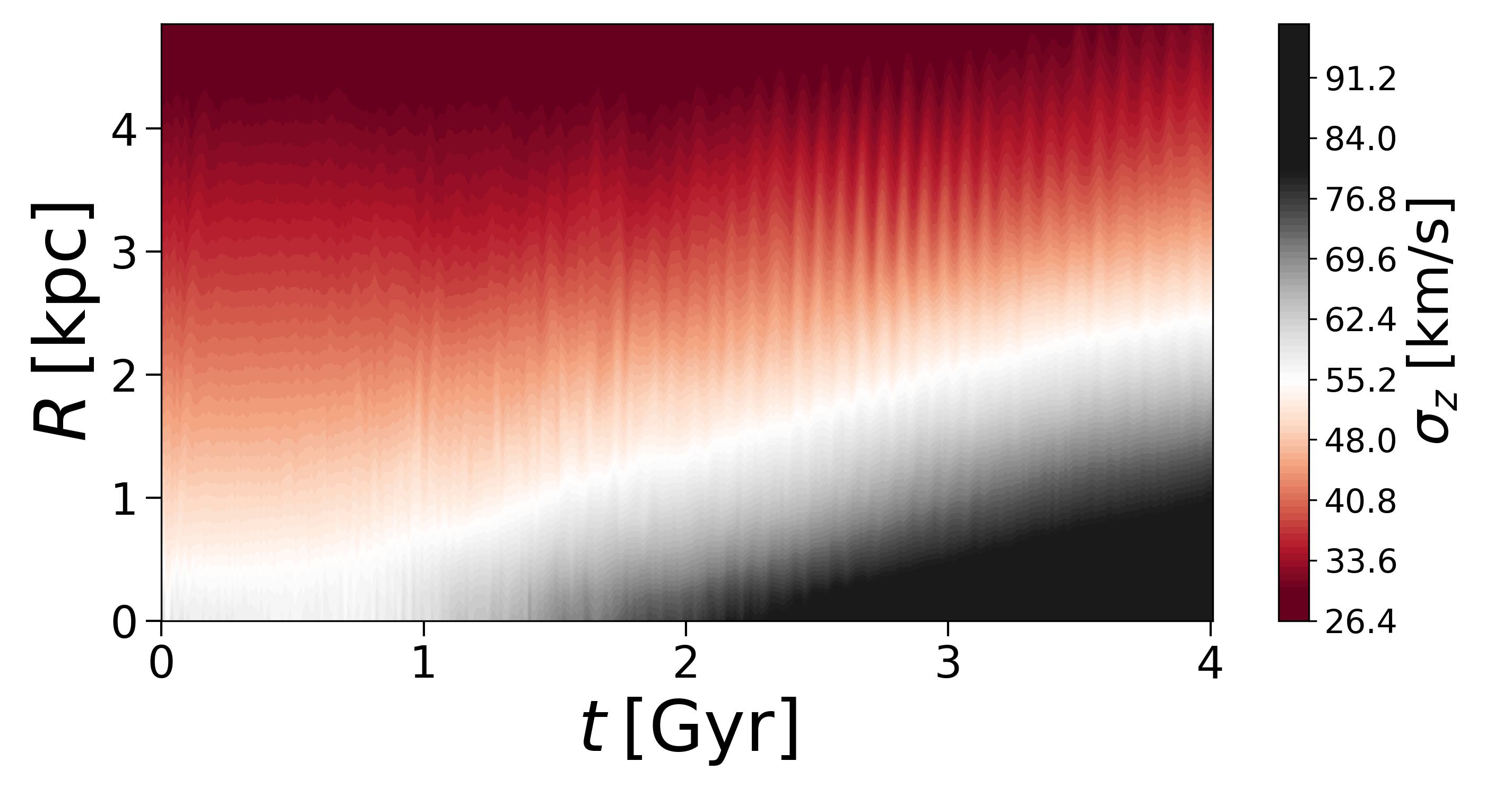} &
        \includegraphics[width=0.33\textwidth]{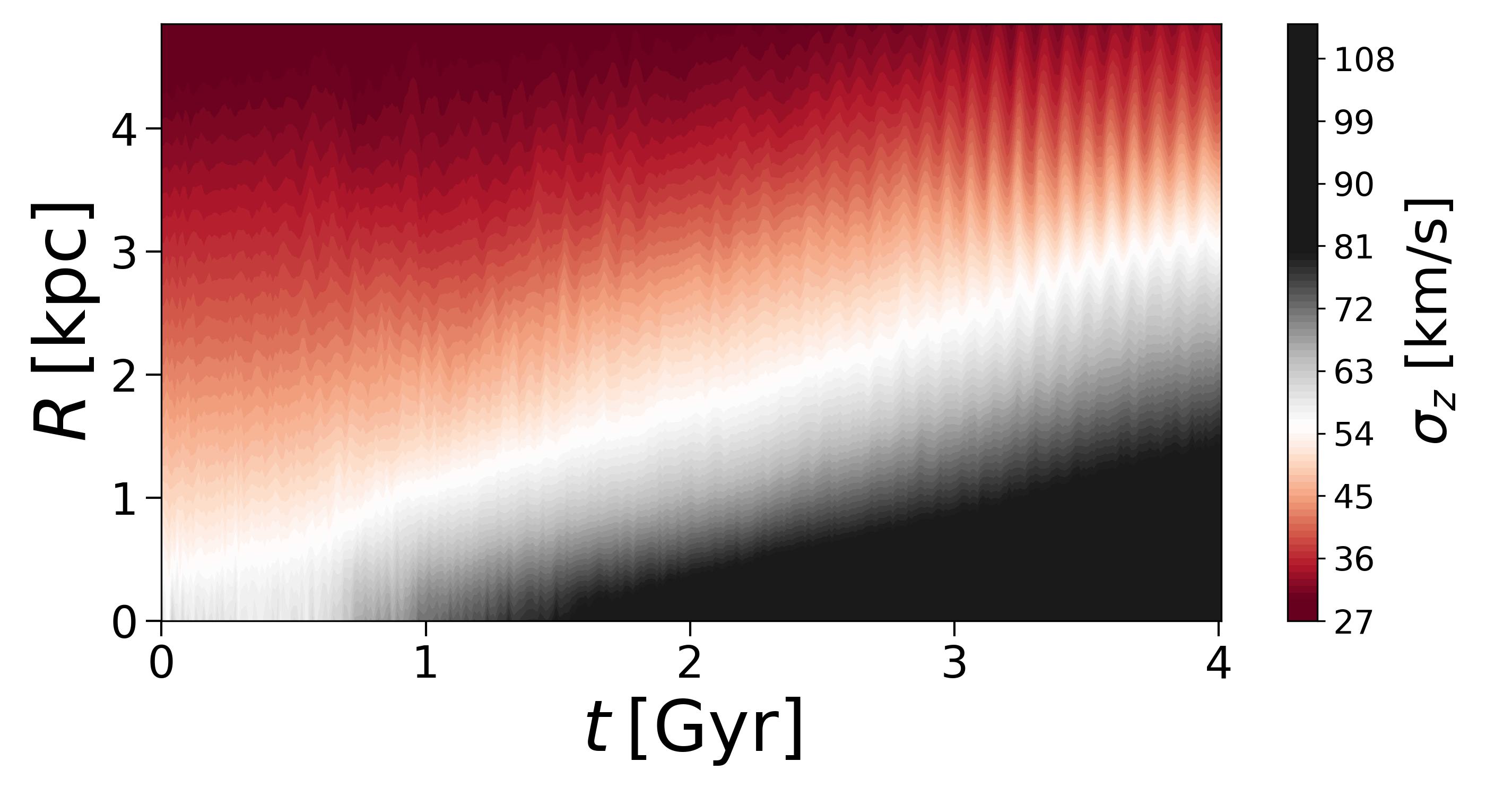} &
        \includegraphics[width=0.33\textwidth]{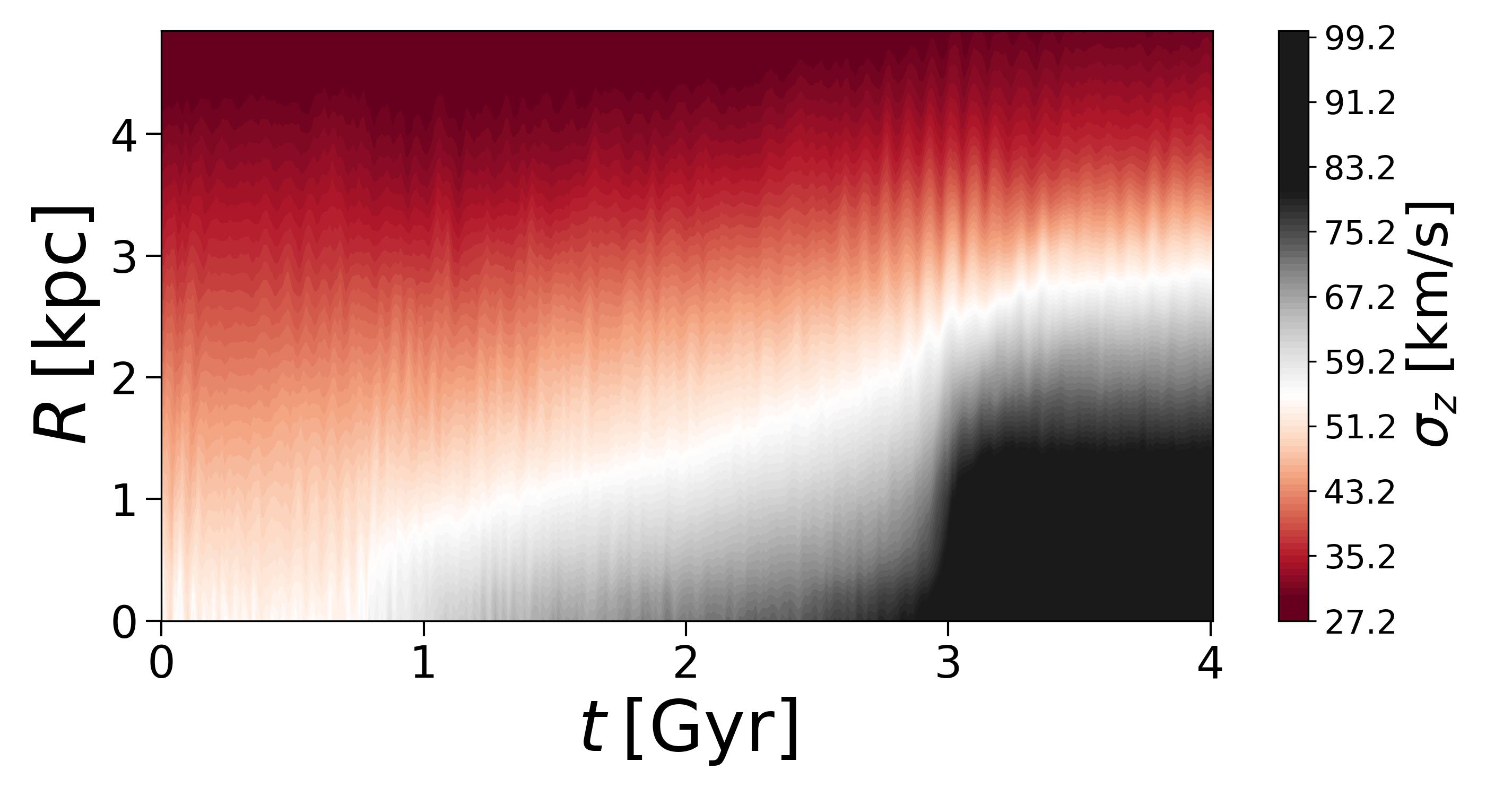} \\
    \end{tabular}
    \caption{Time evolution of the radial distribution of $v_z$ (top panels) and $\sigma_z$ (bottom panels) in the stellar disks over $4 \, \Gyr$. Left, middle, and right: r1c14, r1c14ldm, and r1c14ldmsf06 models, respectively. For the vertical velocity $  v_z  $, the color distribution inside the color bar is fixed across all three models to ensure direct comparison, although the overall range of the color bar is not fixed. Sudden fluctuations in the $  v_z  $ maps indicate the onset of buckling instability. For the vertical velocity dispersion $  \sigma_z  $, the color scale is fixed for values in the range $  30  $--$  80\,\mathrm{km\,s^{-1}}  $. The black regions indicate strong vertical heating of the stellar disk.}
    \label{fig:vz}
\end{figure*}

Figure \ref{fig:face} illustrates the surface density distribution of the stellar disk in all models within a $30 \kpc \times 30 \kpc$ box at the end of the simulation ($4 \, \Gyr$).
The "c16" models share the same disk parameters but differ in DM resolution and softening. These differences yield noticeably distinct outcomes.
Note that the r1c16 model is marginally unstable to non-axisymmetric structure formation, as demonstrated in Paper I. Here, the term ``marginal'' refers to an intermediate stability regime that lies in the middle of the stability spectrum, where resolution effects cause significant divergence in the evolutionary path and formation epoch of non-axisymmetric structures (Paper I).
The bar formation epochs of r1c16 and r1c16ldm, based on $\Ftmax >0.3$, are $1.67 \, \Gyr$ and $1.89 \, \Gyr$, respectively.
Model r1c16 exhibits a prominent bar at $4 \, \Gyr$, as does model r1c16ldm despite its low DM resolution ($\mratio=100$).
The bar in model r1c16ldmsf06 ($\epsdm=0.60 \kpc$), which has larger softening than r1c16 and r1c16ldm ($\epsdm=0.03 \kpc$), appears slightly weaker.
Interestingly, despite the same initial Toomre $Q$ and density structures, model r1c16ldmsf1 ($\epsdm=0.96 \kpc$) fails to form an elongated bar even by the end of the simulation.
One might conjecture that the resolution of its DM halo prevents bar formation.
However, model r1c16sf1, with ten times more DM particles ($\mratio=10$), still does not develop a large-scale bar owing to the large DM softening value ($\epsdm=0.96 \kpc$); only a weak inner oval or short bar is present.
This implies that the choice of an excessively large halo softening length, which is often adopted to mask numerical noise in low DM resolution simulations, exerts a greater impact on bar formation and growth.

The stellar disks in the "c14" models are gravitationally more unstable than those in the "c16" models in terms of the Toomre $Q$ parameter.
The lower halo concentration parameter in the "c14" models results in a smaller DM contribution within the disk region. This results in a lower Toomre $Q$ distribution and makes the disks more susceptible to instabilities and the formation of non-axisymmetric structures.
In fact, as shown in Paper I, model r1c14 forms a visible bar by surpassing $\Ftmax>0.3$ at $1.26 \, \Gyr$. This occurs $0.41 \, \Gyr$ earlier than in model r1c16 with the same resolution.
In Fig. \ref{fig:face}, model r1c14 forms a well-elongated and strong bar after evolving for $4 \, \Gyr$.

Overall, in models with lower resolution or larger softening, the bars are shorter and weaker. Note that the bar in r1c16ldm appears longer in Fig.~\ref{fig:face}, but this is actually a transient moment of bar–spiral mode coupling, during which the bar length can be overestimated \citep{minchev10,quillen11,hilmi20,marques25,kwak25}. For instance, at 3.9 Gyr the bar in the same model appears morphologically shorter owing to the absence of such mode coupling (see Fig.~\ref{fig:face_modecouple}).

Although model r1c16ldmsf1 with $\epsdm=0.96 \, \kpc$ fails to form a bar, model r1c14ldmsf1 does form one. However, this bar appears much smaller than those in the other models.
This implies that bar formation and growth depend on the initial disk stability, softening length, and DM halo resolution.
In a more unstable disk, a stronger seed bar (e.g., \citealt{bekki23}) can form earlier. We conjecture that this enables bar formation in model r1c14ldmsf1 despite the suppressed bar--halo angular momentum exchange due to the large DM softening. However, the growth of the bar is not fully resolved owing to the suppressed DM response (and thus reduced DM-star dynamical friction) in the central bar region. Consequently, the bar is unable to grow stronger and longer in model r1c14ldmsf1.

\subsection{Bar formation and growth}\label{sec:barformation}

To trace the formation and evolution of non-axisymmetric structures, we performed a Fourier analysis:
\begin{eqnarray}
F_0(R)&=& \sum_{j} \mu_{j},\\
F_{m}(R) &=& \frac{\sum_{j} \mu_{j} \, e^{i m \phi_{j}}}{F_{0}(R)}.
\end{eqnarray}
Here, $\mu_{j}$ and $\phi_{j}$ are the mass and azimuthal angle of the $j$th particle in an annulus with a radial bin width of $\Delta R = 0.2\,\mathrm{kpc}$.
The value of $m$ is an integer that denotes the multipole order. We primarily focus on $m=2$ to quantify the bar strength and examine the evolutionary path of bars.

Figure \ref{fig:fmapall} shows the time evolution of the Fourier mode $m=2$ within $8 \, \kpc$ over $4 \, \Gyr$.
The left and right columns correspond to the "c16" and "c14" models, respectively. The scale of the color bar is fixed from 0.01 to 0.40. Red and orange regions indicate strong $m=2$ non-axisymmetry with $F_2 > 0.3$, which we use as an approximate bar-length proxy. However, near the bar end this threshold can be exceeded during episodes of bar-spiral mode coupling \citep{minchev10, quillen11,hilmi20,marques25}. In these intervals, because the bar rotates faster than the spiral pattern, it periodically overlaps with the $m=2$ spiral on the timescale set by their beat frequency. During such alignments the two $m=2$ components add constructively, boosting $F_2$ outside the physical bar and thereby increasing the bar's measurable length in the $F_2$ map. Depending on the strength of the inner spiral structure, the apparent bar length can be overestimated by up to a factor of $\sim2$, both in stellar surface-density maps \citep{hilmi20} and in the centrally induced quadrupole velocity field \citep{vislosky24}. A more conservative length estimate is therefore given by the lower envelope of the $F_2>0.3$ boundary (times when the spiral contribution is weakest).

Overall, the bar formation epochs are not significantly different among models with the same disk stability (see Fig.~\ref{fig:fmapall}), except for the "sf1" models, where bar formation is strongly suppressed. As shown in Paper I, more stable disks are more susceptible to Poisson noise, which causes larger variations in the timing of non-axisymmetric structure formation. For instance, increasing the total resolution (star and DM) of model r1c16 by a factor of 10 delays bar formation by $1.9 \, \Gyr$, whereas the delay is only $0.5 \, \Gyr$ in model r1c14 with the same resolution enhancement (the "r2" models in Paper I). Additionally, the resolution of the DM halo, which is one of the main focuses of this study, slightly delays bar formation as well: (1) decreasing the DM resolution by a factor of 10 (model r1c16ldm) triggers spurious spirals due to massive DM particles, which heat the disk and thus slightly stabilize it; (2) increasing the DM resolution (model r1c16hdm in Paper I) reduces discreteness noise associated with large $\mratio$, resulting in a smoother transition from the spiral phase to the bar phase. Again, all the models have the same disk resolution in this work. In addition to these resolution effects, varying the DM softening lengths introduces further complexity to the numerical effects and the ensuing bar formation. Still, relative to the case of increasing the total resolution by a factor of 10, these differences driven by softening and DM resolution in bar formation epochs are not significant, except for the "sf1" models with $\epsdm=0.96 \kpc$.

Model r1c16 possesses a small bar at $2 \, \Gyr$ (Fig.~1 in Paper I), which grows stronger afterward via angular momentum exchange between stars and DM \citep{athanassoula02,sellwood16,kwak17}, shown until 4 Gyr in Fig. \ref{fig:fmapall}.
Although the bar formation time delay is only $0.41 \, \Gyr$ relative to model r1c14, the two models undergo significantly different evolutionary paths in bar formation and growth.
Model r1c14 forms the strongest bar among all the models. 
Model r1c16 forms a weaker bar and experiences two bar weakening phases around $2.5 \, \Gyr$ and $3.5 \, \Gyr$, due to vertical buckling instability.
The bar in model r1c14 grows continuously without experiencing noticeable buckling instability and forms a well-elongated bar with the region of $F_2 > 0.3$ extending to $4 \, \kpc$ at $4 \, \Gyr$, excluding the bar-spiral coupled regions.
In terms of length and strength, model r1c14 with $\mratio=10$ forms the longest bar.
Note that the vertical buckling instability can occur either gradually with a small amplitude in $v_z$ or suddenly with a strong impact in $v_z$ (e.g., \citealt{kwak17,seo19, lokas20}).
In the following sections, we refer to the gradual and weak heating as vertical instability and the sudden, strong one as buckling instability.

Decreasing the DM halo resolution while keeping the DM softening the same has no significant effect over the $4 \, \Gyr$ evolution period in our idealized disk-halo systems.
Model r1c16ldm forms a bar slightly later than model r1c16 due to the stabilizing effects from early spiral heating by massive DM particles with $\mratio=100$ as shown in Paper I (see their Fig. 4). It also experiences weakening in $F_2$ around $3.5 \, \Gyr$ within the central few kiloparsec region, but not as much as r1c16 does. The overall strength and size of the bar are comparable to those in r1c16.
When comparing models r1c14 and r1c14ldm, their overall evolutionary paths appear similar, although the bar is slightly shorter at 4 Gyr in model r1c14ldm.
There are small differences in minor details when $\mratio=10 \rightarrow 100$, but the overall outcomes appear comparable.

At a glance, Fig. \ref{fig:fmapall} reveals that DM softening exerts the most dramatic effects on bar formation and growth. As the DM softening length increases ($\epsdm = 0.03 \rightarrow 0.60 \rightarrow 0.96 \, \kpc$), the bar strength visibly decays.
In the "sf06" models with $\epsdm=0.60 \, \kpc$, the bar forms more weakly and the bar-driven spirals noticeably diminish relative to models with $\epsdm=0.03 \, \kpc$.
Furthermore, model r1c14ldmsf06 experiences significant bar weakening around $3 \, \Gyr$. Concurrently, bar-driven spirals appear larger temporarily. Note that these spiral structures are distinct from the periodic spirals in other models with $\epsdm=0.03 \, \kpc$ and represent remnants of the buckling instability (see Fig.~8 in \citealt{lokas19}). During the buckling instability, in addition to vertical thickening and the formation of an X-shaped bulge (linked to the bar’s 2:1 vertical resonance; \citealt{quillen14}), the bar dissolves and shortens, forming the barred spirals by ejecting a fraction of stars from the barred orbits.
After the buckling instability, its bar strength remains weak and does not grow stronger as in the pre-buckling phase.

In the "sf1" models, the large DM softening ($\epsdm=0.96 \, \kpc$) significantly suppresses bar formation (Fig.~\ref{fig:fmapall}). The bar strength in model r1c16ldmsf1 does not exceed $F_2=0.3$. In the face-on views (Fig.~\ref{fig:face}), this model develops at most a weak and short inner oval, i.e., it does not form a strong, extended bar by our criterion, even after 4~\Gyr. In model r1c14ldmsf1, which features greater instability, a bar still forms but appears substantially weaker and shorter.
As conjectured in Sect. \ref{sec:overview}, this suggests that the threshold of maximum DM softening length for bar formation varies depending on the disk stability.
However, even when bars form despite an excessive DM softening length, their growth may not be well promoted by the absence of the central angular momentum exchange.

To examine the effects of large DM softening alone ($\epsdm=0.96 \, \kpc$) without mixing the DM resolution effects, we evolved model r1c16sf1, which has the same resolution with model r1c16 ($\ndm=1.14\times10^7$ and $\mratio=10$) but differs only in the DM softening length, increased from $\epsdm=0.03 \, \kpc$ to $0.96 \, \kpc$. In Fig. \ref{fig:fmapsf1}, we calculated the time evolution of the Fourier modes ($m=1$ to 6) and $\Fsum$, where $\Fsum(R) = \sqrt{\sum_{m=1}^{6} \bigl[F_{m}(R)\bigr]^{2}}$. Note that similar figures in Paper I use a color-bar scale ranging from $1\%$ to $15\%$ to highlight spirals in the corresponding modes, whereas here we adopt a scale from $1\%$ to $40\%$ to emphasize the bar mode.

As introduced in Paper I, multimode spirals emerge sequentially from higher to lower modes, accompanied by a decaying epicenter (e.g., $m=5$ and $6$ appear in the outer region of model r2c16, followed by lower modes in the inner region; Fig. 6 in Paper I) This mode cascade proceeds without significant angular momentum loss to the halo until reaching $m=3$, at which point the spiral phase transitions to the bar phase through central mode amplification and subsequent dynamical friction from the DM halo. Capturing this natural mode evolution requires sufficient resolution and an appropriate mass ratio.

Model r1c16sf1 also displays such mode cascading during multimode spiral formation: the $m=6$ spiral appears first around $6 \, \kpc$, followed by lower modes. The $m=3$ mode emerges toward the central region where central angular momentum exchange plays a crucial role in the subsequent bar formation. However, owing to the unresolved star-DM interaction within $R\approx1 \, \kpc$, the amplitude of the $m=2$ mode in the central region remains faint throughout the $4 \, \Gyr$ evolution. Consequently, the model fails to develop a strong bar despite sufficient DM resolution.

Comparing the $F_2$ amplitude in model r1c16sf1 with that of model r1c16ldmsf1 in Fig. \ref{fig:fmapall}, the amplitude appears even more suppressed in r1c16sf1. We conjecture that the massive DM particles in r1c16ldmsf1 ($\mratio=100$) introduce numerical perturbations that contribute to the initial $F_2$ amplification. Still, neither model develops a bar-like structure with $F_2 > 0.3$ by 4 Gyr, likely due to the unresolved angular momentum exchange within $R\approx1 \, \kpc$ when $\epsdm=0.96\,\kpc$.

In Fig. \ref{fig:f2max}, we present the maximum value of $F_2$ within $20 \, \kpc$ as a function of time. To improve visibility, we applied smoothing using a moving average over nine points. The raw data without smoothing are shown in Fig. \ref{fig:f2max_raw}, which exhibit a wide range of oscillations due to bar-spiral periodic overlap \citep{hilmi20, marques25}.
In models r1c16ldmsf1 and r1c16sf1, which do not develop stable, extended bars, these oscillations reflect recurrent episodes of transient inner $m=2$ distortions that do not settle into a coherent bar (Fig.~\ref{fig:f2max_raw}). Morphologically, their inner $  m=2  $ mode does not resemble a bar; rather, it appears as a perturbed disk (Fig.~\ref{fig:face}).
Their stellar disks remain unstable due to the combination of stability and numerical perturbations; however, the lack of angular momentum transfer prevents the phase transition to bar formation, thereby sustaining these fluctuations until the end of their evolution.

Models r1c16 and r1c16ldm with $\epsdm=0.03 \, \kpc$ form the strongest bars among the "c16" models, reaching $\Ftmax \approx 0.5$, whereas models with larger softening attain lower peak values (Fig.~\ref{fig:f2max}). Both "sf1" models fail to grow beyond $F_2 = 0.3$, although r1c16ldmsf1 reaches a slightly higher value owing to enhanced instability from perturbations driven by massive DM particles. Model r1c14ldmsf1 reaches $F_2 \approx 0.3$ but does not grow further due to the absence of central angular momentum exchange.

The "c14" models, except for r1c14ldmsf1, follow similar bar evolution paths until they diverge due to vertical buckling instability.
Note that increasing the DM resolution to $\mratio=1$ in model r1c14hdm does not substantially alter the evolution of bar strength relative to model r1c14 with $\mratio=10$ (see Fig.~\ref{fig:f2max_lz_appendix} and \ref{fig:sigmarz_hdm} in Appendix \ref{appendix:additional}).
Additional unstable disks, such as those in the "c14" models, are less sensitive to resolution effects in terms of the bar formation epoch, making direct comparison between models straightforward. These models follow a similar evolutionary path until around 3 Gyr, when they diverge due to buckling instability. In particular, model r1c14ldmsf06 (with larger softening) undergoes a pronounced buckling instability that significantly weakens the bar strength. We return to the origin of the buckling instability induced by the softening in Sect.~\ref{sec33}. 
In Fig. \ref{fig:f2max}, the bar strength in model r1c14ldmsf06 decreases by 0.2 from $F_2 \approx 0.56$, whereas r1c14 continues to grow and r1c14ldm undergoes gradual bar weakening.
In model r1c14ldmsf03 (Appendix \ref{appendix:additional}), the buckling instability rapidly decreases the bar strength by 0.08 from $F_2 \approx 0.48$ around $2.2 \, \Gyr$ (Fig.~\ref{fig:f2max_lz_appendix}).
This indicates a correlation between the DM softening length and the magnitude of buckling-induced weakening in $F_2$.

To quantify the angular momentum transferred from stars to DM halos, we measured the total disk angular momentum $L_z$ normalized by its initial value $L_{z,0}$ in Fig. \ref{fig:lz}.
The exchange of angular momentum between stars and DM particles serves as the primary mechanism driving bar formation \citep{athanassoula02,athanassoula03,kwak17,kwak19}.
The extent of this angular momentum transfer correlates with the bar's strength and length.
As the stellar bar transfers angular momentum to the DM halo, it decelerates and strengthens, which in turn elevates the radial velocity dispersion of stars and eventually triggers vertical buckling instability \citep{raha91,combes90,martinez04, martinez06,debattista06,kwak17,kwak19,lokas19b, lokas19, lokas25a}.
In the "c16" models, r1c16ldm, which forms the strongest bar, exhibits the greatest angular momentum loss. This may arise because massive DM particles are more effective at facilitating angular momentum exchange during the early phases when the bar is relatively weak.
Compared to r1c16, model r1c16ldmsf06 forms a less strong bar and thus loses less angular momentum, reflecting a correlation between DM softening length and bar strength, as illustrated in Figs. \ref{fig:fmapall} and \ref{fig:f2max}.
In the "c14" models, which form strong bars, the total stellar angular momentum loss reflects the combined effects of resolution and DM softening length: higher resolution and smaller softening boost the angular momentum exchange more effectively.
The difference is small between models r1c14 and r1c14ldm, attributable to gradual bar weakening in r1c14ldm around $3.5 \, \Gyr$, whereas the bar in r1c14 continuously grows longer ($R > 4 \, \kpc$ where $F_2 > 0.3$ in Fig.~\ref{fig:fmapall}).

The DM softening length induces more pronounced changes in bar formation and buckling instability than resolution effects alone. To investigate this, we measured the spherical density profile of the DM halo, $\rho_{\rm DM}$, from $0.1 \, \kpc$ to $5 \, \kpc$ on a log-log scale in Fig. \ref{fig:den}.
We present the "c16" models, as the density profiles of the "c14" models are similar for equivalent resolution and softening lengths (Fig.~\ref{fig:c14density}).
In the "sf1" models ($\epsdm=0.96 \, \kpc$), the central density profiles flatten in the inner $1 \, \kpc$ regardless of DM halo resolution, remaining around $\rho_{\rm DM} \approx 5 \times 10^8 \, \Msun \, \kpc^{-3}$ throughout the evolution.
This flattening is a well-known numerical artifact in cosmological simulations, where Plummer or spline softening mitigates two-body relaxation and noise but underresolves gravitational forces on scales below the softening length, yielding a spurious core instead of the expected cusp \citep{fukushige01, power03, diemand05}. Therefore, the softened potential inhibits tight particle clustering in the center. The effect persists independent of particle number once convergence is reached and scales directly with the softening length.
In our initial conditions, the central cusp hosts most DM particles, but unresolved interactions within the softening length prevent them from staying bound at the very center, thus eroding the cusp rapidly.
In addition to the bar weakening effects, larger softening lengths exacerbate this central flattening. By contrast, models r1c16 and r1c16ldm sustain $\rho_{\rm DM} > 10^9 \, \Msun \, \kpc^{-3}$, with gradual increases over time. 
In general, a reduction in the central mass concentration of the spheroidal component destabilizes the disk and promote bar formation \citep{kwak17, kataria18,  jang23}. However, our results indicate that central mass concentration alone is insufficient to produce a strong bar unless accompanied by efficient angular momentum transfer \citep{athanassoula02, athanassoula03}, particularly in the central region where a nascent bar begins to grow.

\subsection{Resolution and softening on the buckling instability}\label{sec33}
The buckling instability in galactic bars is a vertical bending mode that thickens the bar out of the disk plane, often leading to the formation of boxy or peanut-shaped bulges, as the bar's self-gravity amplifies asymmetries in the velocity distribution \citep{combes90,raha91,kwak17,kwak19, lokas19b,lokas19,lokas25a}. It is often associated with anisotropy in the stellar velocity dispersions, where excessive radial heating from the bar's growth creates an imbalance and triggers out-of-plane oscillations when the vertical support becomes insufficient \citep{merritt94}. A live halo can promote angular momentum exchange and thereby enable stronger bar growth, which can lead to buckling instability \citep{berentzen06}. A common criterion for its onset is when the ratio of vertical to radial velocity dispersion, $\sigma_z / \sigma_R$, drops below 0.5--0.6, although this threshold can vary with factors such as the bar's properties, the presence of a gaseous component, and the halo's dynamical influence \citep{debattista06, kwak17, kwak19, seo19, jang24, jang25}. Moreover, the peanut or X-shape can track the radius of the bar’s 2:1 vertical resonance and migrate outward as the bar slows and the disk thickens \citep{quillen14}. Accordingly, the presence of a peanut or X-shape alone is not uniquely diagnostic of a distinct buckling episode; we therefore relied primarily on kinematic diagnostics, such as $\sigma_z / \sigma_R$.

In Fig. \ref{fig:sigmarz}, we examine the radial profiles of surface density, vertical velocity dispersion $\sigma_z$, and the ratio $\sigma_z / \sigma_R$ for models r1c14, r1c14ldm, and r1c14ldmsf06 to show the impacts of DM resolution and softening on buckling instability. These models were chosen as they follow comparable bar evolutionary paths until divergence arises from vertical buckling instability. To capture the transitions in these properties from the pre- to post-buckling phases, we compute them starting at $2 \, \Gyr$ and extending to $4 \, \Gyr$ with a time step of $0.25 \, \Gyr$. Overall, as evolution proceeds, the surface density in these bar-forming models increasingly contracts and becomes centrally concentrated (e.g., see Fig.~6 in \citealt{kwak17}).

The vertical velocity dispersion increases gradually in most models, except for r1c14ldmsf06, which undergoes a strong buckling instability that vertically perturbs and thickens the stellar components (Fig.~\ref{fig:sigmarz}). In model r1c14, the central $\sigma_z$ ranges from approximately 80 to 100~$\kms$, whereas in r1c14ldm it is about $10\%$ higher (90 to 110~$\kms$), attributable to spurious numerical heating induced by massive DM particles \citep{ludlow21}. In contrast, model r1c14ldmsf06 exhibits a sudden increase in $\sigma_z$ at 2.75~Gyr, when it reaches slightly below 80~$\kms$. This coincides with the epoch of abrupt bar strength decline (Fig.~\ref{fig:f2max}). Although models r1c14ldm and r1c14ldmsf06 share the same DM resolution, the numerical disk heating from massive DM particles is far less effective in r1c14ldmsf06 due to its larger softening length ($\epsdm=0.60 \, \kpc$), resulting in central $\sigma_z$ values more than $20\%$ lower relative to r1c14ldm ($\epsdm=0.03 \, \kpc$). The vertical velocity dispersion $\sigma_z$ acts as a stabilizing factor against the onset and recurrence of buckling instability. For instance, \cite{kwak17} showed that the initial buckling event occurs in the inner bar, elevating local $\sigma_z$ and thereby shifting subsequent recurrences to outer regions less affected by the prior vertical heating.

The ratio $\sigma_z / \sigma_R$ differs markedly among the three models in the bottom panels of Fig. ~\ref{fig:sigmarz}. In the models with lower DM halo resolution (model r1c14ldm and r1c14ldmsf06), this ratio does not drop below $\sim0.5$, whereas in r1c14—which sustains continuous growth of a well-elongated bar extending beyond $4 \, \kpc$ without weakening—the ratio falls below 0.5 for $R > 3 \, \kpc$ by $4 \, \Gyr$. Model r1c14ldm, which experiences gradual vertical instability from 3.25 to $4 \, \Gyr$ (Fig.~\ref{fig:f2max}), shows $\sigma_z / \sigma_R$ rise to $\approx0.6$ after approaching $\approx0.5$. The underlying mechanism driving this gradual bar weakening in the outer region due to DM resolution differences remains unclear and requires further investigation in the future in light of the radial distribution of the vertical instability threshold. As noted in \cite{kwak17}, initially hotter disks undergo vertical instability earlier in the outer regions (see their models DP3 and DP4 in Fig. 14). We therefore conjecture that the elevated disk heating ($\sigma_R$ and $\sigma_z$) in model r1c14ldm, induced by spurious heating from massive DM particles prior to bar formation, likely triggers the vertical instability once the bar extends to outer regions at $R \approx 4 \, \kpc$. Apart from the resolution effects ($\mratio=100$), model r1c14ldmsf06 displays a pronounced increase in $\sigma_z / \sigma_R$ owing to its strong buckling instability.

To trace the occurrence of buckling instability, we present radial distribution maps of $v_z$ and $\sigma_z$ as functions of time in Fig. \ref{fig:vz}. The color bars are fixed, ranging from $-15$ to $15 \, \kms$ for $v_z$ and from $30$ to $80 \, \kms$ for $\sigma_z$, ensuring that identical colors represent equivalent values across models. Models r1c14 and r1c14ldm exhibit no visible changes in $v_z$, whereas r1c14ldmsf06 displays pronounced fluctuations around $3 \, \Gyr$, coinciding with a substantial decline in bar strength (Fig.~\ref{fig:f2max}). The duration of elevated $v_z$ is approximately $0.5 \, \Gyr$, aligning with the period during which the bar strength $F_2$ decreases from $0.56$ to $0.36$ (Fig.~\ref{fig:f2max}).

Given that Fig. \ref{fig:sigmarz} indicates lower pre-buckling $\sigma_z$ in r1c14ldmsf06 compared to r1c14ldm, we also show the $\sigma_z$ distribution in the bottom panels of Fig. \ref{fig:vz}. This clearly reveals that the two models with $\epsdm=0.03 \, \kpc$ undergo gradual vertical heating in the central region during bar formation and growth. Again, model r1c14hdm ($\mratio=1$) undergoes nearly the same bar formation without buckling instability, implying that this is not a mass ratio effect (Fig.~\ref{fig:f2max_lz_appendix} and \ref{fig:sigmarz_hdm}).
In contrast, model r1c14ldmsf06 ($\epsdm = 0.60 \, \kpc$) exhibits no such gradual vertical heating in the central region during the same phase of bar formation.
Notably, all three models maintain nearly equivalent bar strengths $F_2$ until shortly before the buckling instability before $3 \, \Gyr$. Despite these comparable bar strengths, the failure to resolve central $\sigma_z$—a key stabilizing factor against buckling—causes greater imbalance in $\sigma_z / \sigma_R$, thereby triggering a stronger buckling instability in model r1c14ldmsf06 (e.g., \citealt{debattista06,kwak17}).

\section{Discussion and conclusion}
Using N-body simulations, we investigated the effects of resolution and gravitational softening in the DM halo on bar formation and buckling instability. In our dissipationless, isolated disk-halo systems, we fixed the stellar disk parameters and varied the resolution, softening length, and concentration parameter of the DM halo, as changes in the DM fraction alter the stability of the disks.
Across two different levels of stability, we examined the relative effects of $\mratio$ and $\epsdm$ on the bar formation epoch, the evolutionary path of bar properties, and bar weakening via buckling instability.
 
\subsection{DM resolution on bar formation}
The formation of non-axisymmetric structures, such as bars, exhibits a sensitivity to numerical noise that is modulated by the intrinsic stability of the disk. Additional unstable disks show reduced variability in the formation epoch due to shot noise (Paper I). Enhancing the resolution of both the stellar disk and the DM halo by a factor of 10 relative to the "r1" models results in a delay in bar formation (0.5 Gyr for the "c14" models and 1.9 Gyr for the "c16" models). This delay stems from a substantial reduction in Poisson noise within the entire system, which suppresses the generation of noise-induced perturbations and spirals. These spirals otherwise cascade toward the center, augmenting bar formation beyond the disk's intrinsic instability.

Interestingly, as demonstrated in Paper I, both increasing and decreasing the DM halo resolution delay bar formation (defined as $F_{2}>0.3$), albeit for different reasons. For example, increasing the DM halo resolution from $\mratio=10$ to $\mratio=1$ induces a delay in bar formation (0.1 Gyr in the "c14" models and 0.9 Gyr in the "c16" models compared with their hdm counterparts). This occurs because the angular momentum transfer mediated by more massive DM particles is reduced when $\mratio=1$ during the transitional phase from spirals to bars (see Fig. 5 in Paper I). Conversely, decreasing the DM halo resolution from $\mratio=10$ to $\mratio=100$ also delays bar formation but through a different mechanism: the passage of massive DM particles ($\mratio=100$) through the disk plane exerts sporadic gravitational shocks throughout the disk, globally heating the stellar disk and enhancing its stability (see Fig. 4 in Paper I). For instance, the NewHorizon simulation \citep{dubois21} reports the ``missing bar'' problem \citep{reddish22} while adopting $\mdm=1.2\times10^6$ and $\mratio \approx 92$. \cite{ludlow21} points out that such low DM resolution ($\ndm \lesssim 10^6$ and $\mdm \gtrsim 10^6 \, \Msun$) artificially elevates the vertical velocity dispersion due to collisional heating on stellar disks. In addition, our findings suggest that the global increase in $\Fsum$ induced by high $\mratio$ introduces another numerical artifact stabilizing the stellar disks against the bar instability. The impacts of stellar disk and DM halo resolution on the bar formation epoch are intricate with various underlying mechanisms. This complexity demands meticulous attention when configuring galaxy models, particularly those residing in the stable regime of the stability spectrum.

The stable regime, as discussed in Paper I, spans a broad range of disk stabilities. The "c16" models lie approximately in the middle of this spectrum, where resolution effects considerably influence the formation epoch of non-axisymmetric structures (Paper I). Indeed, model r1c16 develops a bar at $1.67 \, \Gyr$, defined by $F_2 > 0.3$, and achieves a peak strength of $F_2 \approx 0.5$ shortly after $2 \, \Gyr$. This timescale is relatively brief for bar formation compared to the Hubble time.

For comparison, \cite{jang25} examined bar formation conditions in galaxies using $\nstar = 10^6$ and $\ndm = 2 \times 10^7$, with particle masses of $\mstar \approx 5 \times 10^4 \, \Msun$ and $\mratio \approx 2$ in their Milky Way-mass models.
Regarding the impact of disk resolution on spiral formation and associated disk heating, \cite{fujii11} analyzed numerical effects on spiral structures and their lifetimes (employing $\nstar = 3 \times 10^5$ to $3 \times 10^7$) and determined that a sufficient number of disk particles (e.g., $\nstar \gtrsim 3 \times 10^6$) is essential to sustain spirals without external cooling. Models with fewer star particles generate spirals predominantly through numerical noise, resulting in earlier development and overamplification (see their Fig.~7). These noise-induced spirals transport perturbations more rapidly and prematurely to the central region, thereby initiating DM-star dynamical friction and accelerating bar formation. This process accounts for the observed $1.9 \, \Gyr$ delay in bar formation when increasing resolution from $\nstar = 5 \times 10^6$ to $5 \times 10^7$ with $\mratio=10$ in model r1c16 and r2c16 in Paper I.
Furthermore, bars in the \cite{jang25} models form between approximately $3.5$ and $6.7 \, \Gyr$, indicating that their disk-halo models are more stable than our r1c16 model and thus more susceptible to numerical noise affecting the timing of bar formation.
Such resolution-dependent delays in bar formation have been also reported by \cite{dubinski09}.
Hence, we conjecture that resimulating these stable models with $\nstar = 5 \times 10^7$ stellar particles and $\mratio<10$, which reduce initial noise levels $\Fsum\approx0\%$ (Paper I), could prevent bar formation within the Hubble time in certain cases.

In practice, employing such a large number of particles to generate nearly noise-free initial conditions is challenging.
In low DM resolution cases with $\mratio=100$, surprisingly, the bar formation timing exhibits no substantial difference, although the evolutionary paths diverge later owing to accumulated numerical effects from massive DM particles. For instance, the bar strength peaks lower and experiences bar weakening in model r1c14ldm compared to the high DM resolution models.
Note that our models represent an idealized disk-halo setup, and $\mratio=100$ would yield more significant effects during the hierarchical formation \citep{ludlow19b,ludlow23}.
Fortunately, the evolutionary path of model r1c14hdm ($\mratio=1$), which employs ten times more DM particles, is comparable to that of model r1c14, as evidenced by their similar evolution of bar strength and angular momentum transfer (Fig.~\ref{fig:f2max_lz_appendix}).
We therefore suggest that a resolution comparable to that of model r1c14 ($\mratio=10$, $\mstar=10^4 \, \Msun$, and $\nstar=5\times10^6$) may not significantly alter the outcomes when simulating such unstable galaxies driven by strong self-instabilities or tidal forcing \citep{lokas14,lokas16, lokas19,  kwak19, bekki23}. For relatively stable models in which bars form gradually after $2 \, \Gyr$, the influence of resolution on bar properties, particularly the formation epoch, must be carefully considered. For instance, the epoch, at which the peak bar strength is reached, is 2.2 Gyr in model r1c16 and 3.1 Gyr in model r1c16ldm (see Fig.~\ref{fig:f2max}).

\subsection{Softening effects on bar strength and buckling instability}
% Softening on bar strength
The DM softening effect is more pronounced than that of DM resolution on bar formation. Note that \cite{iannuzzi13} found no significant differences between fixed and adaptive softening lengths, but their chosen softening length was much smaller than ours (e.g., $\epsilon = 0.05 \, \kpc$ in the fixed model). While the inner $1 \, \kpc$ region may appear insignificant relative to the scale radius and virial radius of the DM halo, the central angular momentum exchange plays a crucial role, particularly during the early phases of bar formation. In these stages, the seed bar remains small and grows via dynamical friction between DM particles and stars. Consequently, insufficient angular momentum transfer arising from an excessively large DM softening length can substantially delay or entirely prevent bar formation.

This may provide an additional factor suppressing bar formation and growth in the NewHorizon simulation \citep{dubois21, reddish22}, which uses $  \epsdm=0.32  $ kpc at $  z=2  $--$0.25$ and $  \epsdm=0.50  $ kpc at $  z=0.25  $--$0$. 
\cite{reddish22} present only one visually confirmed bar, with a length of $R(F_2 > 0.3) \approx 0.6 \, \kpc$, in their Figs. 1 and 8. This small bar persists from $z=1.3$ to $z=0.7$, and the host galaxy is the most massive disk. 
Note that the "sf1" models ($  \epsdm=0.94  $ kpc) develop only an oval or short bar that fails to grow beyond $\sim$3 kpc (roughly $3\times\epsdm$), which implies that the $  \epsdm=0.32  $ kpc adopted in \cite{reddish22} is relatively large compared to their bar length.
In general, low DM mass resolution results in larger and hotter stellar disks with colder halos \citep{ludlow23} and produces fewer halos, owing to the inability to properly resolve dwarf-sized DM halos \citep{revaz18,ludlow19b}. This consequently omits a fraction of mergers during hierarchical formation, potentially leading to more DM-dominated galaxies among lower-mass systems, which is also observed in \cite{reddish22}.
Despite these potentially stabilizing effects associated with their mass resolution, their most massive disk could be gravitationally more unstable and become more prone to bar formation, presumably triggered by tidal forcing.
This may enable the emergence of a nascent bar, but the large adopted $\epsdm$ impedes its subsequent growth via angular momentum transfer, akin to the case in our model r1c14ldmsf1 (Fig.~\ref{fig:face}) and the case in \cite{kaufmann07}.

Sufficient resolution and an appropriate $\epsdm$ are important but cannot be the only explanation for the missing bar problem. For example, the Auriga \citep{grand17} and the Hestia simulations \citep{libeskin20} adopted the same model, yet Hestia still suffers from the missing bar problem despite its higher resolution, presumably owing to different initial conditions. In the fully cosmological context, many interrelated factors come into play. Different merger histories can alter the masses of classical bulges: more massive bulges increase the central mass concentration and suppress bar instability \citep{athanassoula05, kwak17, kataria18, saha18, jang23}. In addition, high gas fractions tend to form weaker and shorter stellar bars by damping disk instabilities and angular momentum transfer \citep{athanassoula13, seo19, lokas20, beane23}. In highly turbulent, gas-rich environments, elevated gas content can even dissolve bars through enhanced dissipation, converting them into central bulges as star formation-driven kinematic heating disrupts non-axisymmetric structures \citep{bland24}. Hence, these effects on the bar formation merit further investigation in future studies.

Overall, the cumulative effects of $\epsdm$ over time from the missing central angular momentum exchange reduce the final bar strength, causing models with larger softening parameters to exhibit weaker bars and lower peak bar strengths throughout their evolution. Considering the nearly identical evolutionary paths of bar strength in models r1c14 and r1c14hdm with $\epsdm=0.03 \, \kpc$, adopting $\epsdm\ge0.3 \, \kpc$ has a much greater impact than adopting $\mratio\ge10$ (Fig.~\ref{fig:f2max_lz_appendix}).

Additionally, larger DM softening lengths (e.g., $\epsdm \ge 0.30 \, \kpc$) may overproduce the buckling instability, which also eventually weakens the bar strength. 
A bar rotates and slows down by transferring its angular momentum to the DM halo, thereby growing stronger and increasing the radial velocity dispersion $\sigma_R$ \citep{athanassoula03,kwak17}. In this context, the location of the bar’s 2:1 vertical resonance (and thus the peanut or X-shape signature) can migrate outward as the bar slows and the disk thickens \citep{quillen14}. Regardless of DM resolution, in bar-forming models with $\epsdm = 0.03 \, \kpc$, the vertical velocity dispersion $\sigma_z$ gradually increases, starting from the very center.
A thicker disk or higher velocity dispersion serves as a stabilizing factor against the onset of buckling instability. In studies examining the effects of disk thickness on the evolution of barred galaxies, \cite{klypin09} and \cite{kwak17} found that thinner disks form shorter bars and undergo buckling instability earlier than their thicker counterparts.
However, large softening lengths prevent the gradual vertical heating in the central region during the bar formation, resulting in a numerically induced larger imbalance in $\sigma_z / \sigma_R$ that triggers a stronger buckling instability.
We find that the decrease in bar strength due to buckling instability intensifies as $\epsdm$ increases from $0.03$ to $0.30$ to $0.60 \, \kpc$ (Fig. \ref{fig:f2max_lz_appendix}). 
Notably, softening lengths in this range ($\epsdm=0.30-0.60 \, \kpc$) are commonly adopted in cosmological and zoom-in simulations, including \textsc{EAGLE} \citep{schaye15}, \textsc{Illustris TNG} \citep{pillepich18a,pillepich18b,pillepich19}, \textsc{Auriga} (level 4) \citep{grand17}, and \textsc{NewHorizon} \citep{dubois21}. Similarly, some studies have also adopted $\epsdm = 0.7 \, \kpc$ in their idealized disk-halo systems \citep{lokas16,semczuk17,lokas18,lokas19}.
We conjecture that the central vertical velocity dispersion $\sigma_z (R < \epsdm)$ may be systematically underestimated with $\epsdm \gtrsim 0.3 \, \kpc$, unless a bar undergoes numerically enhanced buckling, which then eventually lowers the peak bar strength $F_2$.

\bigskip
\bigskip
In conclusion, from a practical standpoint, we recommend adopting $\mratio \le 10$, $\mstar \leq 10^4 \, \Msun$, and $\nstar \ge 5 \times 10^6$ when investigating the formation and evolution of non-axisymmetric structures in Milky Way-mass galaxies. Determining the resolution at which the outcomes of bar instability converge would be interesting, but achieving an effectively noise-free system by indefinitely increasing the number of particles is impractical given current computational resources. In real galaxies, natural noise levels of $\Fsum \gtrsim 1\%$ may arise over time from processes such as stellar feedback \citep{marinacci19, kwak26b, kwak26c}, the presence of numerous globular clusters \citep{donghia13}, and tidal forcing by satellites \citep{lokas14,lokas16, kwak19}. In our N-body models, $\Fsum \approx 1\%$ causes significant divergence in the timing of bar formation. We anticipate that such noise effects on bar timing would be less pronounced in gaseous galaxy models with realistic stellar feedback, although this requires further investigation to confirm. Regarding gravitational softening, our models with $\epsdm \ge 0.30 \, \kpc$ exhibit considerable divergence, producing multiple bar weakening or strong buckling even in the unstable regime. This amplifies the radial-vertical velocity dispersion anisotropy and triggers vertical instabilities. In the future, smaller ($\epsdm<0.30\,\kpc$) or adaptive DM softening lengths \citep{iannuzzi13} in the central region, potentially scaled by the galaxy's effective radius depending on its mass and size, may be needed to properly resolve the central DM-star interactions and to prevent numerical vertical instabilities.

\begin{acknowledgements}
We appreciate the anonymous referee for their positive consideration and constructive comments. IM acknowledges support by the Deutsche Forschungsgemeinschaft under the grant MI 2009/2-1. S.K.Y. acknowledges support from the Korean National Research Foundation (RS-2025-00514475; RS-2022-NR070872).
\end{acknowledgements}

\bibliographystyle{aa} 
\bibliography{ref}

\begin{appendix}

\section{Supplementary figures and additional models}\label{appendix:additional}

In addition to the nine models listed in Table \ref{table:model}, we construct two additional models: r1c14hdm and r1c14ldmsf03. The "hdm" designation indicates high DM resolution, adopting $\mratio=1$ with $\ndm=1.14\times10^8$. In model r1c14, with different masses between DM and star particles, potential numerical effects could arise from using the same softening length for stars and DM particles ($0.03 \, \kpc$). It turns out that such effects have a minimal impact on bar formation of unstable disks when comparing model r1c14 and r1c14hdm. For model r1c14ldmsf03, we test the effect of a DM softening length of $0.30 \, \kpc$ to demonstrate the correlation between the softening length and the buckling effects on bar strength. The results are overlaid for comparison in Fig. \ref{fig:f2max_lz_appendix}.

\begin{figure}[ht]
    %\centering    
    \includegraphics[width=0.245\textwidth]{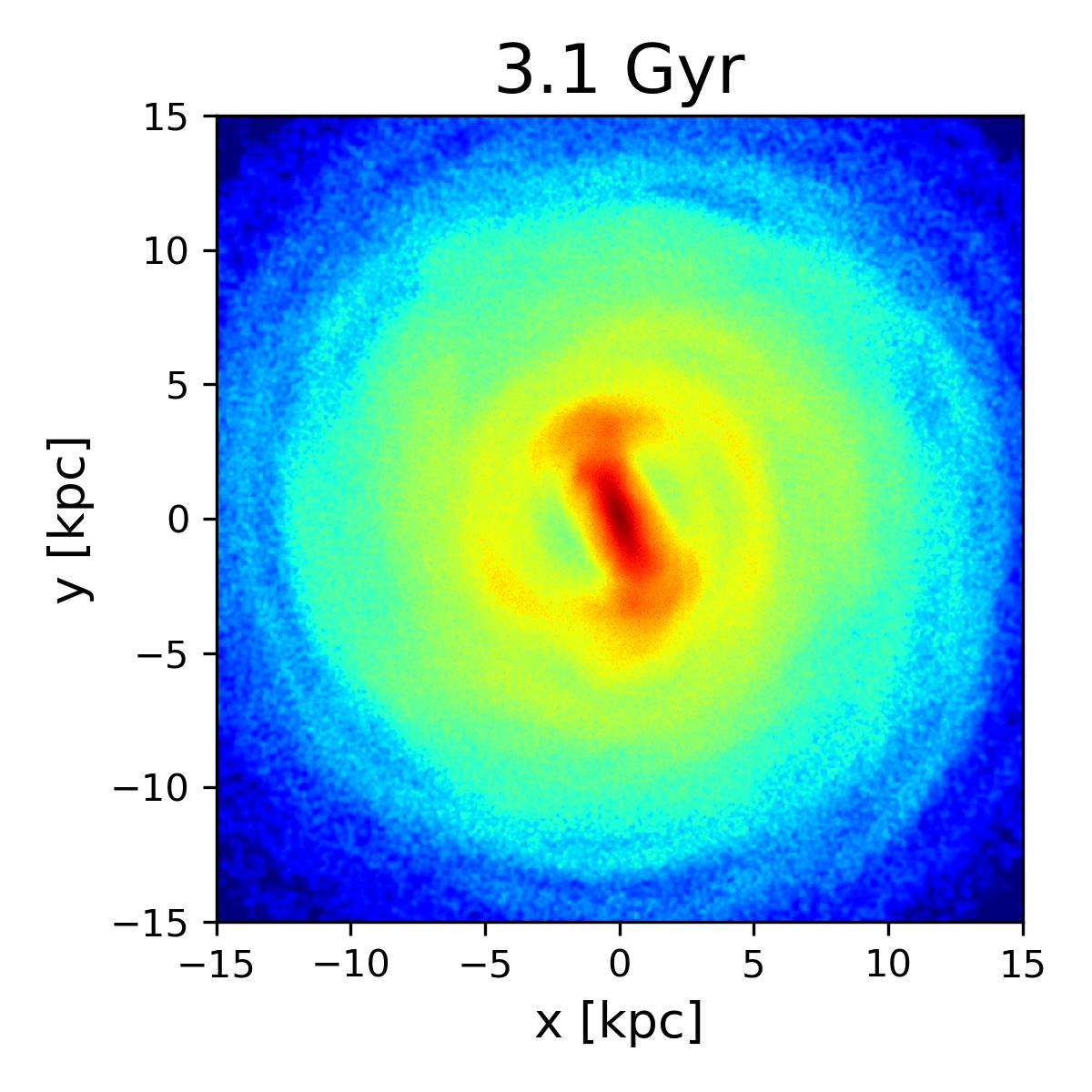}\includegraphics[width=0.245\textwidth]{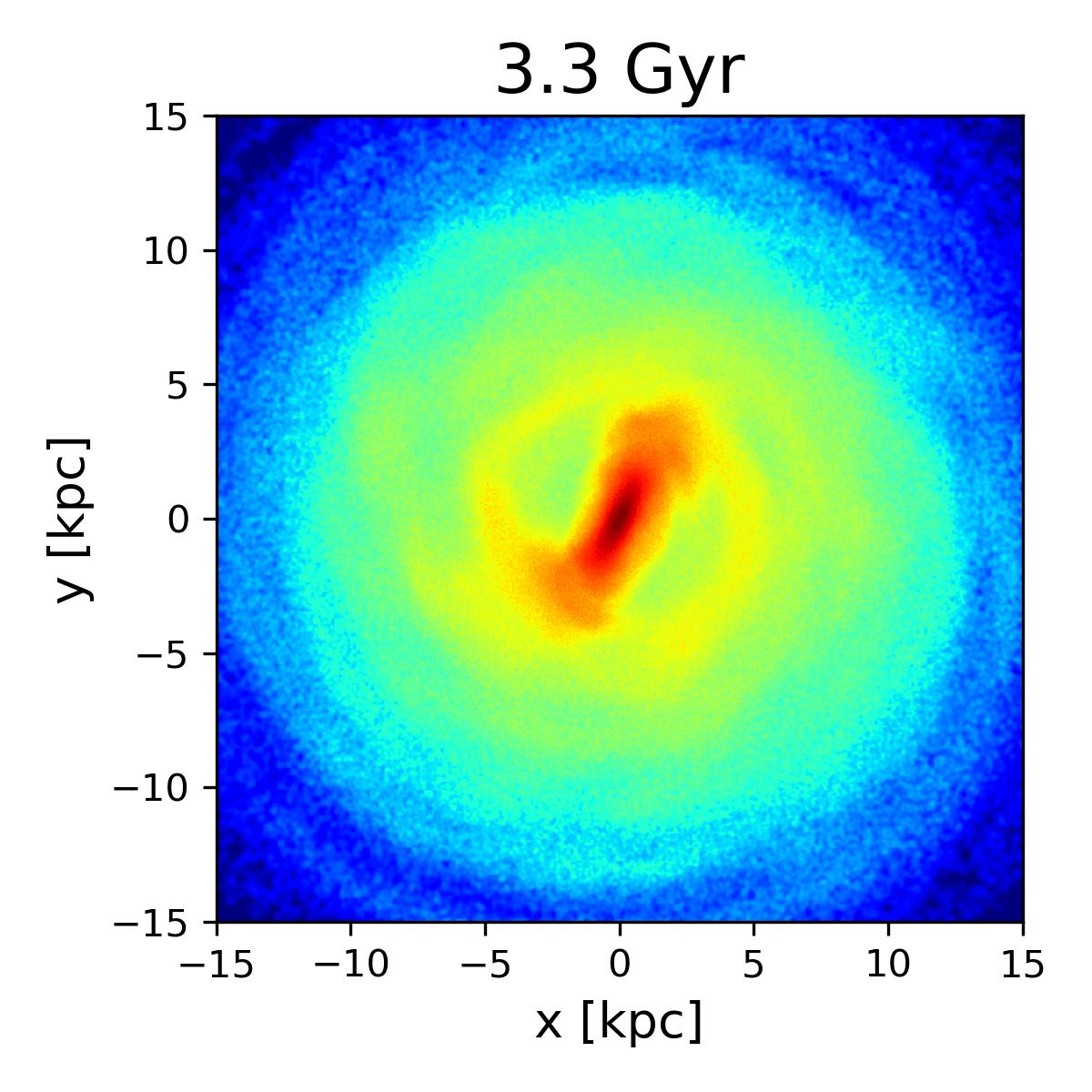}
    \includegraphics[width=0.245\textwidth]{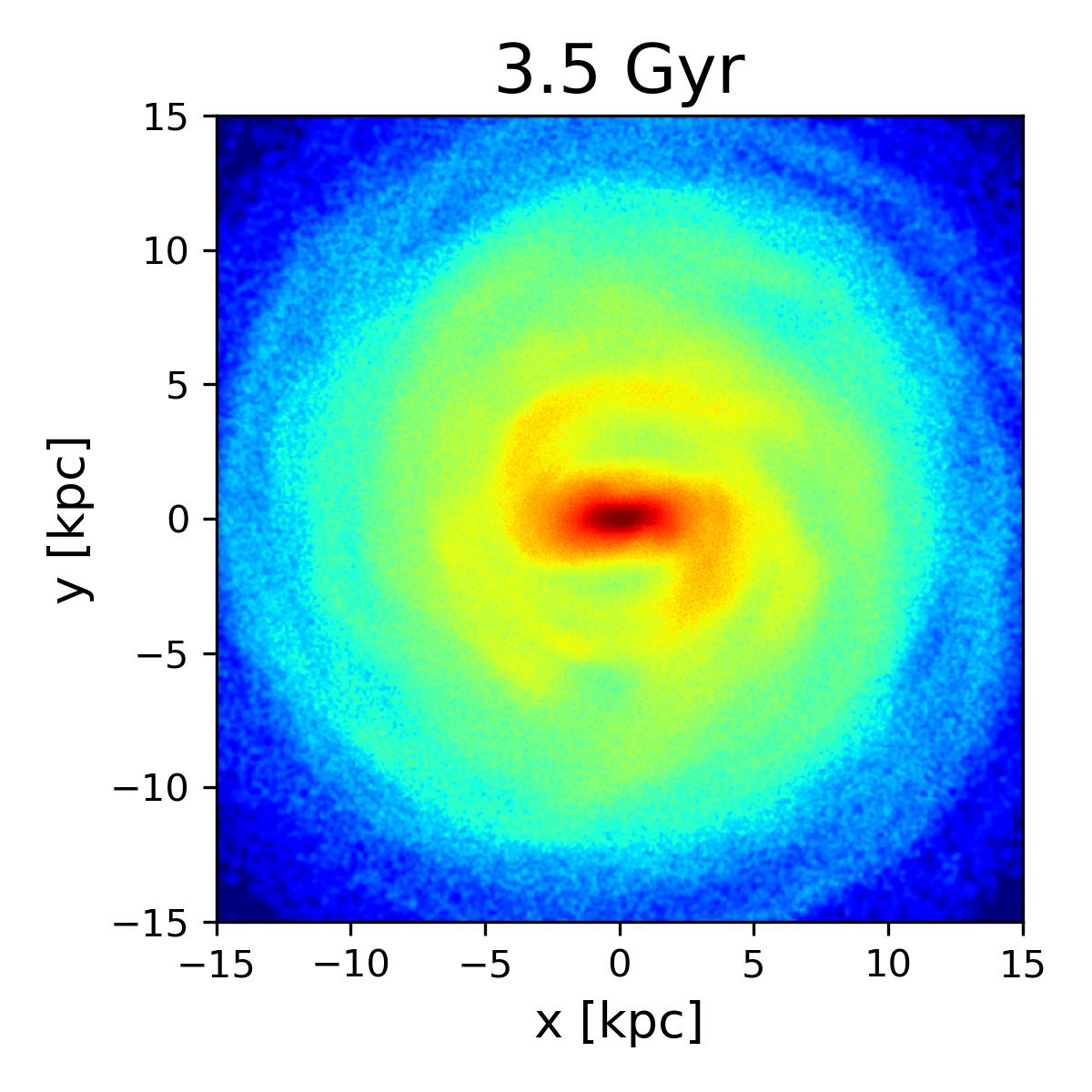}\includegraphics[width=0.245\textwidth]{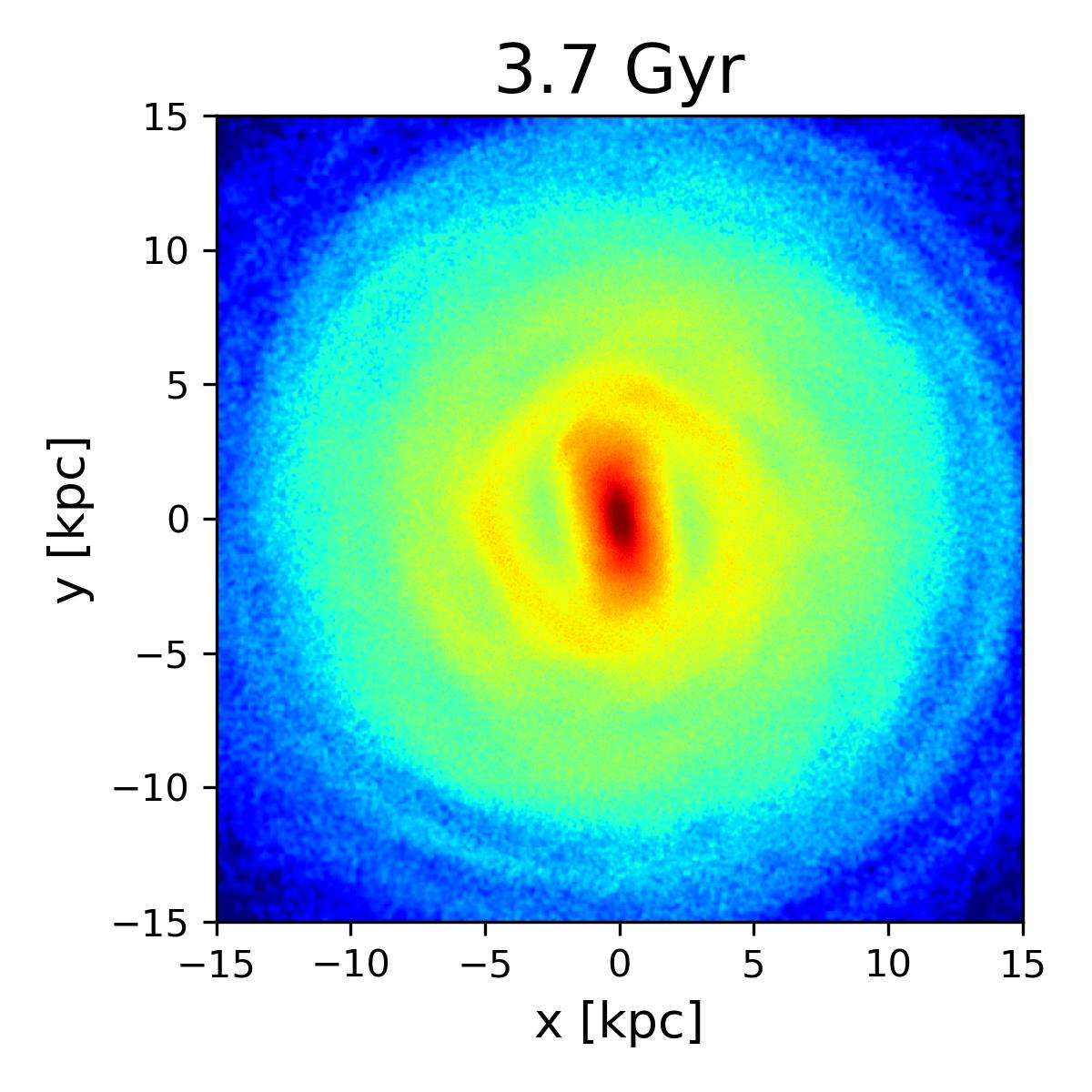}
    \includegraphics[width=0.245\textwidth]{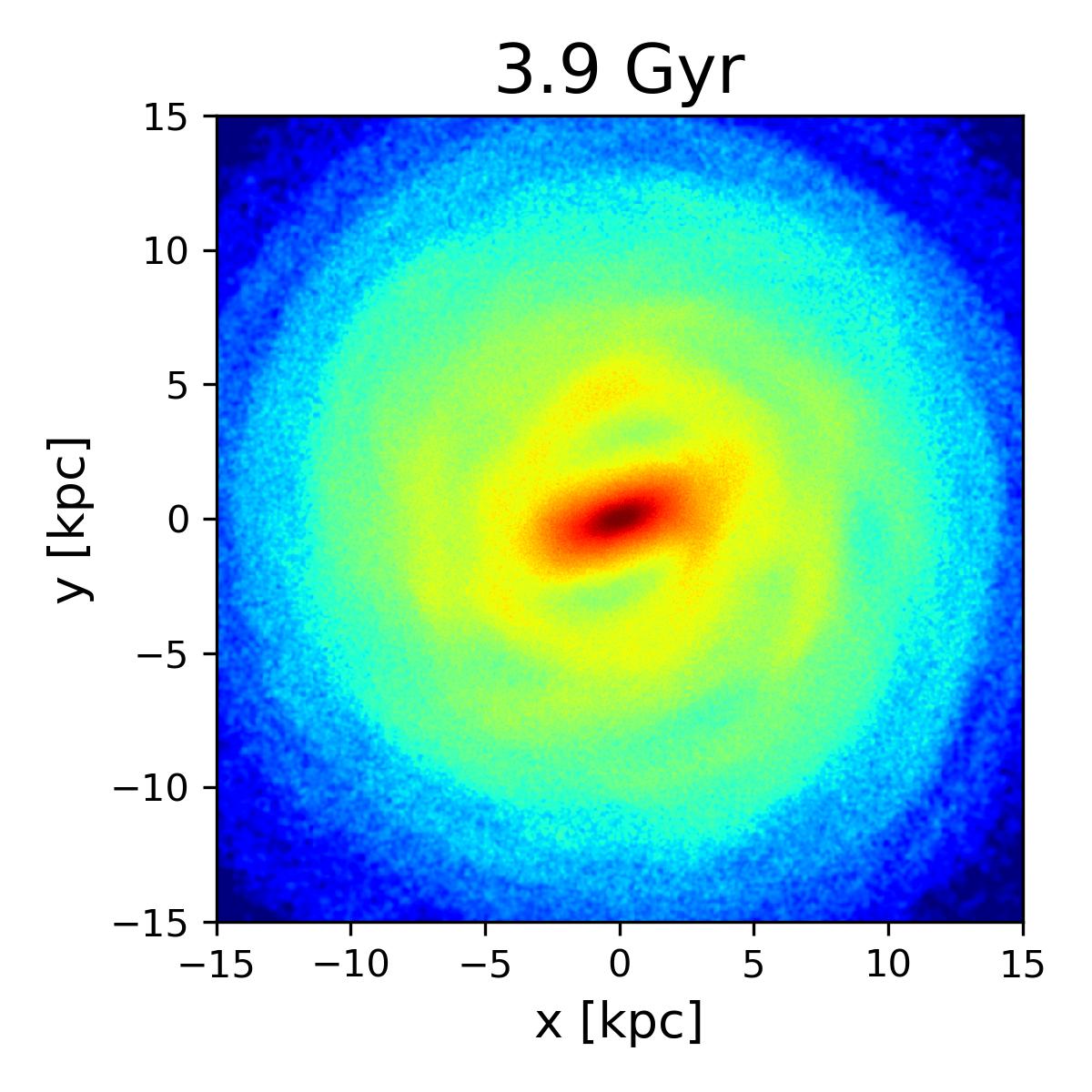} 
    \includegraphics[height=0.245\textwidth]{fig/colorbar2.jpg}
    
    \caption{Face-on projections of the stellar surface density distribution in a $30 \times 30$ kpc box for the r1c16ldm model at 3.1, 3.3, 3.5, 3.7, and 3.9 Gyr. The color bar indicates the surface density in units of solar masses per square kiloparsec, $\Msun\,\mathrm{kpc}^{-2}$.}
    \label{fig:face_modecouple}
\end{figure}

\begin{figure}
    \centering
    \includegraphics[width=0.40\textwidth]{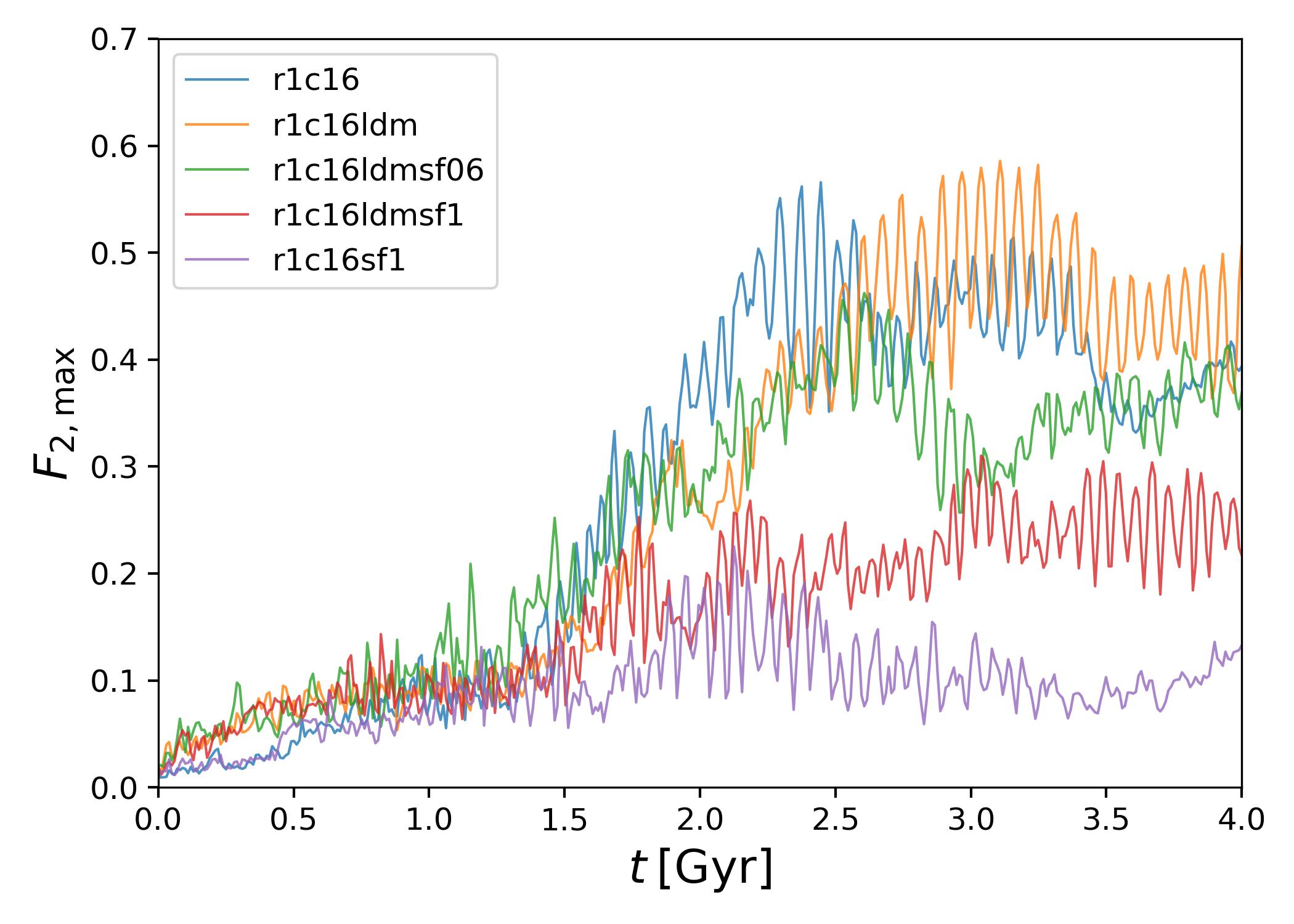}
    \includegraphics[width=0.40\textwidth]{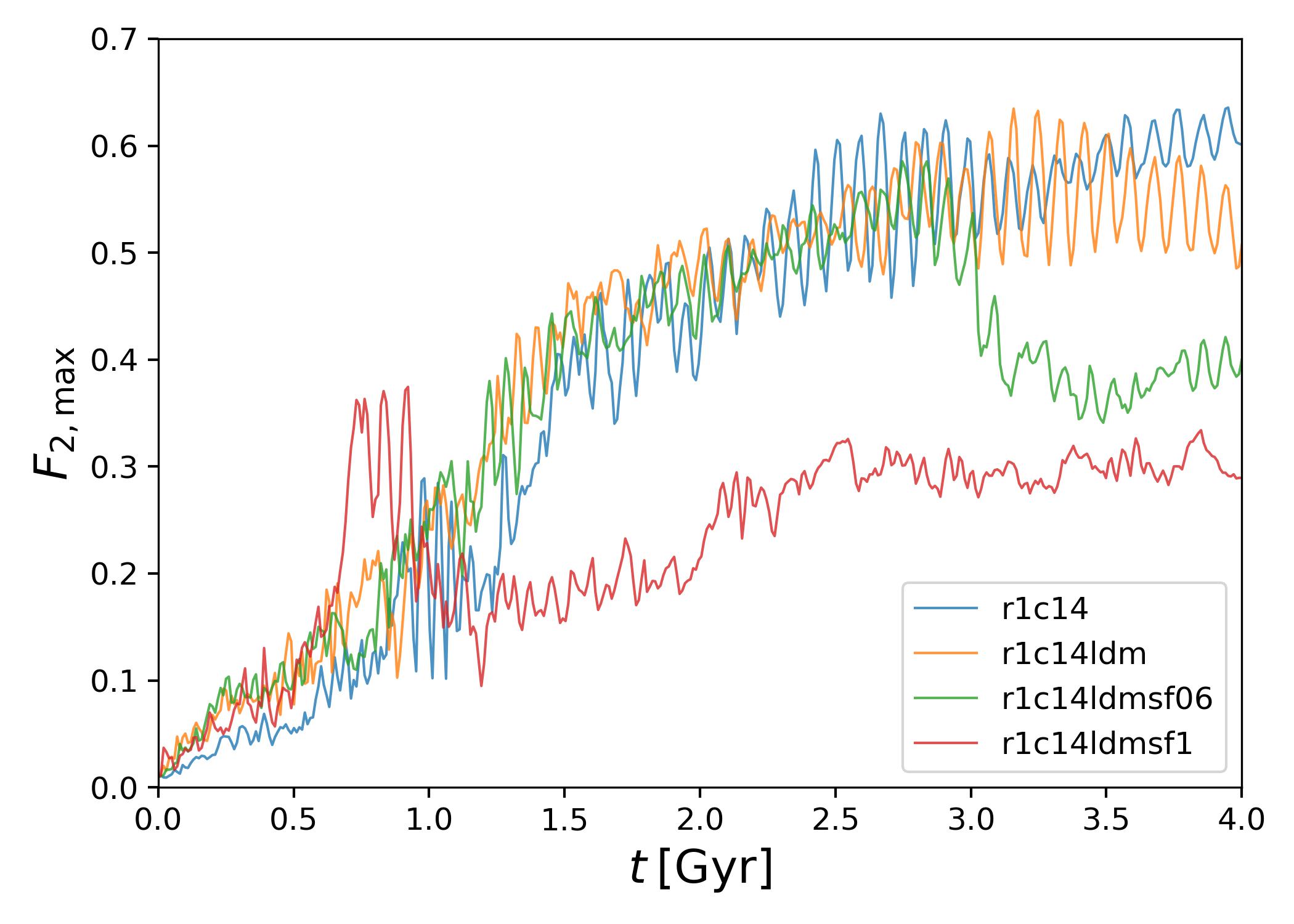}
    \caption{Same as Fig. \ref{fig:f2max} but with all snapshots at 0.01 Gyr shown without smoothing.}
    \label{fig:f2max_raw}
\end{figure}

\begin{figure}[ht]
    \centering
    \includegraphics[width=0.42\textwidth]{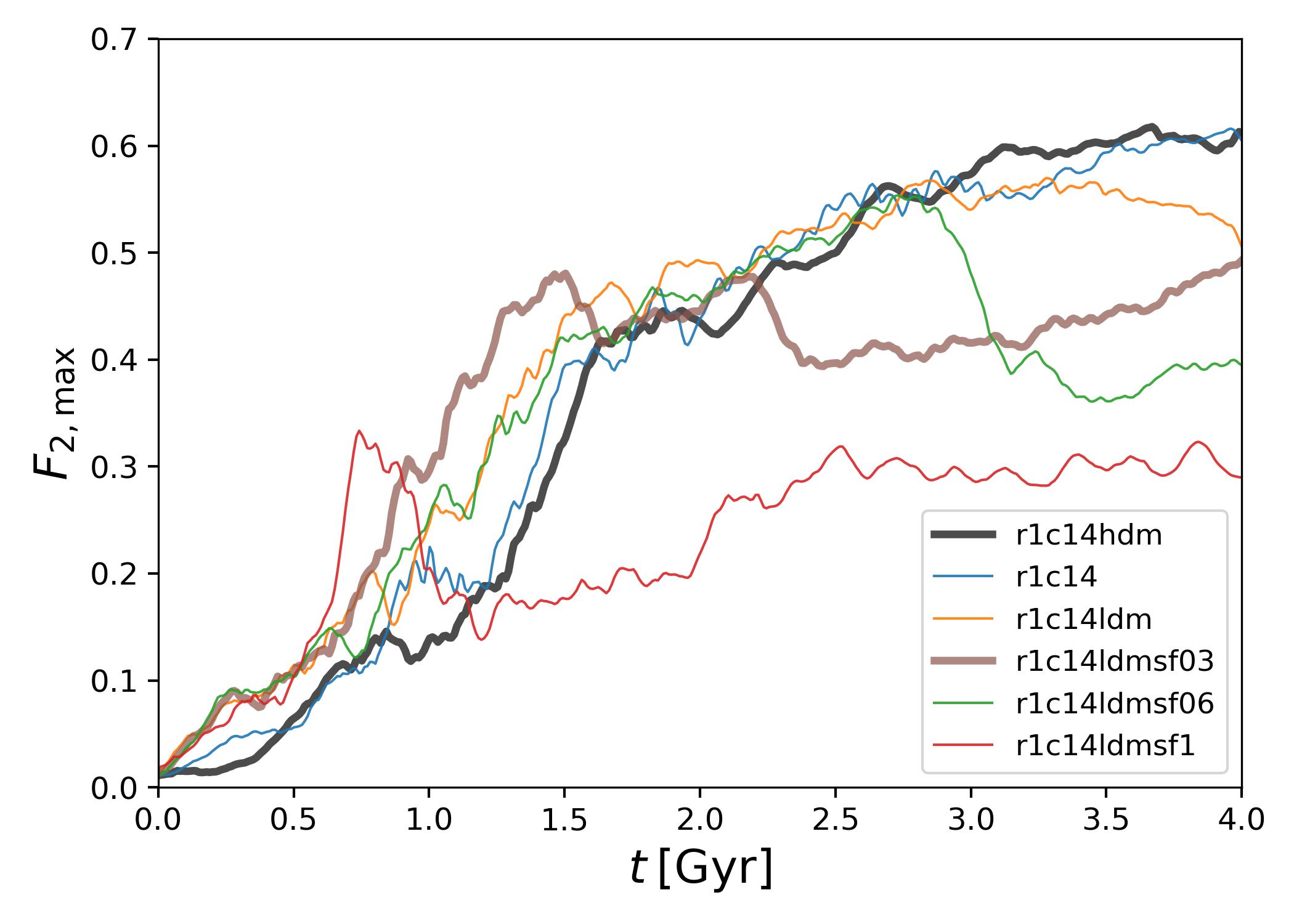}
    \includegraphics[width=0.42\textwidth]{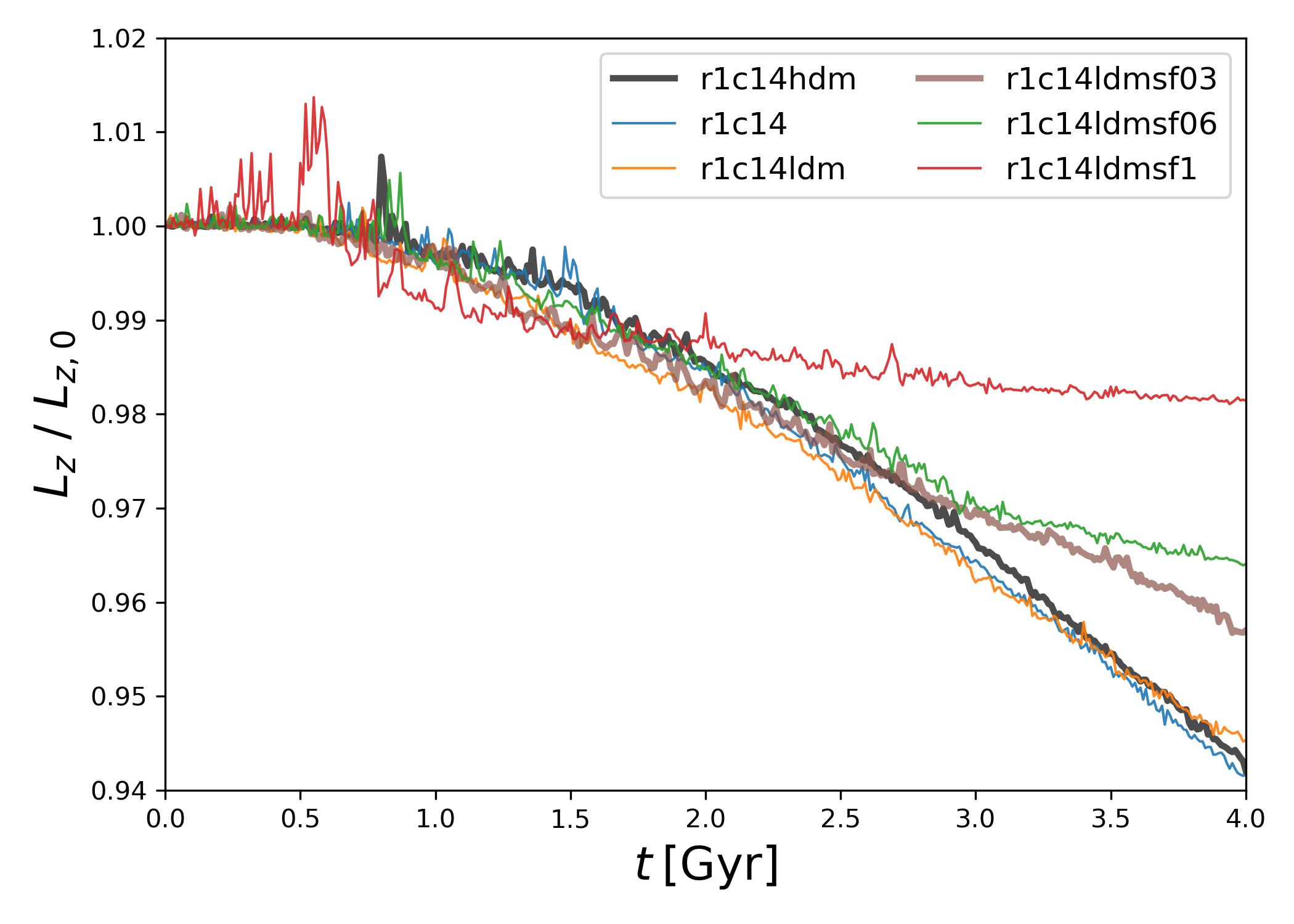}
    \caption{Same as Figs. \ref{fig:f2max} and \ref{fig:lz} but for the "c14" models and including the r1c14hdm ($\mratio=1$) and r1c14ldmsf03 ($\epsdm=0.30 \, \kpc$) models.}
    \label{fig:f2max_lz_appendix}
\end{figure}

\begin{figure}[ht]
    \includegraphics[width=0.37\textwidth]{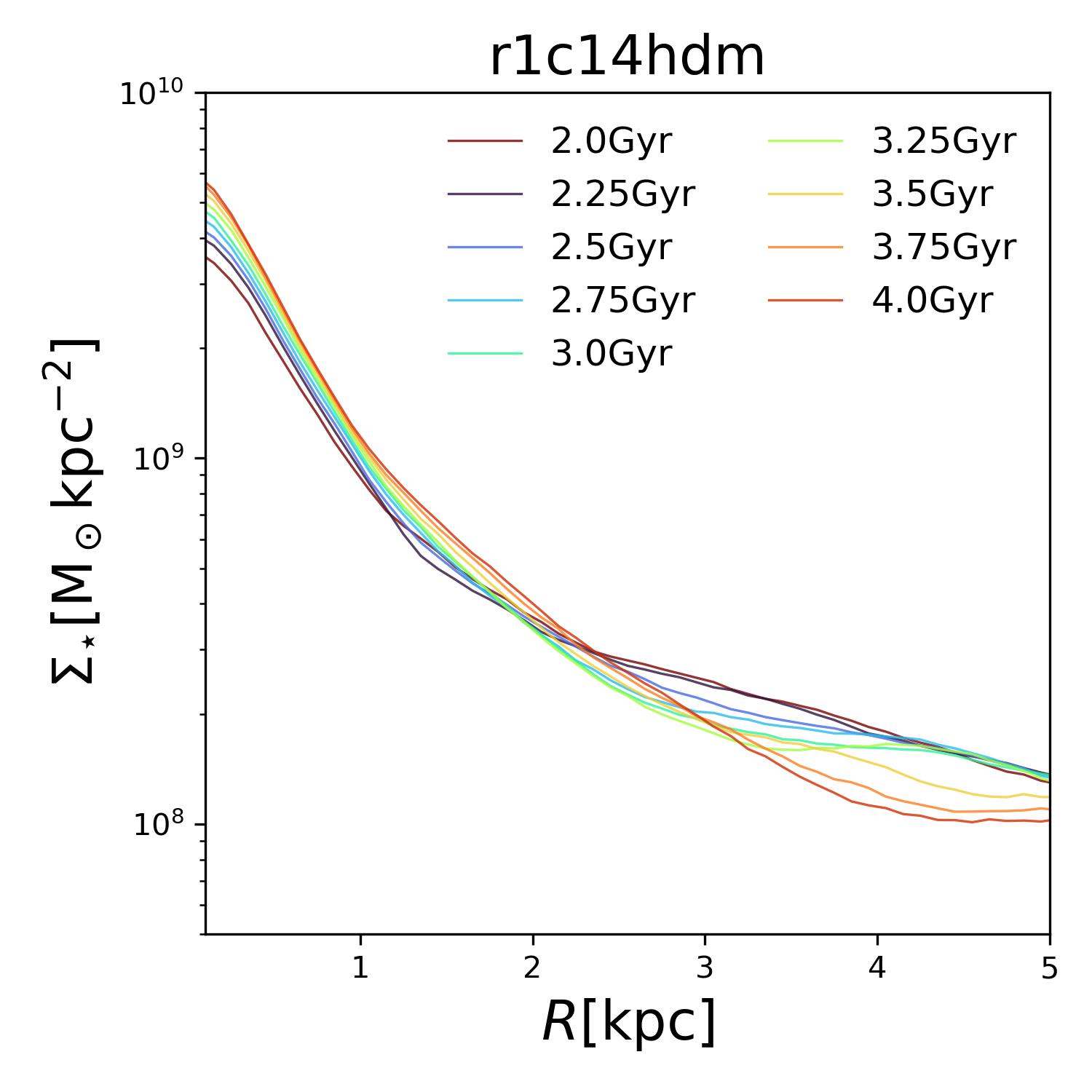}
    \includegraphics[width=0.37\textwidth]{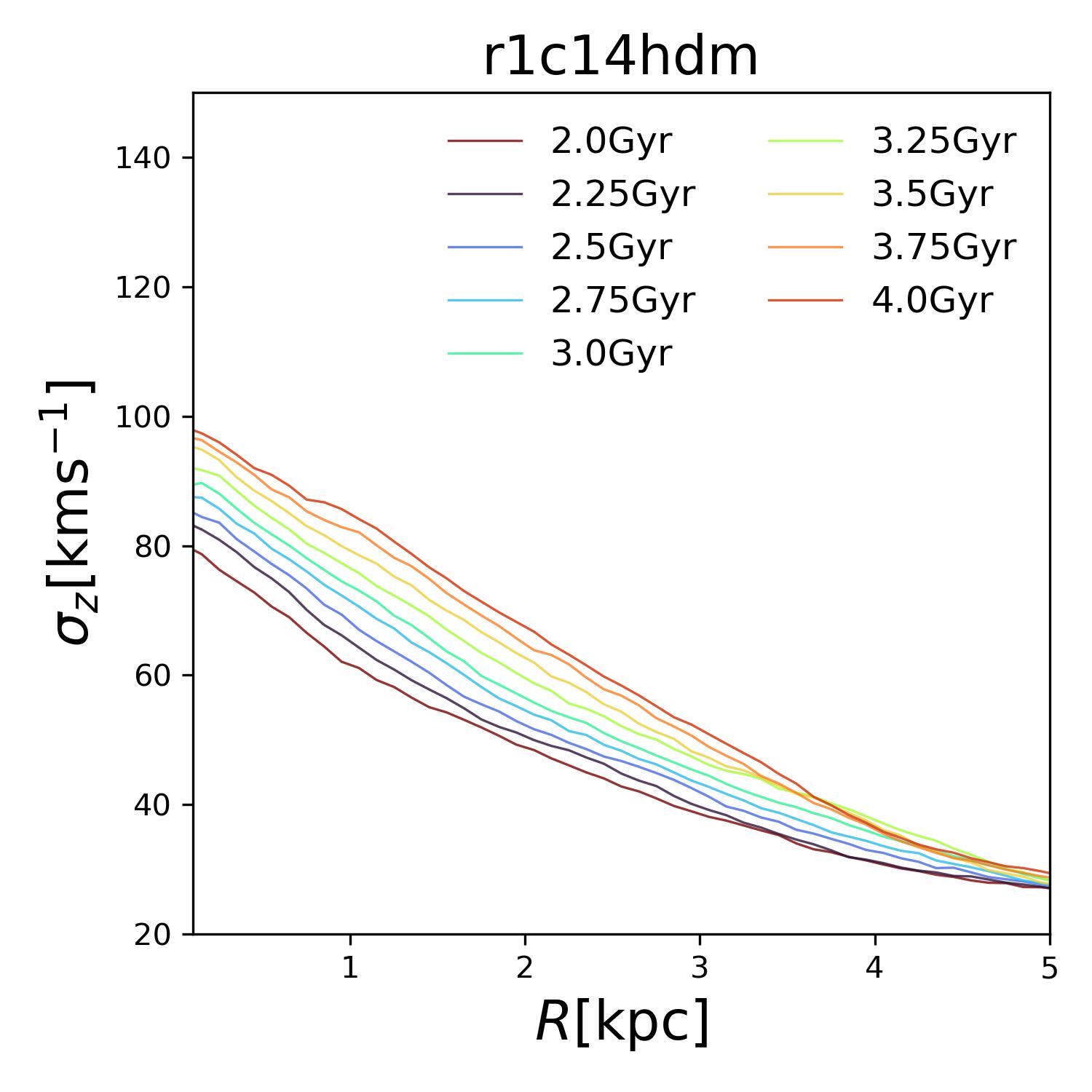}
    \includegraphics[width=0.37\textwidth]{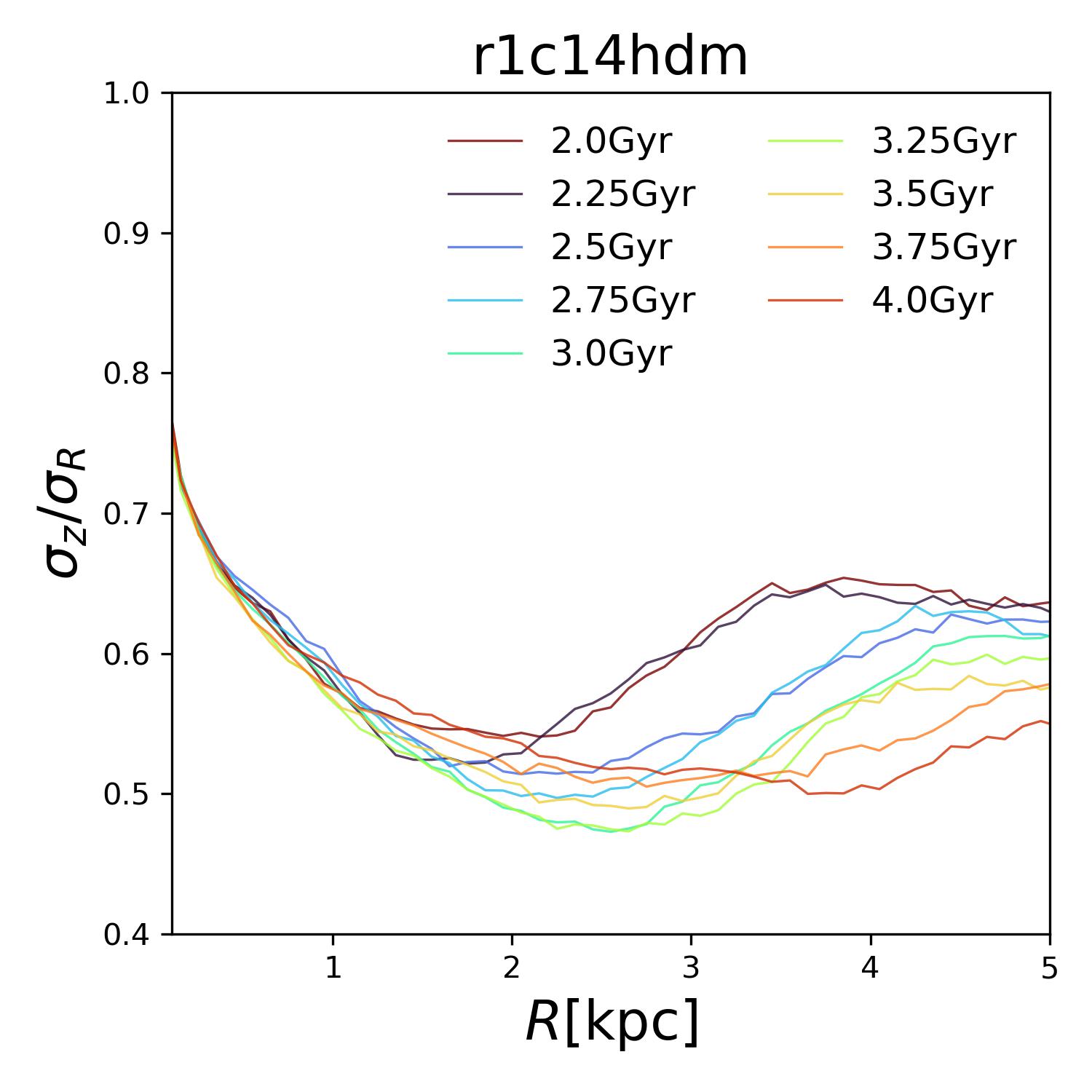}
    
    \caption{Radial profiles of stellar disk properties within $5 \, \kpc$, from $2 \, \Gyr$ to $4 \, \Gyr$, with a time step of $0.25 \, \Gyr$ for the r1c14hdm model. From top to bottom: Stellar density profile, the vertical velocity dispersion of the disk, and the ratio between the vertical and radial velocity dispersions. The corresponding profiles for the r1c14, r1c14ldm, and r1c14ldmsf06 models are shown in Fig.~\ref{fig:sigmarz}.}
    \label{fig:sigmarz_hdm}
\end{figure}

\begin{figure}[ht]

    \centering
    \includegraphics[width=0.25\textwidth]{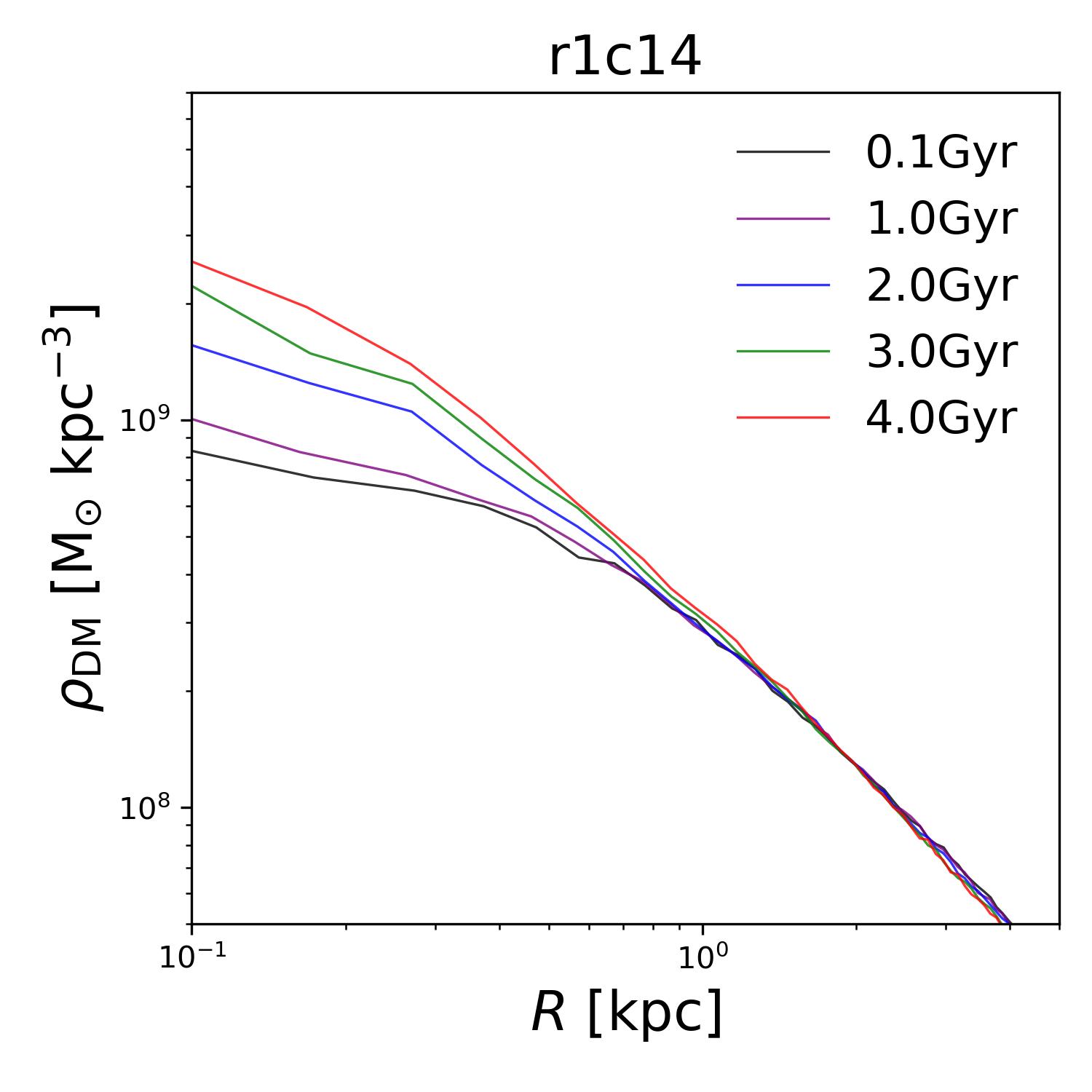}\includegraphics[width=0.25\textwidth]{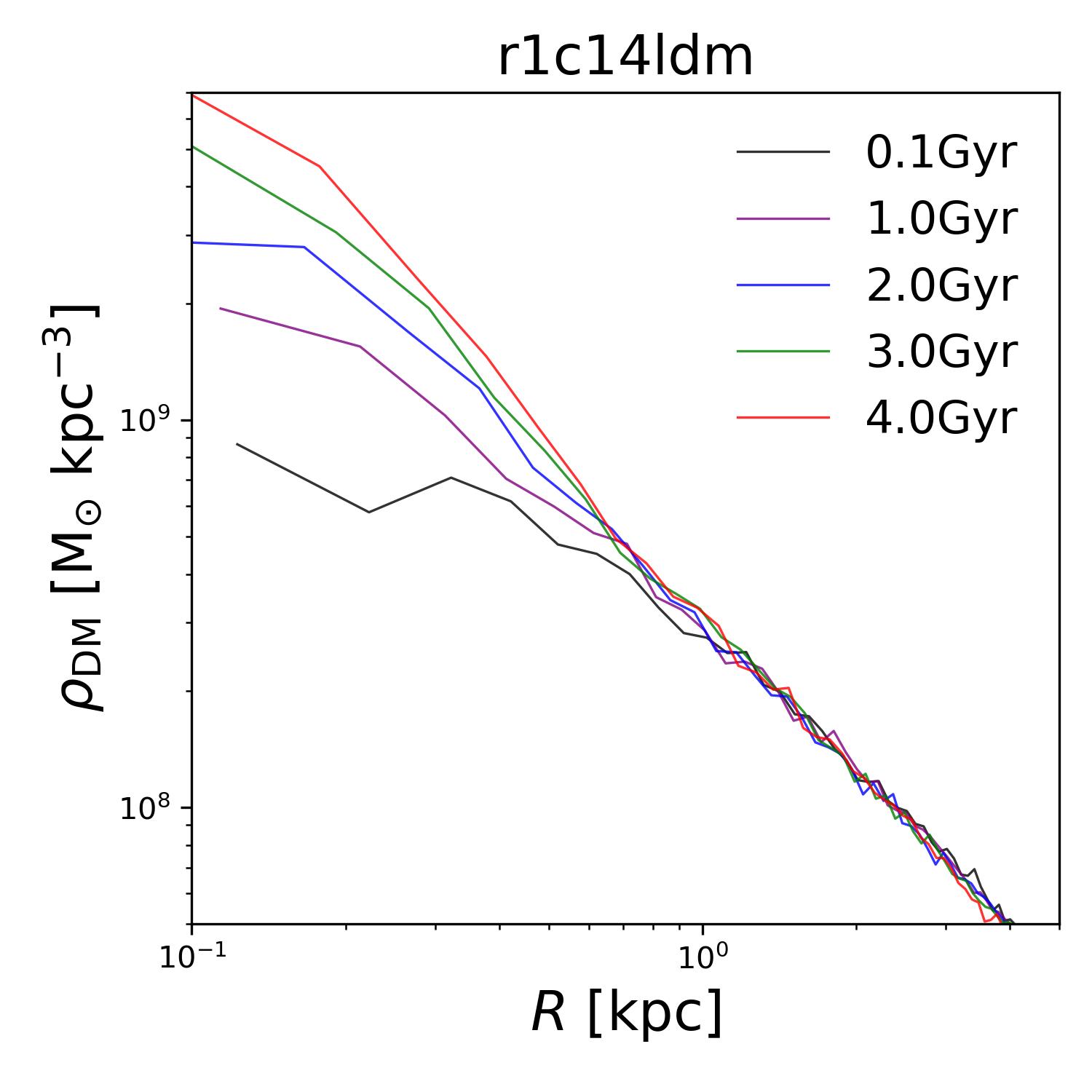}
    
    \includegraphics[width=0.25\textwidth]{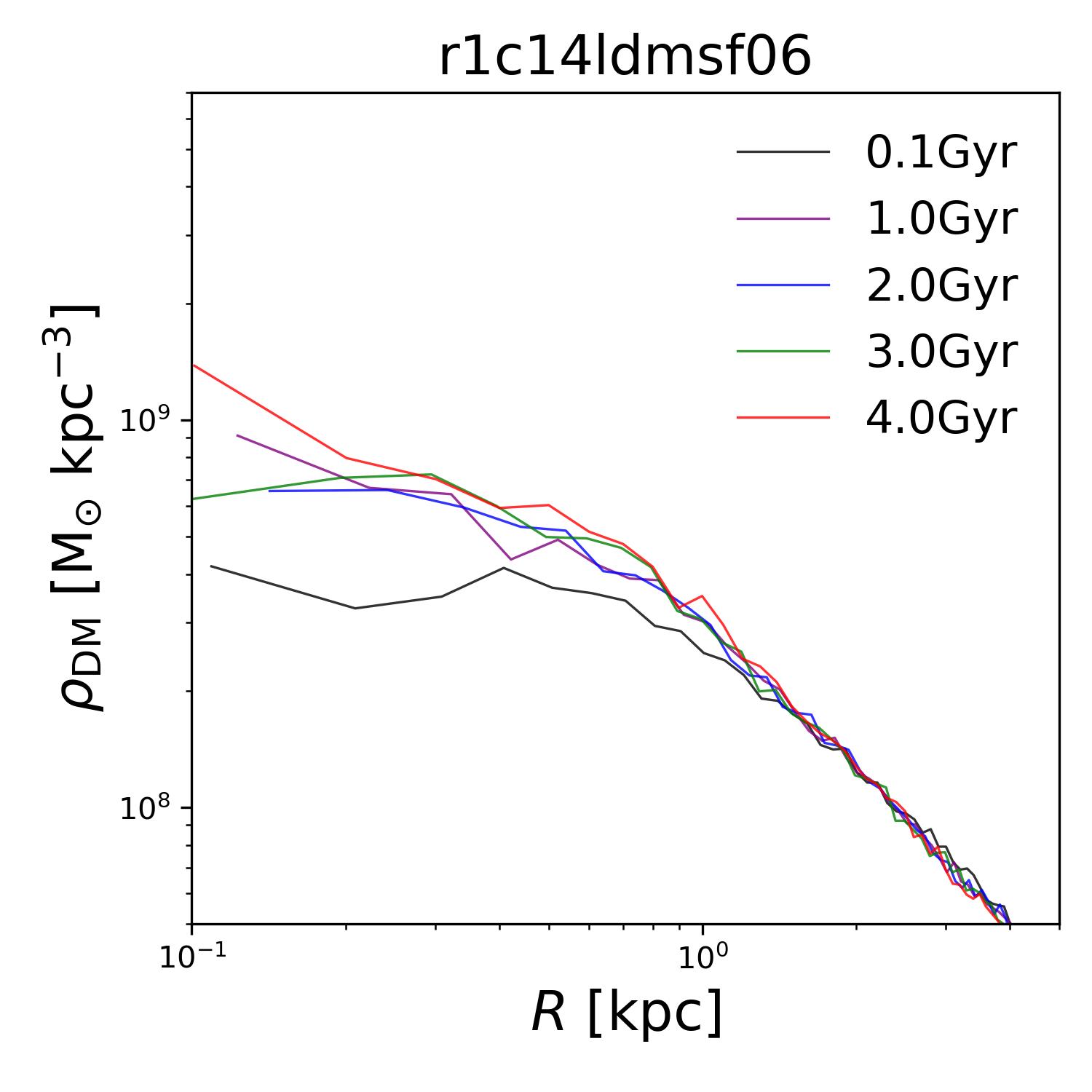}\includegraphics[width=0.25\textwidth]{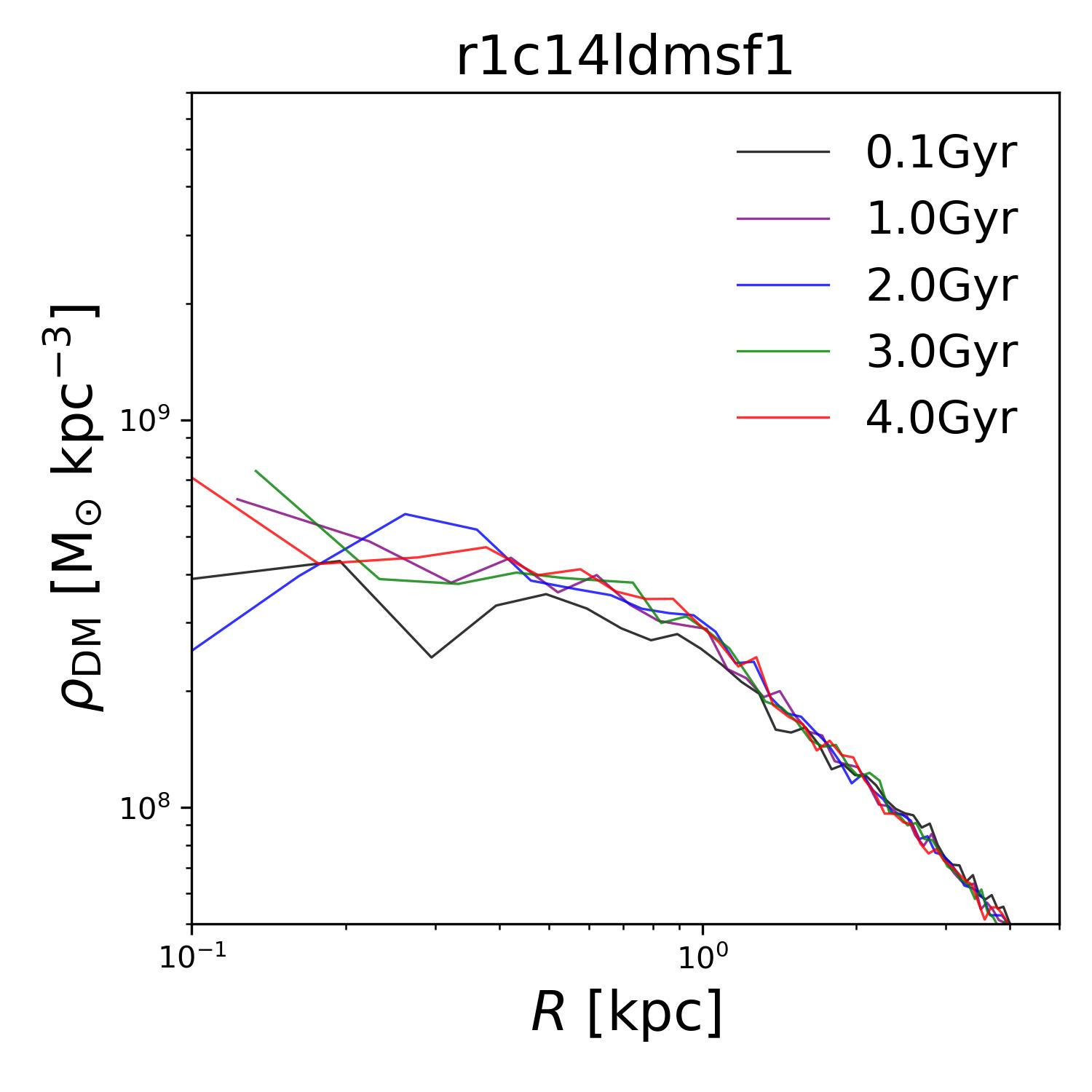}

    \caption{Radial density profiles of the DM halo in the "c14" models from $0.1 \, \Gyr$ to $4 \, \Gyr$, shown between $0.1 \, \kpc$ and $5 \, \kpc$. Both axes are logarithmic. The profiles at different times are color-coded and labeled accordingly.}
    \label{fig:c14density}
\end{figure}

\end{appendix}

\end{document}